\documentclass[a4paper,11pt]{article}
\usepackage{jheppub_mod}
\usepackage[T1]{fontenc}

\usepackage{hyperref}
\usepackage{graphicx}
\usepackage{amsmath,amssymb,mathtools,slashed}
\usepackage{booktabs,tabulary}
\usepackage{dsfont}
\usepackage{color}
\usepackage{multirow}
\usepackage{subcaption}
\usepackage{braket}
\usepackage{adjustbox}
\usepackage{footmisc}

\usepackage{soul}

\newcommand{\cO}{\mathcal{O}}

\newcommand{\mpn}{M_{\pi^0}}
\newcommand{\mpi}{M_{\pi}}
\newcommand{\mpb}{\bar{M}_{\pi}}
\newcommand{\GeV}{\,\text{GeV}}

\newcommand{\chpt}{$\chi$PT\:}
\newcommand{\chptg}{$\chi$PT$_\gamma$}
\newcommand{\Dpi}{\Delta_\pi}

\newcommand{\dpi}{\delta_\pi}
\newcommand{\spin}{s_{00}}
\newcommand{\spi}{s_{+-}}
\newcommand{\spipn}{s_{0+}}

\newcommand{\eps}{\epsilon}

\newcommand{\lbar}{\bar{\ell}}
\newcommand{\IL}{\text{IL}}
\DeclareMathOperator\arctanh{arctanh}

\allowdisplaybreaks[1]

\title{Isospin-breaking in the $\pi \pi$ scattering amplitude I: \\ Effects
  due to the pion mass difference} 
\author[a]{Gilberto Colangelo}
\author[a]{Martina Cottini}
\author[b]{Jacobo Ruiz de Elvira}

\affiliation[a]{Albert Einstein Center for Fundamental Physics, Institute for Theoretical Physics, University of Bern, Sidlerstrasse 5, 3012 Bern, Switzerland}
\affiliation[b]{Departamento de F\'isica Te\'orica and IPARCOS, Facultad de Ciencias F\'isicas, Universidad Complutense de Madrid, Plaza de
  las Ciencias 1, 28040 Madrid, Spain} 

\emailAdd{gilberto@itp.unibe.ch}
\emailAdd{cottini@itp.unibe.ch}
\emailAdd{jacobore@ucm.es}

\abstract{
  This is the first of a series of papers devoted to a detailed analysis of
  isospin-breaking effects in the $\pi \pi$ scattering amplitude and the
  vector form factor of the pion. Isospin breaking originates from the mass difference between up and down quarks and from electromagnetic effects. The latter
  can be further split into effects due to the pion-mass difference and the
  remaining virtual and real photonic effects. In this paper, we derive the
  modifications to the Roy equations for $\pi\pi$ scattering due
  to the mass difference between the charged and the neutral pion. We solve
  the equations numerically after matching them to the chiral
  representation of the $\pi \pi$ scattering amplitude evaluated in the
  same approximation, which is also provided. Numerical results are
  presented and discussed in detail.
}

\begin{document}

\preprint{IPARCOS-UCM-25-059}

\maketitle


\section{Introduction}
The $\pi\pi$ scattering amplitude is understood at a remarkable level of accuracy, particularly at low energy and in the isospin limit~\cite{Ananthanarayan:2000ht,Colangelo:2001df,Caprini:2011ky,GarciaMartin:2011cn,Pelaez:2024uav} (for a short and
beautiful account of the history of $\pi\pi$ scattering, see~\cite{Gasser:2009zz}).  Experimental measurements have provided, on the one hand, necessary input to theoretical calculations~\cite{Hyams:1973zf} and, on the other, have confirmed to the same level of precision the predictions for the $S$-wave scattering lengths~\cite{BNL-E865:2001wfj,NA482:2005wht,NA482:2007xvj,Adeva:2011tc}.
Whenever experimental data are used, either as input or for comparison with theoretical calculations done in the isospin limit, it becomes necessary to
remove from the data isospin-breaking effects. The part that is related to
the emission of real photons (and a minimal part of virtual ones, as needed
to remove infrared divergences) can only be dealt with during data
analysis since it is sensitive to detector geometry and efficiencies. Many
experiments nowadays rely on the use of general-purpose codes like
PHOTOS~\cite{Barberio:1993qi,Nanava:2006vv} which, however, implement
formulae, which are only valid in the soft-photon limit, and treat hadrons as
structureless objects. This should capture the bulk of photon-emission effects, but if high precision is required, this simple treatment is insufficient. 

However, two further isospin-breaking effects should be considered: one is due to the mass difference between up and down quarks, and the other to the mass difference between charged and neutral pions. The latter actually belongs to the broad class of effects generated by virtual photon exchanges, but it is well-defined on its own and can be treated separately from all the others. It turned out that these effects were essential for a precise extraction of the $\pi \pi$ phases from $K_{e4}$ decays~\cite{Colangelo:2008sm,Bernard:2013faa,Bernard:2015vqa}. 

In a series of papers, the first of which is this one, we set ourselves the goal of analyzing these three isospin-breaking effects and obtaining a reliable, model-independent estimate of their size in the region below 1 GeV. The approach we will adopt to achieve this goal is that of dispersion relations, but with an approximation: we will consider intermediate states only up to two pions. The contribution of three or more pions or heavier states ($K\bar{K}$, $\eta \eta$, etc.) with or without a photon to isospin-breaking effects will simply be neglected. A detailed justification of this approximation and an estimate of the neglected effects will be provided in the relevant paper of this series. The main motivation for this analysis is to provide input for a dispersive treatment of the same
isospin-breaking effects for the vector form factor of the pion. A
measurement of this quantity (in particular via the measurement of the
cross-section $e^+e^- \to \pi^+ \pi^-$) provides essential input for the calculation of the hadronic vacuum polarization contribution to the anomalous magnetic moment of the muon
($g-2$). Dealing with isospin-breaking effects in this reaction, especially of final-state radiation, has so far been based on models, in particular on scalar QED, with form factor effects taken into account in an ad hoc manner. A dispersive approach, with the same approximation discussed above, can do better and provides a model-independent estimate of these effects, but it requires the $\pi\pi$ scattering amplitude, including isospin-breaking effects as input.

The present paper is devoted to the study of the effects of the pion-mass difference in the dispersive analysis of the $\pi \pi$ scattering amplitude. In a second one, we will address the (more involved) analysis of real and virtual photonic corrections to the $\pi \pi$ scattering amplitude, excluding from the latter those that contribute to the pion mass difference. A third will apply the same approach to the vector form factor of the pion, for which the $\pi \pi$ scattering amplitude analyzed in the first two will be a necessary input. 
Effects due to the quark-mass difference $m_u-m_d$ will be essentially neglected, based on the following argument.
Strong interactions are flavor blind. The only reason why different quarks
behave differently in QCD is that they have unequal masses. The up-and-down quark masses happen to be much smaller than the QCD scale, so they behave as if they were almost massless. Moreover, their mass
difference, though large in relative terms, is still very small with respect to hadronic scales and is usually neglected: in this limit, the
isospin symmetry emerges. The fact that the quark mass difference is not exactly zero is a source of isospin breaking and needs to be taken into account.
At low energy, one can rely on Chiral Perturbation Theory (\chpt) to estimate the effects of the up and
down quark mass difference. As discussed by Gasser and Leutwyler~\cite{Gasser:1983yg}, in the $SU(2)$ chiral limit the first
effect due to $m_u-m_d$ at low energy is quadratic, and therefore of
$\cO(p^4)$. Thus, it amounts to a tiny shift in the neutral pion mass. If one
considers the $SU(3)$ chiral limit there is also the effect of $\pi^0-\eta$
mixing~\cite{Gasser:1984gg}, as well as isospin-breaking four-meson
vertices, like $\eta-3\pi$, which are both linear in $m_u-m_d$. An effect
in the $\pi \pi$ scattering amplitude is, however, of $\cO((m_u-m_d)^2)$ and thus beyond the accuracy we aim at here.

Another known and non-negligible effect proportional to $m_u-m_d$ is $\rho-\omega$ mixing, which shows up in the isospin $I=1$ $P$-wave. This effect is not accessible in \chpt because it concerns degrees of freedom, which are treated as heavy in this framework. This raises the question of what theoretical tool can be used to estimate effects proportional to $m_u-m_d$ outside the domain of validity of \chpt. Such effects will, in general, be
suppressed by $(m_u-m_d)/E$, with $E$ an arbitrary energy scale satisfying $E \gtrsim \Lambda$, with $\Lambda$ the chiral symmetry breaking scale. Thus, it can, in general, be neglected. An exception, such as $\rho-\omega$ mixing, arises because the denominator can become small---of the order $m_\rho-m_\omega$ in this case---leading to an enhancement of the effect. As a result, within a very narrow energy region, this otherwise small correction can become non-negligible. In conclusion, the suppression of strong isospin-breaking effects at intermediate or high energies can be estimated on the basis of a dimensional argument, but care has to be taken in identifying possible small energy denominators that might enhance the effect.

The structure of the paper is as follows. In Sect.~\ref{sec:QCD+QED}, we discuss the definition of the separation between QED and QCD effects in $\pi\pi$ scattering.
In Sect.~\ref{sec:pion-mass-effects}, we derive a dispersive representation of pion-pion scattering that incorporates the mass difference between neutral and charged pions. Sect.~\ref{sec:chpt} focuses on studying the same dispersive representation within \chptg, which, in turn, provides a numerical estimate of the subtraction constants entering into our formalism.
In Sect.~\ref{sec:roy-beyond}, we detail the derivation of Roy equations when the pion-mass difference is considered, while Sect.~\ref{sec:strategy} outlines our strategy for obtaining a numerical solution. Our final results are discussed in Sect.~\ref{sec:results}, and we provide our conclusions in Sect.~\ref{sec:conclusions}.

\section{\texorpdfstring{The $\pi \pi$ scattering amplitude in QCD$+$QED}{The pipi scattering amplitude in QCD+QED}}\label{sec:QCD+QED}

The main source of isospin breaking for the $\pi \pi$ scattering amplitude is represented by photonic radiative corrections, which can be split into real and virtual ones. As is well known, even if one measures a process in which external photons are not seen by the detector, the finite energy resolution of the latter implies that the measured process has to
involve real photons too, though with an upper cut-off on their energy.
This cut-off corresponds to the minimal energy of the photons seen by the
detector involved. It is also well known that, individually, both the cross-section including virtual photon exchanges as well as that involving real
photon emission are infrared divergent, whereas their sum is finite---for any value of the energy cut-off. Physical results are only obtained as a sum of both contributions. As soon as isospin is broken, it is not meaningful to work with amplitudes of definite isospin. So, if we aim to perform an analysis of these effects, it is necessary to switch to a charge basis to describe all possible channels of the $\pi \pi$ scattering amplitude, which are the
following:
\begin{eqnarray}
T^{c}(s,t,u)&:=&T(\pi^+ \pi^- \rightarrow \pi^+ \pi^-)\;,\nonumber\\ 
T^{x}(s,t,u)&:=&T(\pi^+ \pi^- \rightarrow \pi^0 \pi^0)\;,\nonumber \\ 
T^{n}(s,t,u)&:=&T(\pi^0 \pi^0 \rightarrow \pi^0 \pi^0)\;, 
\label{eq:chargedT1}
\end{eqnarray}
with the usual definition of the Mandelstam variables $s$, $t$ and $u$.
There are two further amplitudes, which are, however, related by crossing
transformations to the previous ones, even in the presence of isospin
breaking:
\begin{eqnarray}
T^{++}(s,t,u):=T(\pi^+ \pi^+ \rightarrow \pi^+ \pi^+)&\!\!=\!\!&
T^c(u,t,s)\;,\nonumber \\  
T^{+0}(s,t,u):=T(\pi^+ \pi^0 \rightarrow \pi^+ \pi^0)&\!\!=\!\!&
T^x(t,u,s) \; .
\label{eq:chargedT2}
\end{eqnarray}
We will denote the amplitudes in the charged basis collectively as
$T^k(s,t,u)$, with $k=c,\,x,\,n,\,++,\,+0$, whereas the amplitudes in fixed isospin channels are denoted as $T^I(s,t,u)$, with $I=0,1,2$. For each of the five amplitudes $T^k$, we also define the corresponding ones with real photon emission as:
\begin{equation}
T^c_\gamma:=T(\pi^+ \pi^- \rightarrow \pi^+ \pi^- \gamma) \; ,
\end{equation}
and analogously for all the other $T^k$.

When discussing these amplitudes within the framework of QCD+QED at low
energy, we should not apply any perturbative expansion for QCD (other than the chiral one at low energy, where needed). Still, we can use perturbation theory for QED effects and expand all amplitudes in the electromagnetic coupling $\alpha$. Defining such an expansion is not completely trivial since the separation of QED effects from pure QCD ones is not unambiguous~\cite{Gasser:2003hk}. This also implies that the definition of the order zero term is not unique. In addition to the issue of principle, there is also a matter of convenience; since QED corrections also affect hadron masses, if one were to adopt the pure QCD limit of the QCD+QED theory (assuming for a moment that this can be unambiguously defined) as the order zero in the expansion $\alpha$, one would have the consequence that the kinematics of the process would change after considering effects of $\cO(\alpha)$. We will therefore define the term $\cO(\alpha^0)$ as the one where all hadrons have their physical masses in the full theory (QCD+QED), but all other effects of $\cO(\alpha)$ are switched off. So, for example, we write the $\pi \pi$ scattering amplitudes as
\begin{equation}
T^k(s,t,u)=T^k_0(s,t,u)+ \alpha T^k_1(s,t,u)+\cO(\alpha^2)\;,
\end{equation}
where $T^k_0(s,t,u)$ is the amplitude with physical pion masses but no
other effects due to photon exchanges.

The ambiguity in the definition of the theory around which the expansion is performed does not affect the amplitudes $T^k_\gamma$, since these concern a process that can only happen in the presence of QED, for $\alpha\neq 0$, and we are interested in calculating these only to leading order in $\alpha$, 
\begin{equation}
T^k_\gamma=e \left[ T^k_{\gamma,0}+ \cO(\alpha) \right] \; . 
\end{equation}

\section{\texorpdfstring{Effects due to $M_{\pi^+}-\mpn$}{Effects due to mpi-mpi0}}\label{sec:pion-mass-effects}

In the chiral expansion, taking into account the effects due to the pion mass difference is straightforward, but doing this for the $\pi \pi$ interaction in a dispersive framework that extends beyond the chiral regime is significantly more involved. In $K_{e4}$ decays the allowed range
for the di-pion mass in the final state is limited by the kaon mass, so
that it is possible to rely on \chpt (or on a non-relativistic expansion) to
estimate the corresponding isospin-breaking
effects~\cite{Colangelo:2008sm,Bernard:2013faa,Bernard:2015vqa}. The problem was still nontrivial because if one wants to extract the $\pi \pi$
phase shift from the measurement, one has to be able to calculate the
isospin-breaking corrections without making assumptions, even indirectly,
on the phase shift.

Another process in which such a difference matters and has been estimated
is $\eta \to 3 \pi$~\cite{Colangelo:2018jxw}. This decay allows in
principle, a precision determination of the up and down quark mass
difference~\cite{Colangelo:2016jmc}, but to do so, one needs to
accurately evaluate strong-interaction effects beyond \chpt, i.e., in a
dispersive approach, and also take into account isospin-breaking
effects. The dispersive treatment of final-state interactions is in this
case particularly
involved~\cite{Khuri:1960zz,Kambor:1995yc,Anisovich:1996tx,Gasser:2018qtg}
and has been performed in the isospin limit. Extending the treatment to take into account the pion-mass difference would be highly nontrivial.
Instead, a simplified approach has been adopted in~\cite{Colangelo:2018jxw}, which essentially consists of making a map of kinematic variables between the isospin-symmetric and real-world cases, thereby allowing one to use the isospin-symmetric amplitude for real-world kinematics. 

In view of the precision requirements for the muon $g-2$, we will aim here at a full dispersive treatment of these effects, with the approach, which we are going to spell out in the rest of this section. The kinematic map introduced in~\cite{Colangelo:2018jxw} will also be used here, but only as a parameterization of the isospin-breaking effects, which we are aiming to determine dispersively here.

\subsection{\texorpdfstring{Dispersive representation of the $\pi \pi$ scattering amplitude
in the isospin limit}{Dispersive representation of the pipi scattering amplitude in the isospin limit}}\label{sec:disp_rep}

We start from a representation of the isospin-invariant $\pi \pi$
scattering amplitude in which the contributions generated by the imaginary
parts of the $S$- and $P$-waves below a certain energy $\sqrt{s_2}= 2$ GeV
is separated from the rest:
\begin{equation}
\label{eq:ASPd}
A(s,t,u) = A(s,t,u)_{SP}+A(s,t,u)_d \; .
\end{equation}
At energies below the upper limit of validity of the Roy equations, $s_1=68
\mpi^2$, the background amplitude $A(s,t,u)_d$, which collects the contribution of the imaginary parts of higher waves as well as of high energies from $S$- and $P$-waves, can be well described by a polynomial.
Given the nature of this part of the amplitude, we expect it to be
little sensitive to the pion mass difference. This will be borne out by our result: although a sensitivity to this mass difference is visible, its size is small compared to the uncertainties with which the background amplitude is currently known.

The amplitude $A(s,t,u)_{SP}$ can be expressed in terms of three functions
of a single variable:
\begin{eqnarray}
\label{eq:ASP} 
A(s,t,u)_{SP} 
&=& 32\pi\left\{ \frac{1}{3}W^0(s)+
\frac{3}{2}(s-u)W^1(t)
+\frac{3}{2}(s-t)W^1(u)\right. \nonumber \\
&& \qquad \left.+\frac{1}{2} W^2(t)+ \frac{1}{2} W^2(u)
-\frac{1}{3} W^2(s) \right\}\;.
\end{eqnarray}
The functions $W^0(s)$, $W^1(s)$, $W^2(s)$ have only a right-hand cut and
admit a simple dispersive representation in terms of the imaginary parts of
the $S$- and $P$-waves as well as by the two $S$-wave scattering lengths
$a_0^0,\,a_0^2$: 
\begin{eqnarray}
\label{eq:W} 
W^0(s)&=& \frac{a_0^0\, s}{4\mpi^2} +
\frac{s(s-4\mpi^2)}{\pi}\int_{4\mpi^2}^{s_2}
ds'\;\frac{\mathrm{Im}\, t_0^0(s')}{s'(s'-4\mpi^2)(s'-s)} \; , \nonumber \\
      W^1(s) &=& \frac{s}{\pi}\int_{4\mpi^2}^{s_2}
ds'\;\frac{\mathrm{Im}\, t_1^1(s')}{s'(s'-4\mpi^2)(s'-s)} \; , \\
       W^2(s)&=& \frac{a_0^2\, s}{4\mpi^2}  +
\frac{s(s-4\mpi^2)}{\pi}\int_{4\mpi^2}^{s_2}
ds'\;\frac{\mathrm{Im}\, t_0^2(s')}{s'(s'-4\mpi^2)(s'-s)} \; .  \nonumber
\end{eqnarray}

This representation is, of course, valid in the isospin limit, and it is in this limit that the imaginary parts have been determined as a solution of the Roy equations below $s_1$, supplemented by phenomenological input between $s_1$ and $s_2$. In the following, we will redo the Roy-equation analysis, allowing for a mass difference between the neutral and the charged pions. However, this will require some changes to the formulation of the Roy equations, as we will specify in the next subsection.

We first recall the definition of the isospin amplitudes in terms of the
isospin-invariant amplitude $A(s,t,u)$:
\begin{align}
\label{eq:AtoTI}
T^0(s,t,u)&=3 A(s,t,u)+A(t,u,s)+A(u,s,t)\;,\nonumber \\
T^1(s,t,u)&=\hskip 2 cm  A(t,u,s)-A(u,s,t)\;, \nonumber \\
T^2(s,t,u)&=\hskip 2 cm A(t,u,s)+A(u,s,t)\;.
\end{align}
We then need to switch from the isospin basis to the charge basis, and
define the scattering amplitudes accordingly. We do so in the isospin limit,
where the relation between the amplitudes in the two bases reads as
follows:
\begin{eqnarray}
T^{c}(s,t,u)&\!\!=\!\!& \frac{1}{3}T^0(s,t,u) + \frac{1}{2}T^1(s,t,u) +
\frac{1}{6}T^2(s,t,u)\;,\nonumber\\  
T^{x}(s,t,u)&\!\!=\!\!& \frac{1}{3}T^0(s,t,u)-\frac{1}{3}T^2(s,t,u)\; ,\nonumber \\ 
T^{n}(s,t,u)&\!\!=\!\!& \frac{1}{3}T^0(s,t,u)+\frac{2}{3}T^2(s,t,u)\; . 
\label{eq:isospinT}
\end{eqnarray}
The two further amplitudes $T^{++}(s,t,u)$ and $T^{+0}(s,t,u)$, which are related by the crossing transformations in~\eqref{eq:chargedT2} to the previous ones, will be discussed too, sometimes in alternative to those in the direct channels, depending on convenience.

Inserting~\eqref{eq:AtoTI} into~\eqref{eq:isospinT}, one trivially obtains the expression of each of these amplitudes $T^k$, $k=c,x,n$, in terms of the
isospin-invariant amplitude $A$,
\begin{eqnarray}
T^{c}(s,t,u) &\!\!=\!\!& A(s,t,u)+A(t,u,s)\;,\nonumber \\ 
T^{x}(s,t,u)&\!\!=\!\!& A(s,t,u)\;,         \nonumber \\ 
T^{n}(s,t,u) &\!\!=\!\!& A(s,t,u)+A(t,u,s)+A(u,s,t) \; , 
\label{eq:AtoTi}
\end{eqnarray}
thereby inheriting from~\eqref{eq:ASPd} the splitting
between the $SP$ and the background amplitude, 
\begin{eqnarray}
T^{k}(s,t,u) = T^{k}_{SP}(s,t,u)  + T^{k}_{d}(s,t,u) \; .
\end{eqnarray}
As argued above, only the amplitude $T^{k}_{SP}$ is expected to be sensitive to the pion mass difference, and we will concentrate on this in the following. 

\subsection{Effects due to the pion mass difference}\label{sec:single-variable_decomp}
The Roy equations have been solved in the isospin limit. The definition of the latter is arbitrary, however, and in the choice adopted in~\cite{Ananthanarayan:2000ht} all pion masses are set equal to the physical charged pion mass $\mpi=139.57039$ MeV.\footnote{In~\cite{Ananthanarayan:2000ht} the old PDG value of $\mpi=139.57018$ MeV was used, but the difference is irrelevant for the present analysis.} Away from the isospin limit, the Roy equations take a very different form since different processes get contributions from different intermediate states in each channel. Even when considering only $S$- and $P$-waves, instead of coupled integral equations for three different partial waves, one ends up with coupled integral equations for seven different partial waves (of the five different channels in~\eqref{eq:chargedT1} and~\eqref{eq:chargedT2}, only $T^c$ and $T^{+0}$ admit a $P$-wave). 

To estimate the effect in the region below $s_1$, we proceed as follows.
First, we concentrate only on the $T^k_{SP}$ amplitudes and insert the
expression~\eqref{eq:ASP} of $A_{SP}$ in terms of single-variable functions. This provides a dispersive representation of the physical amplitudes, but still in the isospin limit. As a next step, we express the imaginary parts of the fixed-isospin partial waves appearing in the definition of the $W^I$ functions~\eqref{eq:W} in terms of the imaginary parts of the physical channels and correspondingly define new
single-variable functions labeled accordingly. In the following, we detail this procedure for each amplitude. 

\paragraph{The $\pi^0 \pi^0 \to \pi^0 \pi^0$ amplitude:}
First, let us consider the $T^n$ amplitude, which is completely crossing-symmetric in the three Mandelstam variables, i.e., in each channel, the same intermediate states, $\pi^0 \pi^0$ or $\pi^+ \pi^-$, are possible and with the same weight. Doing the steps outlined above, one ends up with the following representation:
\begin{eqnarray}\label{eq:neutral_decomp}
T^n_{SP}(s,t,u) = 32 \pi \left( W_{S}^{n,00}(s) + W_{S}^{n,+-}(s)+(s
  \leftrightarrow t) + (s \leftrightarrow u) \right)\;,
\end{eqnarray}
with
\begin{eqnarray}
W_{S}^{n,00}(s) &=& \frac{a_n^{00} \, s}{4 \mpn^2} +
\frac{s (s-4\mpn^2)}{\pi} \int_{4 \mpn^2}^{s_1} \, ds'
\frac{\text{Im}\,t_{S}^{n,00}(s')}{s'(s'-4\mpn^2)(s'-s)}\;, \nonumber \\
W_{S}^{n,+-}(s) &=& \frac{s (s-4\mpn^2)}{\pi} \int_{4 M_{\pi}^2}^{s_1}
\, ds' \frac{\text{Im}\,t_{S}^{n,+-}(s')}{s'(s'-4\mpn^2)(s'-s)} \; ,
\label{eq:Wns}
\end{eqnarray}
where $\mpi$ and $\mpn$ denote the charged- and neutral-pion mass, respectively, and, as anticipated, the superscripts $00$ and $+-$ in the $W$-functions
indicate the charges of the intermediate pions contributing to the imaginary part in the unitarity relation. 

\paragraph{The amplitude  $\pi^+\pi^+ \to \pi^+\pi^+$:}
For this amplitude, the $P$-wave is again absent in the $s$ channel, but
allowed in the $t$ and $u$ channels. We, therefore, need to define 
the scattering angle (or the cosine thereof, $z_{t,u}$) in the
$t$- or $u$-center-of-mass frame:  
\begin{eqnarray}
s &=& \frac{1}{2}(t-4M_{\pi}^2)(z_t - 1), \quad u=
\frac{1}{2}(t-4M_{\pi}^2)(-z_t - 1)\;, \nonumber \\
z_t &=& \frac{s-u}{t-4M_{\pi}^2} \; .
\end{eqnarray}
The analogous expression to~\eqref{eq:ASP} for the $T^{++}$ amplitude reads
\begin{eqnarray}
T^{++}_{SP}(s,t,u) &=& 32 \pi \left( W_{S}^{++}(s) + W_{S}^{c,00}(t) +
  W_{S}^{c,+-}(t) + W_{S}^{c,00}(u) +  W_{S}^{c,+-}(u) \right. \nonumber \\
&& \qquad\left. +(u-s) W_{P}^{c,+-}(t) + (t-s) W_{P}^{c,+-}(u)
  \right)\;,
\label{eq:T++SP}
\end{eqnarray}
where 
\begin{eqnarray}
W_{S}^{++}(s) &=& \frac{a^{++} \, s}{4 M_{\pi}^2}
+\frac{s(s-4M_{\pi}^2) }{\pi} \int_{4 M_{\pi}^2}^{s_1} \, ds'
\frac{\text{Im}\,t_{S}^{++}(s')}{s'(s'-4M_{\pi}^2)(s'-s)}\;, \nonumber \\
W_{S}^{c,+-}(s) &=& \frac{a_c^{+-}\, s}{4 M_{\pi}^2} +
\frac{s (s-4M_{\pi}^2)}{\pi} \int_{4 M_{\pi}^2}^{s_1} \, ds'
\frac{\text{Im}\,t_{S}^{c,+-}(s')}{s'(s'-4M_{\pi}^2)(s'-s)}\;, \nonumber \\
W_{S}^{c,00}(s) &=& \frac{s (s-4M_{\pi}^2)}{\pi} \int_{4 \mpn^2}^{s_1}
\, ds' \frac{\text{Im}\,t_{S}^{c,00}(s')}{s'(s'-4M_{\pi}^2)(s'-s)}\;, \nonumber\\ 
W_{P}^{c,+-}(s) &=& \frac{s}{\pi} \int_{4 M_{\pi}^2}^{s_1} \, ds' \frac{3
  \text{Im}\,t_{P}^{c,+-}(s')}{s'(s'-4M_{\pi}^2)(s'-s)} \; .
\label{eq:T++SPW}
\end{eqnarray}
Via the crossing relation~\eqref{eq:chargedT2}, \eqref{eq:T++SP}
also provides a representation for $T^c$.

\paragraph{The amplitude $\pi^+ \pi^- \to \pi^0 \pi^0$:} 
For this amplitude, there is no $P$-wave in the $s$ channel, because of the
two identical particles in the final state, but a $P$-wave is present in
the $t$ and $u$ channels, for the crossed process $\pi^+ \pi^0 \to \pi^+
\pi^0$. In the $t$ channel, for example, the scattering angle can be
expressed in terms of the Mandelstam variables as follows:
\begin{equation}
z_t =
\frac{t(s-u)+\Delta_\pi^2}{\lambda(t,\mpi^2,\mpn^2)} \;, \;
\Delta_\pi:=\mpi^2-\mpn^2 \;,
\end{equation}
with $\lambda(a,b,c)=a^2+b^2+c^2-2(a\,b+b\,c+c\,a)$ the K\"allen function.

This is all we need to derive the expression of the $T^x_{SP}$ amplitude in
terms of single-variable functions:
\begin{eqnarray}
T^x_{SP}(s,t,u) &=& 32 \pi \left[ W_{S}^{x,+-}(s)+  W_{x,S}^{00}(s) 
+ W_{S}^{+0}(t)+ W_{S}^{+0}(u) \right. \\
&& \qquad\left. + \left(t(s-u)+\Delta_\pi^2 \right) \ W_{P}^{+0}(t) 
+  \left(u(s-t)+\Delta_\pi^2 \right) W_{P}^{+0}(u)\right]\;, \nonumber
\end{eqnarray}
which are defined as
\begin{eqnarray}
W_{S}^{x,+-}(s) &\!\!\!=\!\!\!& \frac{a_1 \, s}{4 M_{\pi}^2} +
\frac{s (s-4M_{\pi}^2)}{\pi} \int_{4 M_{\pi}^2}^{s_1} \, ds'
\frac{\text{Im}\,t_{x,S}^{+-}(s')}{s'(s'-4M_{\pi}^2)(s'-s)}\;,\nonumber \\
W_{S}^{x,00}(s) &\!\!\!=\!\!\!& 
\frac{s (s-4M_{\pi}^2)}{\pi} \int_{4 \mpn^2}^{s_1} \!\!\! ds'
\frac{\text{Im}\,t_{x,S}^{00}(s')}{s'(s'-4M_{\pi}^2)(s'-s)}\;,\nonumber \\
W_{S}^{+0}(s) &\!\!\!=\!\!\!& \frac{a_2 \,
  s}{4 \mpb^2} +\frac{s (s-4\mpb^2)}{\pi}
\int_{4\mpb^2}^{s_1}\!\!\! ds' \frac{\text{Im}\,t_{S}^{+0}(s')}{s'(s'-4\mpb^2)(s'-s)}\;,\nonumber \\ 
W_{P}^{+0}(s)&\!\!\!=\!\!\!& \frac{1}{\pi}
\int_{4\mpb^2}^{s_1}\!\!\! ds' \frac{3 \text{Im}\,t_{P}^{+0}(s')}{\lambda(s',\mpi^2,\mpn^2)(s'-s)}  \; ,
\end{eqnarray}
where $\mpb=(M_{\pi}+\mpn)/2$ and the two scattering-length-like quantities $a_{1,2}$ are related to the true scattering lengths by
\begin{eqnarray}
a_1&=&a_x^{+-}+2 \eps_\pi a_2- 8 \Dpi \left[ \mpb^2(1+\eps_\pi)
  f_s^{+0}(-\Dpi) -\mpi^2f_p^{+0}(-\Dpi) \right],\label{eq:a1rel} \\
a_2&=&\frac{a_c^{+0}}{1+\eps_\pi^2}+\Dpi^2
\left(\frac{1-\eps_\pi^2}{1+\eps_\pi^2} \right)f_s^{+0}(\eps_\pi \Dpi),\label{eq:a2rel}
\end{eqnarray}
and where we have introduced
\begin{align}
\eps_\pi:= \frac{\Dpi}{4 \mpb^2} ,& \;\; \; f_s^{+0}(s)=\frac{1}{\pi}\int_{4\mpb^2}^{\infty}\!\!\! ds' \frac{\text{Im}\,t_{S}^{+0}(s')}{s'(s'-4\mpb^2)(s'-s)},\;\;\nonumber\\
f_{p}^{+0}(s)&=\frac{1}{\pi}
\int_{4\mpb^2}^{\infty}\!\!\! ds' \frac{3 \text{Im}\,t_{P}^{+0}(s')}{\lambda(s',\mpi^2,\mpn^2)(s'-s)}\,.
\end{align}
The introduction of the parameters $a_1$ and $a_2$ is necessary in order to ensure that
\begin{align}
    W_S^{x,+-}(4M_\pi^2)=a_1\quad\text{and}\quad W_S^{+0}(4\bar{M}_\pi^2)=a_2\ .
\end{align}
while 
\begin{align}
    t_S^{x}(4M_\pi^2)=a_x^{+-}\quad\text{and}\quad t_S^{+0}(4\bar{M}_\pi^2)=a_c^{+0}\ .
\end{align}
\subsection{\texorpdfstring{Unitarity relations for $\Delta M_\pi \neq 0$}{Unitarity relation for the Delta mpi not zero}}\label{sec:unitarity}

We now want to estimate the effect of the pion mass difference in each
physical channel and shift $M_{\pi^0}$ to its physical value. This affects the physical ranges of the kinematic variables and, correspondingly, the thresholds of the various processes.
Moreover, it is also reflected in the analytic structure of the $T^k$ amplitudes for which, in addition to the branch cut at $4 \mpi^2$ in the
$s$ channel, a second one develops at $4 \mpn^2$, or at $4\mpb^2$ in the $t$ channel, as has been
discussed in detail in~\cite{Colangelo:2008sm}. Indeed, the expressions obtained above do show these different cuts already, provided one takes for $\mpn$ its physical value.  


Nevertheless, in order to solve the integral equations derived in Sect.~\ref{sec:single-variable_decomp}, there is still one piece of information missing: the expression of the imaginary parts of the amplitudes that appear in the dispersive integrals above, as fixed by unitarity. As an initial step, we only consider here intermediate two-pion states, postponing the inclusion of inelastic contributions to Sect.~\ref{sec:imag_inelastic}. 
For the $S$-wave of the $T^{n,x,c}$ amplitudes, we have a coupled-channel problem in the $s$ channel, where the unitarity relation is expressed at best in matrix form~\cite{Colangelo:2008sm}: 
\begin{eqnarray}\label{eq:imTs}
\mathrm{Im}T_S(s)&=& T_S(s) \Sigma(s) T_S^*(s) \; , \; \; \; \mbox{with} \; \;  T_S=  \left(
  \begin{array}{ll} t_{S}^n(s) & - t_{S}^x(s) \\ -t_{S}^x(s) &
    t_{S}^c(s)\end{array}
  \right) \; , \nonumber \\
 \qquad \Sigma(s)&=& \left( \begin{array}{ll}
      \sigma_0(s)\theta(s-4 \mpn^2) & 0 \\
      0 & 2 \sigma(s) \theta(s-4 M_{\pi}^2) \end{array}
  \right) \; ,
\end{eqnarray}
with $\sigma(s)=\displaystyle\sqrt{1-\frac{4M_{\pi}^2}{s}}$ and $\sigma_0(s)=\displaystyle\sqrt{1-\frac{4M_{\pi^0}^2}{s}}$ the phase-space factors for the charged and neutral pions, respectively.

From this, one can read off the imaginary parts for each of the components (to simplify the notation, we absorb $\theta$-functions in the
definition of the $\sigma$'s):
\begin{eqnarray}
\mathrm{Im}\,t_{S}^n(s)&=&\sigma_0(s)|t_{S}^n(s)|^2+ 2
\sigma(s)|t_{S}^x(s)|^2\;, \nonumber \\ 
\mathrm{Im}\,t_{S}^x(s)&=&\sigma_0(s)t_{S}^n(s)t_{S}^{x}(s)^*+ 2
\sigma(s)t_{S}^x(s)t_{S}^{c}(s)^*\;, \nonumber \\ 
\mathrm{Im}\,t_{S}^c(s)&=&\sigma_0(s)|t_{S}^x(s)|^2+ 2
\sigma(s)|t_{S}^c(s)|^2 \; . 
\label{eq:ImTS}
\end{eqnarray}
For the remaining partial waves of relevance here, again considering only intermediate pion states, the unitarity relation takes the form of the standard optical theorem: $\mathrm{Im}t=\rho |t|^2$, with $\rho$ the relevant phase space factor. 
In this way, we obtain the following expressions of the imaginary parts we need to insert in the dispersive integrals.
\paragraph{Amplitude $T^n$:}

\begin{align}\label{eq:uni-tn}
\text{Im}\,t_{S}^{n,00}(s) =&\sigma_0(s) \left|t_{S}^n(s)\right|^2,
  \nonumber \\
\text{Im}\,t_{S}^{n,+-}(s) =& 2\sigma(s)
\left|t_{S}^x(s)\right|^2.
\end{align}

\paragraph{Amplitudes $T^c$ and $T^{++}$:}
\begin{align}\label{eq:uni-tn-and-tpp}
\text{Im}\,t_{S}^{c,00}(s) =&\sigma_0(s) \left|t_{S}^x(s)\right|^2\,,
\nonumber \\ 
\text{Im}\,t_{S}^{c,+-}(s) =& 2\sigma(s) \left|t_{S}^c(s)\right|^2\,,
  \nonumber \\
\text{Im}\,t_{P}^{c,+-}(s)=&2\sigma(s) \left|t_{P}^c(s)\right|^2\,,
\nonumber \\ 
\text{Im}\,t_{S}^{++}(s)=& \sigma(s) \left| t_S^{++}(s)\right|^2\,.
\end{align}

\paragraph{Amplitudes $\pi^+ \pi^- \to \pi^0 \pi^0$ and $\pi^+ \pi^0 \to \pi^+ \pi^0$:}

\begin{align}
\text{Im}t_{S}^{x,+-}(s) =& 2\sigma(s) t_{S}^x(s)t_{S}^c(s)^*\,,
\nonumber \\ 
\text{Im}t_{S}^{x,00}(s) =& \sigma_0(s)
t_{S}^n(s) t_{S}^x(s)^*\,,
\nonumber\\ 
\text{Im}\,t_{S}^{+0}(s) =&
\frac{2\,\lambda^{1/2}(s,M_{\pi}^2,\mpn^2)}{s} \left|t_{S}^{+0}(s)\right|^2\,,
\nonumber \\  
\text{Im}\,t_{P}^{+0}(s) =&\frac{2\,\lambda^{1/2}(s,M_{\pi}^2,\mpn^2)}{s} \left|
  t_P^{+0}(s)\right|^2\,.
\end{align}

\section{\texorpdfstring{The $\pi \pi$ scattering amplitude in \chpt$_\gamma$:
  pion-mass difference effects}{The pipi scattering amplitude in chpt}}\label{sec:chpt}
  
A crucial step in the procedure just outlined is to provide input values for the subtraction constants, which are all expressed in terms of scattering lengths. Like in the isospin-limit case~\cite{Colangelo:2001df}, they will be fixed by matching to \chpt at low energy. To do this, we first need to bring the $\pi\pi$ scattering amplitude, including isospin-breaking corrections, into a form that matches the dispersive one just presented. The $\pi\pi$ scattering amplitudes in the presence of isospin breaking have been calculated in two papers: the amplitude $T^x$ by Knecht and Urech~\cite{Knecht:1997jw} and later extended to $T^c$ by Knecht and Nehme~\cite{Knecht:2002gz}. These two papers contain both the effect of the pion-mass difference as well as the effects due to virtual photons, but in a form that allows them to be easily disentangled.
Here, we will make this separation explicit while recasting the amplitude in a form that allows for an easy matching to our dispersive representation.

Let us start with the $T^n$ amplitude: the \chpt representation
of the functions $W^{n,00}_{S}$ and $W^{n,+-}_{S}$ up to order $p^4$ reads:
\begin{align}
W_{S}^{n,00}(s)&=\frac{s}{4M_{\pi^0}^2}\left[a_n^{00}+b_n^{00}
  \frac{\left(s-4M_{\pi^0}^2\right)}{M_\pi^2}\right]+ \frac{1}{64 \pi
  F_\pi^4} D_0^0\left[M_{\pi^0}^4\right](s)\,,\nonumber\\ 
W_{n,S}^{+-}(s)&=\frac{1}{32 \pi F_\pi^4}D^0\left[(x-M_{\pi^0}^2)^2\right](s)\,,
\end{align}
where $F_\pi$ is the pion decay constant, and $a_n^{00}$ and $b_n^{00}$ denote the scattering length and slope parameter, respectively, in the threshold expansion of the neutral amplitude. The function
\begin{equation}
D_i^j[p(x)](s):=\frac{s^2(s-4 M_{\pi^j}^2)}{16 \pi^2}\int_{4 M_{\pi^i}^2}^\infty
\!\!\!\! dx \frac{\sigma_i(x)p(x)}{x^2(x- 4 M_{\pi^j}^2)(x-s)}\;,\quad\text{with}\quad i,j=0, 
\pm 
\label{eq:Dijd}
\end{equation}
is the triply-subtracted dispersive integral (twice at $s=0$ and once at $4
M_{\pi^j}^2$ as indicated by the superscript), where the discontinuity is given by the polynomial $p(x)$ multiplied by the two-pion phase space $\sigma_i(s)$. Note that, in the following, the sub- and superscripts referring to charged pions will be omitted, i.e., $D[p]:=D^\pm_\pm[p]$, $D^0[p]:=D^0_\pm[p]$, and so on.

The explicit expressions for the threshold parameters $a_n^{00}$ and $b_n^{00}$ are listed in the App. ~\ref{app:subconst}. Here, we show only the expanded results up to $\mathcal{O}(\dpi^2)$ (after having verified that this is a good approximation),
\begin{align}\label{eq:an00_chpt}
a_n^{00}=&\frac{M_\pi^2}{32\pi F_\pi^2}\Bigg\{1-\dpi+\xi\Bigg[ 
    \left(4\lbar_1+8\lbar_2-\frac{3}{2}\lbar_3+2\lbar_4\right)\left(1-\dpi\right)^2 +
      \frac{13}{2}- 4
    \dpi\left(1+\frac{\dpi}{2}\right)\nonumber\\
      &\hspace{1.3 cm}+\dpi\left(1-\dpi\right)\left(\frac{\Bar{k}_{31}}{9}-\frac{10}{9}\Bar{k}_2-\Bar{k}_4\right)\nonumber-9 \pi\sqrt{\dpi}\left(1-\frac{3}{2}\dpi\right)\Bigg]\Bigg\}\;,\nonumber\\     
    b_n^{00}=&\frac{M_\pi^2}{32\pi F_\pi^2}\,\xi
    \left[\frac{4}{3}(\lbar_1+2\lbar_2)(1-\dpi)+\frac{1}{2}
      +\frac{1}{6}\dpi(1+2 \dpi)
      -\frac{9\pi}{4}\sqrt{\dpi}\left(1-\frac{\dpi}{2}\right) \right]\;,
\end{align}
where
\begin{equation}
\dpi:=\frac{\Dpi}{\mpi^2}\;,\quad\text{and}\quad\xi:=\frac{M_\pi^2}{16 \pi^2 F_\pi^2} \; .
\end{equation}

The threshold parameters are expressed in terms of the mesonic $l_i$ and electromagnetic $k_i$ low-energy constants (LECs), with $k_{31}$ given by the combination 
\begin{equation}
k_{31}=-\frac{5}{9}k_1+k_3\;.
\end{equation}
The renormalization of the LECs at one-loop order was studied in~\cite{Knecht:1997jw} and is given by
\begin{equation}
l_i=l_i^r(\mu)+\gamma_i\lambda\;,\quad
k_i=k_i^r(\mu)+\sigma_i\lambda\;,
\end{equation}
where $\gamma_i$ and $\sigma_i$ are the corresponding renormalization group $\beta$ functions~\cite{Gasser:1982ap,Knecht:1997jw}, $\mu$ is the renormalization scale, and $\lambda$ reads,
\begin{equation}
\lambda=\frac{\mu^{d-4}}{32\pi^2}\left(\frac{2}{d-4}-\log(4\pi)+\gamma-1\right)\;,
\end{equation}
with $\gamma$ the Euler-Mascheroni constant.

The $\bar l_i$ and $\bar k_i$ are defined from the renormalized LECs as
\begin{align}
l^r_i=\frac{\gamma_i}{32\pi^2}\left(\bar l_i+\log{\frac{\mpi^2}{\mu^2}}\right)\;,\quad
k^r_i=\frac{\sigma_i}{32\pi^2}\left(\bar k_i+\log{\frac{\mpi^2}{\mu^2}}\right)\;,
\end{align}
so that the combination $\bar k_{31}$ reads
\begin{equation}
\Bar{k}_{31}:=\frac{9}{4Z}\left[\left(3+\frac{4}{9}Z\right)\Bar{k}_1-3\Bar{k}_3\right]\;.
\end{equation}

The constants $\sigma_i$ renormalize both photon and pion-mass difference divergences, with the latter encoded in the factor
\begin{equation}
   Z:=\frac{\Delta_\pi}{2e^2F^2}\; .
\end{equation}
The relevant $\sigma_i$ values here are
\begin{equation}
\sigma_1=-\frac{27}{20}-\frac{1}{5}Z\;,\quad \sigma_2=2Z\;,\quad\sigma_3=-\frac{3}{4}\;,\quad \sigma_4=2Z;\,\quad \sigma_6=\frac{1}{4}+2Z\;,\quad \sigma_8=\frac{1}{8}-Z\;.
\end{equation}

The combination $k_{31}$ is defined such as $\sigma_{31}=Z/9$, implying that the three counterterms appearing in $a_n^{00}$ in~\eqref{eq:an00_chpt}---namely, $k_2$, $k_4$, and $k_{31}$---only renormalize effects due to the pion-mass difference. Of course, this does not imply that the finite parts also only contain effects due to the pion mass difference---in other words, that the finite parts would vanish in the limit $Z\to 0$. This is an intrinsic ambiguity of the present calculation which cannot be resolved. This is not a problem, however, because the analysis presented here only concern a part of the isospin-breaking corrections, and once all of them will be evaluated and combined in our final result, the ambiguity will disappear.

For the triply-subtracted integral~\eqref{eq:Dijd} to converge, the polynomial $p(x)$ has to be at
most second order. This additional subtraction---one more than in~\eqref{eq:Wns}---is necessary because, in \chpt, the
discontinuities are valid only at low energy, where they are expanded in powers of momenta and take the form of second-degree polynomials (times the phase space). 
To achieve a proper matching, we must apply an additional subtraction to the general dispersive expressions in~\eqref{eq:Wns}, thereby obtaining a sum rule for $b_n^{00}$. For $a_n^{00}$, the \chpt representation provided here is an essential input to the dispersive approach. 

For the $T^{++}$ amplitude, we obtain the following expressions for the
single-variable functions at $\cO(p^4)$:
\begin{align}
W_S^{++}(s)=&\frac{s}{4M_\pi^2}\left[a^{++}+\left(b^{++}+c^{++}s\right)
  \frac{\left(s-4M_\pi^2\right)}{M_\pi^2}\right]+
\frac{1}{128 \pi F_\pi^4}D\left[(x-2M_{\pi}^2-2\Delta_\pi)^2\right](s)\;,\nonumber\\  
W_{c,S}^{+-}(s)=&\frac{s}{4M_\pi^2}\left[a_c^{+-}+\left(b_c^{+-}+c_c^{+-}s\right)
  \frac{\left(s-4M_\pi^2\right)}{M_\pi^2} \right]+ \frac{1}{128 \pi
  F_\pi^4}D\left[(x+4\Delta_\pi)^2\right](s)\;,\nonumber\\ 
W_{c,S}^{00}(s)=&\frac{1}{64 \pi F_\pi^4}D_0\left[(x-M_{\pi^0}^2)^2\right](s)\;,\\
W_{c,P}^{+-}(s)=&-\frac{1}{384 \pi F_\pi^4}D\left[(x-4M_{\pi}^2)^2\right](s)\;,
\end{align}
together with the following ones for the expanded constants $a$, $b$, and $c$:
\begin{align}\label{eq:a++_chpt}
a^{++}=&-\frac{M_\pi^2}{16 \pi
  F_\pi^2}\Bigg\{1-\dpi-\xi\Bigg[\frac{4}{3}\left(\lbar_1+2\lbar_2\right)-\frac{1}{2}\left(\lbar_3+4\lbar_4\right)\left(1-\dpi\right)^2\nonumber\\
  &+\frac{1}{2}\left(1+3\dpi+\frac{88}{9}\dpi^2\right)-\dpi
  (1-\dpi)\left(\frac{\Bar{k}_{31}}{9}-4\Bar{k}_{32}+\frac{62}{9}\Bar{k}_2+5\Bar{k}_4\right)\Bigg] \Bigg\}\;,\nonumber\\ 
b^{++}=&\frac{M_\pi^2}{48\pi F_\pi^2} \xi \left(4\lbar_2-\frac{23}{9}-4\dpi+\dpi^2\right)\;,\nonumber\\
c^{++}=&-\frac{\xi}{864 \pi F_\pi^2}\;,\nonumber\\
\mathrm{Re}[a_c^{+-}]=&\frac{M_\pi^2}{16 \pi F_\pi^2} \Bigg\{1+\dpi+
\xi\Bigg[3+
\frac{4}{3}(\lbar_1+2\lbar_2)-\frac{1}{2}\lbar_3+2\lbar_4+ \dpi \left(2+\lbar_3\right)\nonumber\\
&+\dpi^2
\left(\frac{62}{9}-\frac{1}{2}\lbar_3-2\lbar_4\right)\nonumber\\
&+\dpi \left(\frac{\Bar{k}_{31}}{9}(1+\dpi)+4\Bar{k}_{32}(1-\dpi)-\frac{2}{9}\Bar{k}_2\left(5-31\dpi\right)+\Bar{k}_4(1+5\dpi)\right)\Bigg] \Bigg\}\;,\nonumber\\
\mathrm{Re}[b_c^{+-}]=&\frac{M_\pi^2 }{24 \pi F_\pi^2} \xi
\left[\frac{73}{72}+\lbar_1+\lbar_2+4\dpi+\dpi^2\right]\;,\nonumber\\ 
c_c^{+-}=&\frac{\xi}{1728 \pi F_\pi^2}\;,  
\end{align}
where $k_{32}$ is the combination,
\begin{equation}
k_{32}:=k_3+2(k_6+k_8) \; \; \Rightarrow \;
\Bar{k}_{32}:=-\frac{1}{8Z}\left[3\Bar{k}_3-2\left(1+8Z\right)\Bar{k}_6-\left(1-8Z\right)\Bar{k}_8\right]
\; .
\end{equation}
The renormalization group $\beta$ function for this combination is $\sigma_{32}=2Z$, ensuring once again that all LECs in~\eqref{eq:a++_chpt} only absorb divergences proportional to the pion-mass difference. 
Note that~\eqref{eq:a++_chpt} provides the exact expressions of $a^{++}$, $b^{++}$, $c^{++}$, and $c_c^{+-}$, while $a_c^{+-}$ and $b_c^{+-}$ are expanded up to $\mathcal{O}(\dpi^2)$, with their exact expressions given in App.~\ref{app:subconst} for completeness (the difference between the two expressions is numerically irrelevant).

The amplitude $T^x$ is more complicated because of the presence of
particles with unequal masses in the $t$- and $u$-channels. 
This leads to particularly cumbersome expressions for the subtraction constants. However, an expansion in the pion-mass difference significantly simplifies these expressions while maintaining numerical accuracy. Below, we present the single-variable functions:
\begin{align}
W_{x,S}^{+-}(s)&=\frac{s}{4M_\pi^2}\left[a_1+\left(b_1+c_1
    s\right)\frac{\left(s-4M_\pi^2\right)}{M_\pi^2}\right]-\frac{1}{32
  \pi F_\pi^2}D\left[\left(\frac{s}{2}+2\dpi\right)\left(s-M_{\pi^0}^2\right)\right](s)\;,\nonumber\\
W_{x,S}^{00}(s)&=-\frac{1}{64 \pi F_\pi^2}D_0\left[M_{\pi^0}^2\left(s-M_{\pi^0}^2\right)\right](s)\;,\nonumber\\
W_S^{+0}(s)&=\frac{s}{4\Bar{M}_\pi^2}\left[a_2+\left(b_2+c_2 s\right)
  \frac{\left(s-4\Bar{M}_\pi^2\right)}{M_\pi^2}\right] - \frac{1}{128
  F_\pi^4}D_{+0}^{+0}\left[\left(s-2M_\pi^2\right)^2\right](s)\;,\nonumber\\ 
W_P^{+0}(s)&=-\frac{\lambda(s,M_{\pi^0}^2,M_\pi^2)}{384 F_\pi^4}\Tilde{J}_{+0}^{(2)}(s)\;,
\end{align}
where $\displaystyle\Tilde{J}_{+0}^{(2)}(s):=\frac{\Bar{J}_{+0}(s)-s\Bar{J}_{+0}^{'}(0)}{s^2}$, which is regular at $s=0$.

The expanded subtraction constants are
\begin{align}\label{eq:a1_chpt}
\mathrm{Re}[a_1]=&-\frac{M_\pi^2}{32 \pi
  F_\pi^2}\Bigg\{3+\frac{\dpi}{4}(2+\dpi)+\xi\Bigg[\frac{11}{2}
+\frac{4}{3}(\lbar_1+2\lbar_2)-\frac{1}{2}\lbar_3+6 \lbar_4\nonumber\\  
 &+\dpi
 \left(\frac{247}{108}+2\lbar_1-\frac{4}{3}\lbar_2+\frac{3}{4}\lbar_3-5\lbar_4\right) -\dpi^2
\left(\frac{1309}{216}+\frac{\lbar_1}{3}+\frac{\lbar_3}{8}+\frac{\lbar_4}{2}\right)\nonumber\\
&+\dpi \left(\frac{\Bar{k}_{31}}{18}\left(6+\dpi\right)+\Bar{k}_{32}(2-\dpi)+\frac{\Bar{k}_2}{9}\left(6+13\dpi\right)+\Bar{k}_4(2+\dpi)\right)\Bigg]\Bigg\}\;,\nonumber\\
\mathrm{Re}[b_1]=&-\frac{M_\pi^2}{96 \pi F_\pi^2}\xi
\left[\frac{119+72\lbar_1}{18}+\frac{38}{9}\dpi-\frac{\dpi^2}{5}\right]\;,\nonumber\\
c_1=&-\frac{\xi}{288 \pi F_\pi^2}\left(\frac{1}{3}+\frac{\dpi}{6}+\frac{\dpi^2}{10}\right),\nonumber\\
a_2=&\frac{M_\pi^2}{32 \pi F_\pi^2}
\Bigg\{1-\dpi\left(1+\frac{\dpi}{16}\right)-\xi\Bigg[\frac{1}{2}+ 
\frac{4}{3}(\lbar_1+2\lbar_2)-\frac{1}{2}\lbar_3-2\lbar_4  \nonumber\\  
&\qquad-\dpi \left(3+\frac{4}{3}(\lbar_1+2\lbar_2)- \lbar_3-4\lbar_4 \right) + \frac{\dpi^2}{12}
\left( \frac{2669}{72}-\lbar_1+2\lbar_2-\frac{45}{8} (\lbar_3+4\lbar_4 )\right)\nonumber\\
&\qquad-\dpi \left(1-\dpi\right) \left(\frac{\Bar{k}_{31}}{9}-2\Bar{k}_{32}+\frac{26}{9}\Bar{k}_2+2\Bar{k}_4\right)\Bigg] \Bigg\}\;,\nonumber\\
b_2=&\frac{M_\pi^2}{864 \pi F_\pi^2}\xi 
\left[23-36\lbar_2+\dpi \left(18\lbar_2+\frac{49}{2}\right)+\dpi^2 \frac{3}{16}\left(12\lbar_2-5\right)\right],\nonumber\\
c_2=&\frac{\xi}{1728 \pi F_\pi^2}\left(1-\frac{\dpi^2}{80}\right)\;.
\end{align}
Note that, once again, all LECs in~\eqref{eq:a1_chpt} solely renormalize divergences proportional to the pion-mass difference. 

\section{\texorpdfstring{Derivation of the Roy equations for $\Delta M_\pi \neq 0$}{Derivation of Roy equations}}\label{sec:roy-beyond}

Roy equations beyond the isospin limit can be obtained directly by projecting the single-variable decomposition of the pion-pion amplitudes, defined in Sect.~\ref{sec:single-variable_decomp}, into partial waves.
This projection is defined by
\begin{equation}
t^k_J(s)=\frac{1}{64\pi}\int_{-1}^{1}{\text{d}z\,P_J(z)\,T^k(s,t(s,z),u(s,z))}\;,
\end{equation}
where $k$ labels any of the $\pi\pi$ channels in the charge basis defined in~\eqref{eq:chargedT1}~and~\eqref{eq:chargedT2}, $J$ denotes the angular momentum, $P_J$ stands for the Legendre polynomial of degree $J$, and $z$ is the cosine of the scattering angle in the corresponding $s$-channel center-of-mass frame.

For example, the $S$-wave projection of the neutral channel amplitude reads
\begin{equation}
t^n_S(s)=\frac{1}{64\pi}\int_{-1}^{1}{\text{d}z\,T^n(s,t(s,z),u(s,z))}\;,
\end{equation}
where in this case
\begin{align}
 t(s,z)=\frac{(s-\spin)(z-1)}{2}\;,\quad
 u(s,z)=-\frac{(s-\spin)(z+1)}{2}\;,\quad \spin=4M^2_{\pi^0}\;.
\end{align}

Including explicitly the single-variable decomposition of the $\pi^0\pi^0\to\pi^0\pi^0$ amplitude in~\eqref{eq:neutral_decomp}, and performing the angular integration analytically, the neutral-channel $S$-wave can be recast as
\begin{align}\label{eq:Roy_neutral}
t_S^n(s)=&\;a_n^{00}+\int_{\spin}^{\spi}{\text{d}s'\, K_n(s,s')\,\text{Im}t_{S}^{n,00}(s')}\nonumber\\
&+\int_{\spi}^{s_1}
{\text{d}s'\, K_n(s,s')\,\left(\text{Im}t_{S}^{n,00}(s')+\text{Im}t_{S}^{n,+-}(s')\right)}+d^n_S(s)\;,
\end{align}
where $\spi=4M^2_{\pi^\pm}$, $\text{Im}t_{S}^{n,00}$ and $\text{Im}t_{S}^{n,+-}$ are defined in~\eqref{eq:ImTS}, and $d^n_S$ denotes the driving-term contribution, i.e., the partial-wave projection of the background integral
\begin{equation}
d^n_S(s)=\frac{1}{64\pi}\int_{-1}^{1}{\text{d}z\,T^n_d(s,t(s,z),u(s,z))}\;,
\end{equation}
accounting for the contribution of the imaginary part of higher $J>1$ partial waves as well as of the $S$- and $P$-waves above $s_1$.  

The kernel $K_n(s,s')$ is defined as
\begin{equation}
    K_n(s,s')=\frac{1}{\pi}\left[\frac{1}{s'-s}-\frac{2}{s'}-\frac{1}{s'-\spin}+\frac{2}{s-\spin}\log\left(1+\frac{s-\spin}{s'}\right)\right]\;,
\end{equation}
and encodes the whole analytic structure of the $\pi^0\pi^0\to\pi^0\pi^0$ amplitude; the first term in brackets, $1/(s'-s)$, accounts for the right-hand cut discontinuity generated by both $\pi^0\pi^0$ and $\pi^+\pi^-$ intermediate states, while the log term includes the $t$- and $u$-channel discontinuities. 
Note that, on the one hand, 
\begin{equation}
\lim_{s'\to \spin} K(s,s')=\frac{1}{\pi}\frac{1}{s'-\spin}+{\cal O}\left[(s'-\spin)^0\right]\;,
\end{equation}
which, together with the threshold behavior of $\text{Im}\,t_{S}^n$, makes the integrand integrable around $s'\sim\spin$.
On the other hand, in the $\lim s\to \spin$ the kernel behaves as $K(s,s')={\cal O}(s-\spin)$ and hence, the dispersion relation in~\eqref{eq:Roy_neutral} fulfills $t_S^n(\spin)=a_n^{00}$, i.e., the neutral-channel partial-wave amplitude is fixed from the neutral-channel scattering length, which in practice is taken from the \chpt expression in~\eqref{eq:an00_chpt}.  

It is also particularly illuminating to discuss the partial-wave projection of the charged-channel amplitude, which can be obtained from~\eqref{eq:T++SP} using the crossing-symmetry relation in~\eqref{eq:chargedT2}, i.e.,  
\begin{equation}
t^c_J(s)=\frac{1}{64\pi}\int_{-1}^{1}\text{d}z\,P_J(z)\,T^{++}(u(s,z),t(s,z),s)\;,
\end{equation}
where in this case, we have contributions from both $S$- and $P$-waves. Performing once again the partial-wave projection analytically, we get, for example, for the $S$-wave
\begin{align}\label{eq:Roy-charged-S}
    t_S^{c}(s)=&\frac{1}{2}\left(1+\frac{s}{\spi}\right)a_c^{+-}+\frac{1}{2}\left(1-\frac{s}{\spi}\right)a^{++}+\int_{\spin}^{\spi}\text{d}s' K_{s,S}^{+-}(s',s)\,\text{Im}t_{S}^{c,00}(s')\nonumber\\
    +&\int_{\spi}^{s_1}\text{d}s'\,\bigg[K_{s,S}^{+-}(s',s)\,\left(\text{Im}t_{S}^{c,00}(s')+\text{Im}t_{S}^{c,+-}(s')\right)+K_{s,P}^{+-}(s',s)\,\text{Im}t_{P}^{c,+-}(s')\nonumber\\
    &\hspace{1.8cm}+K_{s,++}^{+-}(s',s)\,\text{Im}t_{S}^{++}(s')\bigg]+d_S^c(s)\;,
\end{align}
where $\text{Im}t_{S}^{c,00}$, $\text{Im}t_{S}^{c,+-}$, $\text{Im}t_{P}^{c,+-}$, and $\text{Im}t_{S}^{++}$ are defined in~\eqref{eq:uni-tn-and-tpp}, $d_S^c(s)$ is the driving-term contribution, and the kernels read
\begin{align}
     K_{s,S}^{+-}(s',s)=&\frac{1}{\pi}\left[\frac{1}{s'-s}-\frac{s'+s+3(s'-\spi)}{2s'(s'-\spi)}+\frac{1}{s-\spi}\ln{\left(1+\frac{s-\spi}{s'}\right)}\right]\;,\nonumber\\
    K_{s,P}^{+-}(s',s)=&\frac{3}{\pi}\left[\,-\frac{3s+2s'-\spi}{2s'(s'-\spi)}+\frac{2s+s'-\spi}{(s-\spi)(s'-\spi)}\ln{\left(1+\frac{s-\spi}{s'}\right)}\right]\;, \nonumber\\
    K_{s,++}^{+-}(s',s)=&\frac{1}{\pi}\left[\,\frac{s-2s'+\spi}{2s'(s'-\spi)}+\frac{1}{s-\spi}\ln{\left(1+\frac{s-\spi}{s'}\right)}\right]\;.
\end{align}

Once again, all kernels are suppressed in the $s\to\spi$ limit and~\eqref{eq:Roy-charged-S} satisfies $t^c_S(\spi)=a_c^{+-}$. Nevertheless, due to the $K_{s,S}^{+-}(s',s)\,\text{Im}t_{S}^{c,00}(s')$ contribution both in the first and second integrals in~\eqref{eq:Roy-charged-S}, the $s'\to\spi$ limit must be studied carefully. On the one hand, close to the charged-pion threshold the $K_{s,S}^{+-}$ kernel behaves as 
$$\lim_{s'\to\spi}{K_{s,S}^{+-}(s',s)}=-\frac{s+\spi}{2\pi\spi(s'-\spi)}+{\cal O}\left[\left(s'-\spi\right)^0\right]\;.$$
On the other hand, $\text{Im}t_{S}^{c,00}$ only vanishes at the neutral-pion threshold, hence leading to an end-point singularity. Although by construction this singularity is integrable, its numerical implementation requires special treatment. First, the $K_{s,S}^{+-}$ can be decomposed in terms of a regular and singular part as
\begin{align}
K_{s,S}^{+-}(s',s)=&\;K_{s,S}^{+-,\,I}(s',s)+K_{s,S}^{+-,\,II}(s',s)\;,\nonumber\\
K_{s,S}^{+-,\,I}(s',s)=&\;\frac{1}{\pi}\left[\frac{1}{s'-s}-\frac{2}{s'}+\frac{1}{s-\spi}\ln{\left(1+\frac{s-\spi}{s'}\right)}\right]\;,\nonumber\\
K_{s,S}^{+-,\,II}(s',s)=&\;\frac{-(s+\spi)}{2\pi s'(s'-\spi)}\;,
\end{align}
where I and II denote the regular and singular contributions, respectively.
Second, the singular piece can be recast as
\begin{align}\label{eq:end-point}
K_{s,S}^{+-,\,II}(s',s)\,\text{Im}t_{S}^{c,00}(s')=&\frac{1}{\sigma_0(\spi)}K_{s,S}^{+-,\,II}(s',s)\left(\sigma_0(\spi)\,\text{Im}t_{S}^{c,00}(s')\!-\!\sigma_0(s')\,\text{Im}t_{S}^{c,00}(\spi)\right)\nonumber\\
+&\frac{\sigma_0(s')}{\sigma_0(\spi)}K_{s,S}^{+-,\,II}(s',s)\,\text{Im}t_{S}^{c,00}(\spi),
\end{align}
so that the first term in~\eqref{eq:end-point} vanishes exactly in the $s'\to\spi$ limit and can be computed using standard integration routines. 
Third, the second term in~\eqref{eq:end-point} can be integrated analytically. Namely, 
\begin{align}
&\int_{\spin}^{s_1}\text{d}s' K_{s,S}^{+-,II}(s',s)\,\sigma_0(s')= \\
&\quad\frac{2}{\spi}\left\{ \sigma_0(s_1)-\sigma_0(\spi) \left[\arctanh\left(\frac{\spi-s_1\left(1-\sigma_0(s_1)\right)}{\spi \sigma_0(\spi)}\right)\!+\!\frac{1}{2}\log\left(\frac{1-\sigma_0(\spi)}{1+\sigma_0(\spi)}\right)\right]\right\}\;.
\nonumber
\end{align}

Similar end-point singularities appear in the partial-wave projection of the $T^{++}$, $T^x$, and $T^{+0}$ amplitudes, for which the very same procedure should be applied. Expressions of the Roy equations, including explicit expressions for all the kernels, for the remaining channels in the charge basis are collected in App.~\ref{app:kernels}.

\section{\texorpdfstring{How to solve the Roy equations for $\Delta M_\pi \neq 0$}{How to solve Roy equations}}\label{sec:strategy}

In order to solve the Roy equations, we must first fix all input quantities that appear in~\eqref{eq:Roy_neutral},~\eqref{eq:Roy-charged-S}, and App.~\ref{app:kernels}---namely, the values of the scattering lengths, the imaginary parts of  $S$- and $P$-waves above $s_1$ and the driving terms. For the former, we will make use of the \chptg~ predictions obtained in Sect.~\ref{sec:chpt}. For the latter two, as already anticipated, we will use the same phenomenological estimates given in~\cite{Colangelo:2001df,Caprini:2011ky}, even though those references used these in an isospin-symmetric context. The reason for this strategy is two-fold. First, pion-mass difference effects are expected to be more relevant at low energies, where, on the one hand, the contribution of the $\pi^0\pi^0$, $\pi^+\pi^0$, and $\pi^+\pi^-$ threshold-shift effect is still sizable, and, on the other hand, the impact of the different masses and phase-space factors, entering both the kernels and the imaginary parts, is enhanced. Second, the Roy-equation solutions discussed in~\cite{Colangelo:2001df,Caprini:2011ky} were obtained in the isospin limit, but the high-energy and higher partial-wave contributions were fixed from experimental data without attempting to remove isospin-breaking effects. In that context, this was also motivated by the expectation that isospin-breaking effects are small at high energy. But all this means that, in the present context, we can just adopt those estimates without the need to apply any correction to them.

\subsection{\texorpdfstring{Partial-wave parameterizations for $\Delta M_\pi \neq 0$}{Partial-wave parameterizations}}\label{sec:IB_param}

In the physical region, the Roy equations beyond the isospin limit derived in Sect.~\ref{sec:roy-beyond}
relate the real part of the $\pi\pi$ partial waves with an integral over their imaginary parts, which, in turn, are also related through the unitarity relations collected in Sect.~\ref{sec:unitarity}, hence providing a coupled system of integral equations.  

Partial waves are customarily expressed in terms of their phase shift and elasticity, i.e., the phase and modulus of the $S$-matrix element, respectively.  In the isospin limit, the unitarity relations for the three isospin amplitudes $I=0,1,2$ are diagonal, and $\pi\pi$ scattering remains elastic below the first inelastic threshold, which formally starts at the four-pion threshold. Nevertheless, in practice, inelastic effects become experimentally noticeable only at higher energies~\cite{Hyams:1973zf} and $\pi\pi$ scattering can be considered elastic up to energies around 1 GeV\footnote{In the isospin limit, the S0 wave is elastic up to the $K\bar K$ threshold, the P wave inelasticity starts at the $\pi\omega$ threshold,  and the S2 wave remains elastic up to energies around 1 GeV~\cite{Pelaez:2024uav}.}. Hence, in this elastic region, Roy equations translate into a coupled system of integral equations for the phase shifts. Above this energy, one also needs to include the elasticity, which, once again, is obtained from experimental data. 

Beyond the isospin limit, one has to switch to the charge basis, which, as discussed in Sect.~\ref{sec:unitarity}, no longer diagonalizes the unitarity relation in the case of the $T^n$, $T^x$, and $T^c$ $S$-waves, and requires introducing an additional internal $\pi\pi$ elasticity parameter along with a non-$\pi\pi$ inelasticity contribution. For the latter, we once again rely on the results in~\cite{Colangelo:2001df,Caprini:2011ky}, which, since they are extracted directly from data, already account for pion-mass difference effects. For the former, we will perform a coupled-channel analysis.  

Thus, following the example in~\cite{Ananthanarayan:2000ht,Colangelo:2001df,Caprini:2011ky,GarciaMartin:2011cn,Pelaez:2024uav,Buettiker:2003pp,Hoferichter:2015hva}, to solve this system of equations, we parameterize the phase shift at low energies in a suitable form, whose free parameters are obtained by minimizing the difference between the left- and right-hand side of Roy equations. In the following, we will discuss each of these parameterizations.

\subsubsection{Elastic case}

We will start with the $\pi^+\pi^+\to\pi^+\pi^+$ $S$-wave, which in terms of its phase shift and elasticity is parameterized as
\begin{equation}
t^{++}_S(s)=\frac{\eta^{++}_S(s)\,\text
{e}^{2i\,\delta^{++}_S(s)}-1}{2i\sigma(s)}\;.
\end{equation}

For the elasticity $\eta_S^{++}$, we consider the isospin-limit results in~\cite{Colangelo:2001df,Caprini:2011ky} so that $\eta_S^{++}(s)=\eta_0^2(s)$. In the case of the phase shift, we use
\begin{equation}\label{eq:++param}
\delta_S^{++}(s)=\delta_0^2(s)\left(1+\dpi\sum_{i=0}^7 c_i^{++}\,\left(\frac{s-\spi}{\spi}\right)^i\right)\;,
\end{equation}
i.e., it is parameterized in terms of the isospin-limit phase shift result plus a correction proportional to the pion-mass difference squared $\Delta_\pi=M_\pi^2 \dpi$, which we express as a polynomial factor multiplying the phase shift.

At the charged-pion threshold $\spi$, the parametrization in~\eqref{eq:++param} must match the $a^{++}$ scattering length, so that the $c_0^{++}$ coefficient is fully determined by the relation
\begin{equation}
c^{++}_0=\frac{\Delta a^{++}}{a^{++}_\text{IL}\dpi}\;,
\end{equation}
with $\Delta a^{++}=a^{++}-a^{++}_\text{IL}$.

Additionally, to ensure a well-defined integrand, the parametrization in~\eqref{eq:++param} and its first derivative at $s_1$, the maximum energy of validity of Roy equations, must match the isospin-limit results employed in the driving term. This imposes two additional constraints, allowing us to fix the values of the coefficients $c_6^{++}$ and $c_7^{++}$ as follows
\begin{align}\label{eq:matching_conditions}
    c_6^{++}=&-\left(\frac{\spi}{s_1-\spi}\right)^6\sum_{k=0}^5{(7-k)\,c_k^{++}\left(\frac{s_1-\spi}{\spi}\right)^k}\;,\nonumber\\
    c_7^{++}=&-\left(\frac{\spi}{s_1-\spi}\right)^7\sum_{k=0}^6{c_k^{++}\left(\frac{s_1-\spi}{\spi}\right)^k}\;,
\end{align}
leaving $c_{1-5}^{++}$ as the only free parameters.

Similar constraints apply to the $\pi^+\pi^-\to \pi^+\pi^-$ $P$-wave, which we parameterize as
\begin{equation}
t^{c}_P(s)=\frac{\eta^{c}_P(s)\,\text
{e}^{2i\,\delta^{c}_P(s)}-1}{4i\sigma(s)}\;, 
\end{equation}
with $\eta^{c}_P(s)=\eta^1_1(s)$ and $\delta^{c}_P(s)$
\begin{equation}\label{eq:+-Pparam}
\delta_P^{c}(s)=\delta_1^1(s)\left(1+\dpi\sum_{i=0}^7 c_i^{c,P}\,\left(\frac{s-\spi}{\spi}\right)^i\right)\;,
\end{equation}
where $c_0^{c,P}$ is now a free parameter, since for this wave there is no relation to a subtraction constant, and $c_{6-7}^{c,P}$ are fixed by the matching conditions at $s_1$ analogously to~\eqref{eq:matching_conditions}.

The $\pi^+\pi^0\to\pi^+\pi^0$ channel is slightly more involved since, in this case, the threshold opens at $\spipn=(M_{\pi^+}+M_{\pi^0})^2$. Starting with the $S$-wave, its parameterizations in terms of its phase shift and elasticity reads
\begin{equation}\label{eq:tp0S_param}
t^{+0}_S(s)=\frac{\eta^{+0}_S(s)\,\text
{e}^{2i\,\delta^{+0}_S(s)}-1}{4i\sigma_{+0}(s)}\;,
\end{equation}
with $\sigma_{+0}(s)=\lambda^{1/2}(s,M_{\pi}^2,\mpn^2)/s$. Once again, we assume that the elasticity is given by its isospin-limit value in~\cite{Colangelo:2001df,Caprini:2011ky},
and thus:
\begin{equation}
\eta_S^{+0}(s)=\bigg\{\begin{array}{lr}
1& s\le \spi\;,\\
\eta_0^2(s)&s>\spi\;.
\end{array}
\end{equation}
The phase shift is parameterized in terms of the isospin-limit phase \(\delta_0^2\), which is, however, only defined for energies above the charged–pion threshold $\spi$. To extend the isospin-limit phase down to the physical $\pi^+\pi^0$ threshold we use the kinematic map (see also~\cite{Colangelo:2018jxw})
\begin{equation}\label{eq:map+0}
\hat s(s)=\frac{s_1(s-\spipn)-\spi(s-s_1)}{s_1-\spipn}\;,
\end{equation}
which ensures that $\hat s(\spipn)=\spi$ and $\hat s(s_1)=s_1$.

Thus, in terms of this map, we parameterize the phase shift as
\begin{equation}\label{eq:+0Sparam}
\delta_S^{+0}(s)=\delta_0^2(\hat s(s))\left(1+\dpi\sum_{i=0}^7 c_i^{+0,S}\,\left(\frac{s-\spipn}{\spipn}\right)^i\right)\;.
\end{equation}

In order to ensure that at $\spipn$, the parametrization in~\eqref{eq:+0Sparam} coincides with the $a_c^{+0}$ scattering length, the coefficient $c_0^{+0,S}$ is fixed to the value
\begin{equation}
c_0^{+0,S}=\frac{1}{\dpi}\left[\frac{\spi}{\spipn}\sqrt{\frac{\mpn}{\mpi}}\sqrt{\frac{s_1-\spipn}{s_1-\spi}}\left(1+\frac{\Delta a_c^{+0}}{a_{c,\IL}^{+0}}\right)-1\right]\;,
\end{equation}
with $\Delta a_c^{+0}=a_c^{+0}-a_{c,\IL}^{+0}$. 
In the same way, the matching conditions with the isospin-limit results in~\cite{Colangelo:2001df,Caprini:2011ky} fix $c_6^{+0,S}$ and $c_7^{+0,S}$ analogously to~\eqref{eq:matching_conditions}. Their explicit expression is provided in Appendix~\ref{app:bcs1}.

The $\pi^+\pi^0\to\pi^+\pi^0$ $P$-wave is parameterized similarly. Namely, 
\begin{equation}\label{eq:tp0P_param}
t^{+0}_P(s)=\frac{\eta^{+0}_P(s)\,\text
{e}^{2i\,\delta^{+0}_P(s)}-1}{4i\sigma_{+0}(s)}\;,
\end{equation}
with 
\begin{equation}
\eta_P^{+0}(s)=\bigg\{\begin{array}{lr}
1& s\le \spi\;,\\
\eta_1^1(s)&s>\spi\;,
\end{array}
\end{equation}
and 
\begin{equation}\label{eq:+0Pparam}
\delta_P^{+0}(s)=\delta_1^1(\hat s(s))\left(1+\dpi\sum_{i=0}^7 c_i^{+0,P}\,\left(\frac{s-\spipn}{\spipn}\right)^i\right)\;,
\end{equation}
where $c_0^{+0,P}$ is a free parameter and the coefficients $c_6^{+0,P}$ and $c_7^{+0,P}$ are provided in Appendix~\ref{app:bcs1}.

\subsubsection{Coupled-channel case}

As already discussed in Sect.~\ref{sec:unitarity}, in the charge basis, the unitarity relation for the $T^n$, $T^c$, and $T^x$ $S$‑waves (see~\eqref{eq:ImTS}) does not become diagonal. Therefore, these waves must be parameterized through a coupled‐channel formalism. Labeling the $\pi^0\pi^0$ state as 1 and the $\pi^+\pi^-$ as 2, we can define a two-by-two inelastic $S$-matrix as:
\begin{equation}
S=\left(\begin{array}{cc}
\eta_S^n(s)\,\eta_S^x(s)\,\text{e}^{2i\delta_S^n(s)}&i\sqrt{1-\eta_S^x(s)^2}\,\text{e}^{i\left(\delta_S^n(s)+\delta_S^c(s)+\delta_S^x(s)\right)}\\
i\sqrt{1-\eta_S^x(s)^2}\,\text{e}^{i\left(\delta_S^n(s)+\delta_S^c(s)+\delta_S^x(s)\right)}&\eta_S^c(s)\,\eta_S^x(s)\,\text{e}^{2i\delta_S^c(s)}
\end{array}\right)\;,
\end{equation}
with the $T$-matrix elements connected to the $S$-matrix by standard coupled-channel relations
\begin{align}
S_{11}(s)=S^n(s)=&1+2i\sigma_0(s)t_{S}^n(s)\;,\nonumber\\
S_{12}(s)=S^x(s)=&2i\sqrt{2\sigma_0(s)\sigma(s)}t_{S}^x(s)\;,\nonumber\\
S_{22}(s)=S^c(s)=&1+4i\sigma(s)t_{S}^c(s)\;.
\end{align}
or, equivalently,
\begin{align}\label{eq:Ts}
t_{S}^n(s)=&\frac{\eta_S^n(s)\,\eta_S^x(s)\,\text{e}^{2i\delta_S^n(s)}-1}{2i\sigma_0(s)}\;,\nonumber\\
t_{S}^x(s)=&\frac{\sqrt{1-\eta_S^x(s)^2}\,\text{e}^{i\left(\delta_S^n(s)+\delta_S^c(s)+\delta_S^x(s)\right)}}{2\sqrt{2\sigma_0(s)\sigma(s)}}\;,\nonumber\\
t_{S}^c(s)=&\frac{\eta_S^c(s)\,\eta_S^x(s)\,\text{e}^{2i\delta_S^c(s)}-1}{4i\sigma(s)}\;.
\end{align}

Thus, this inelastic $S$-matrix parametrization is defined via three phase shifts and three inelasticities:
\begin{itemize}
\item $\delta_S^n$ and $\delta_S^c$ are the phase shifts of the neutral and charged channels, respectively.
\item $\eta_S^x$ denotes the $\pi^+\pi^-\to\pi^0\pi^0$ inelasticity, which, below the first non-$\pi\pi$ inelastic threshold, provides the pion-pion inelasticity in the charge basis.
\item $\eta_S^n$ and $\eta_S^c$ account for the non-$\pi\pi$ inelastic contributions of the neutral and charged channels, respectively.
\item $\delta_S^x$ provides the charge-exchange inelastic phase shift.
\end{itemize}

The neutral phase shift is parameterized as
\begin{equation}\label{eq:deltanS}
\delta_S^{n}(s)=\delta_S^{n,\text{IL}}(\bar s(s))\left(1+\dpi\left[\sum_{i=0}^7 c_i^{n}\,\left(\frac{s-\spin}{\spin}\right)^i+\sigma(s)\sum_{i=0}^4{\tilde c_i^n\left(\frac{s-\spin}{\spin}\right)^i}\right]\right)\;,
\end{equation}
where, analogously to~\eqref{eq:map+0}, the map
\begin{equation}\label{eq:sbarmap}
\bar s(s)=\frac{s_1(s-\spin)-\spi(s-s_1)}{s_1-\spin}\;,
\end{equation}
extends the isospin-limit results below $\spi$, and the coefficients $\tilde c_i^n$ allow the imaginary part of $t_S^n(s)$ to develop a new contribution starting at the charged-pion threshold. Indeed, as already noted in Sect.~\ref{sec:unitarity}, the phase-space factor $\sigma(s)$ in~\eqref{eq:deltanS} comes multiplied by $\theta(s-\spi)$, so it only contributes to $\delta_S^{n}(s)$ above $\spi$. 

The coefficient $c_0^n$ is fixed by the condition that the $t_n^S$ partial wave coincides at threshold with the isospin-breaking $a_n^{00}$ scattering length from \chptg. Namely, evaluating $t_n^S$ in~\eqref{eq:Ts} at the neutral-pion threshold one gets
\begin{equation}    t_S^n(\spin)=a_{n,\IL}^{00}\left(1+c_0^{n}\,\dpi\right)\sqrt{\frac{\spin\;\left(s_1-\spi\right)}{\!\!\!\spi\left(s_1-\spin\right)}}\;,
\end{equation}
so that, defining $\Delta a_n^{00}=a_{n}^{00}-a_{n,\IL}^{00}$, one gets
\begin{equation}
c_0^{n}=\frac{1}{\dpi}\left[\sqrt{\frac{\!\spi\left(s_1-\spin\right)}{\spin\;\left(s_1-\spi\right)}}\left(1+\frac{\Delta a_n^{00}}{a_{n,\IL}^{00}}\right)-1\right]\;.
\end{equation}
Finally, taking into account that $\spi=\spin+4\Delta_\pi$, at leading order in the pion-mass difference, this reduces to
\begin{equation}
c_0^{n}\simeq\frac{1}{\dpi}\frac{\Delta a_n^{00}}{a_{n,\IL}^{00}}\;.
\end{equation}

Once again, the coefficients $c_6^n$ and $c_7^n$ are fixed by imposing a continuous and differentiable matching at $s_1$. Their explicit expression, expanded at leading order in $\Delta_\pi$, is provided in Appendix~\ref{app:bcs1}.

In the same way, the charged $S$-wave phase shift is expressed as
\begin{equation}
\delta_S^{c}(s)=\delta_S^{c,\text{IL}}(s)\left(1+\dpi\left[\sum_{i=0}^7 c_i^{c,S}\,\left(\frac{s-\spi}{\spi}\right)^i+\sigma_0(s)\sum_{i=0}^3\,\tilde c_i^c\left(\frac{s-\spi}{\spi}\right)^i\right]\right)\;, 
\end{equation}
where, as in the $t_S^n$ case, the coefficients $\tilde c_i^c$ allow $t_S^c$ to develop an imaginary part starting at the neutral-pion threshold, as demanded by~\eqref{eq:ImTS}.

The coefficient $c_0^{c,S}$ is given by
\begin{equation}\label{eq:c0cS_val}
c^{c,S}_0=\frac{\Delta a^{+-}_c}{a^{+-}_{c,\text{IL}}\dpi}-\tilde c_0^c\,\sigma_0(\spi)\;,
\end{equation}
with $\Delta a^{+-}_c=a^{+-}_{c}-a^{+-}_{c,\text{IL}}$, and where the second term in~\eqref{eq:c0cS_val} is suppressed in $\dpi$.

Finally, the matching conditions at $s_1$ fix the coefficients $c_6^{c,S}$, $c_7^{c,S}$ as usual, and their expression is again provided in Appendix~\ref{app:bcs1}.

Below the first inelastic non-$\pi\pi$ threshold $s_\text{in}$\footnote{In this case, $s_\text{in}$ is assumed to correspond to the $\bar KK$ threshold, i.e., $s_\text{in}=4M_K^2$.}, $\eta_S^n=\eta_S^c=1$ and $\delta_S^x$=0, hence recovering the standard two-channel $S$-matrix parametrization. In this energy region, the elasticity parameter $\eta_S^x$ provides the $\pi\pi$-inelasticity among the three $T^n$, $T^c$ and $T^x$ scalar channels, so that
\begin{equation}
\eta_S^x=\vert S_{11}\vert=\vert S_{22}\vert=\sqrt{1-\vert S_{12}\vert^2}\quad\text{for}\quad s<s_\text{in}\;.
\end{equation}
Above $s_\text{in}$, $\eta_S^x$ accounts for the whole charge-exchange inelasticity, and we still allow for isospin-breaking corrections up to $s_1$. Above this energy, we impose once again the isospin-limit results in~\cite{Colangelo:2001df,Caprini:2011ky}.

To ensure the correct threshold behavior of the $t_S^x$ partial wave, as well as to describe the cusp effect of its imaginary part at both the charge- and neutral-pion thresholds, we parameterize this elasticity parameter as
\begin{equation}\label{eq:etax_param}
\eta_S^x(s)=\tilde\eta_S^{x}(s)\left( 1+\dpi\frac{\sqrt{(s-\spi)(s-\spin)}}{\spi}\sum_{i=0}^7c_i^x\left(\frac{s-\spi}{\spi}\right)^i\right)\;,
\end{equation}
with $\tilde\eta_S^{x}(s)$ defined in~\eqref{eq:iso_defs_in_S}.

Once again, the value of the coefficient $c_0^x$ is fixed by the $T^x$ isospin-breaking scattering length value $a_x^{+-}$ as computed in \chptg, i.e., 
\begin{align}
c_0^x=&\frac{4}{\dpi}\Bigg[(a_x^{+-})^2-(a_x^{+-}+\Delta a_x^{+-})^2\;\text{exp}\bigg\{i\left(1+ \dpi \sum_{k=0}^72^{2k}c_k^x \left(\frac{\Delta_\pi}{s_{00}}\right)^{k}\right) \\
&\times\tan^{-1}{\frac{4(a_x^{+-}+a_0^2)\sqrt{(s_1-s_{+-})(s_1(2s_{+-}-s_{00})-s_{+-}^2)\Delta_\pi}}{(s_1-s_{+-})^2-s_1(s_1-s_{00})+8(3a_x^{+-}+2a_x^{+-}a_0^2+2(a_0^2)^2)(s_1-s_{+-})\Delta_\pi}}\bigg\}\Bigg]\;,
\notag
\end{align}
with $\Delta a_x^{+-}=a_x^{+-}-a_{x,\IL}^{+-}$. At leading order in $\dpi$ this matching condition simplifies to
\begin{equation}
   c_0^x= -8\,a_{x,\IL}^{+-}\left(\frac{\Delta a_x^{+-}}{ \dpi}+\frac{2a_{n,\IL}^{00\;\;2}a_{x,\IL}^{+-}}{s_{+-}}\right)\;.
\end{equation}
Continuity and differentiability with the isospin-limit solution at $s_1$ fix the coefficients $c_6^x$ and $c_7^x$, and their expression is provided in Appendix~\ref{app:bcs1}.

In the same way, above $s_\text{in}$ the elasticity parameters $\eta_S^{n}$ and $\eta_S^c$ fulfill $0\le \eta_S^{n},\,\eta_S^c\le 1$, and their values are fixed once again from the isospin-limit results in~\cite{Colangelo:2001df,Caprini:2011ky}. Namely, 
\begin{align}
\eta_S^n(s)=&\left\{\begin{array}{lr}1&\quad s\le\spi\;,\\\displaystyle\frac{\eta_S^{n,\IL}(s)}{\eta_S^{x,\IL}(s)}=\frac{\vert 1+2i\sigma(s)\,\left(t_0^0(s)+2t_0^2(s)\right)/3}{\sqrt{1-8\sigma(s)^2\left\vert \left(t_0^0(s)-t_0^2(s)\right)/3\right\vert^2}}&\quad s>\spi\;,\end{array}\right.\\
\nonumber\\
\eta_S^c(s)=&\frac{\eta_S^{c,\IL}(s)}{\eta_S^{x,\IL}(s)}=\frac{\vert 1+4i\sigma(s)\,\left(2t_0^0(s)+t_0^2(s)\right)/6 \vert}{\sqrt{1-8\sigma(s)^2\left\vert \left(t_0^0(s)-t_0^2(s)\right)/3\right\vert^2}}\;.
\end{align}

Finally, $\delta_S^x$ is non-vanishing only above $s_\text{in}$ and provides the charge-exchange phase shift beyond the elastic approximation, i.e., 
\begin{equation}
\delta_S^x=\arg\left(2\sqrt{2}\,i\,\sigma(s)\left(t_0^0(s)-t_0^2(s)\right)/3\right)-\delta_S^{n,\IL}(s)-\delta_S^{c,\IL}(s)\;.
\end{equation}

\subsection{Imaginary parts in the inelastic regime}\label{sec:imag_inelastic}

Below the first non-$\pi\pi$ inelastic threshold, the isospin-breaking parameterizations derived in the previous section, combined with the unitarity relations discussed in Sect.~\ref{sec:unitarity}, provide enough information to compute the imaginary part of the amplitudes in the charge basis. However, the Roy equations beyond the isospin limit, as derived in Sect.~\ref{sec:roy-beyond}, require these imaginary parts to be defined up to $s_1$, an energy that lies above $s_\text{in}$. 
Thus, in the kinematic region $s_\text{in}<s\le s_1$, the unitarity relations from Sect.~\ref{sec:unitarity} have to be extended to incorporate inelastic contributions. Namely, 

\paragraph{Amplitude $T^n$:}

\begin{equation}\label{eq:uni-tn_ine}
\text{Im}t_{S}^n(s)=\text{Im}t_{S}^{n,00}(s)+\text{Im}t_{S}^{n,+-}(s) =\sigma_0(s) \left|t_{S}^n(s)\right|^2+ 2\sigma(s)\left|t_{S}^x(s)\right|^2+\frac{\eta_s^x(s)^2(1-\eta_S^n(s)^2)}{4\sigma_0(s)}.
\end{equation}

\paragraph{Amplitudes $T^c$ and $T^{++}$:}
\begin{align}\label{eq:uni-tn-and-tpp_ine}
\text{Im}t_{S}^c(s) =&\text{Im}t_{S}^{c,00}(s)+\text{Im}t_{S}^{c,+-}(s) =\sigma_0(s) \left|t_{S}^x(s)\right|^2+2\sigma(s) \left|t_{S}^c(s)\right|^2+\frac{\eta_s^x(s)^2(1-\eta_S^c(s)^2)}{8\sigma(s)},\nonumber \\ 
\text{Im} t_{P}^{c,+-}(s)=&2\sigma(s) \left|t_{P}^c(s)\right|^2+\frac{1-\eta_P^c(s)^2}{8\sigma(s)},\nonumber \\ 
\text{Im} t_{S}^{++}(s)=& \sigma(s) \left| t_S^{++}(s)\right|^2+\frac{1-\eta_S^{++}(s)^2}{4\sigma(s)}.
\end{align}

\paragraph{Amplitudes $\pi^+ \pi^- \to \pi^0 \pi^0$ and $\pi^+ \pi^0 \to \pi^+ \pi^0$:}

\begin{align}
\text{Im}t_{S}^x(s) &=\text{Im}t_{S}^{x,+-}(s)+\text{Im}t_{S}^{x,00}(s) \nonumber\\
&= 2\sigma(s) t_{S}^x(s)t_{S}^c(s)^*+\sigma_0(s)
t_{S}^n(s) t_{S}^x(s)^*\nonumber \\
&\quad+\frac{i\,\eta_S^x(s)\sqrt{1-\eta_S^x(s)^2}}{4\sqrt{2\sigma_0(s)\sigma(s)}}\text{e}^{i(\delta_S^n(s)-\delta_S^c(s))}\left(\eta_S^n(s)\text{e}^{i\,\delta_S^x(s)}-\eta_S^c(s)\text{e}^{-i\,\delta_S^x(s)}\right)\;,\nonumber \\
\text{Im}t_{S}^{+0}(s) &=
2\sigma_{+0}(s)\left|t_{S}^{+0}(s)\right|^2+\frac{1-\eta_S^{+0}(s)^2}{8\sigma_{+0}(s)}\;,\nonumber \\  
\text{Im}t_{P}^{+0}(s) &=2\sigma_{+0}(s) \left|
  t_P^{+0}(s)\right|^2+\frac{1-\eta_P^{+0}(s)^2}{8\sigma_{+0}(s)}\;.
\end{align}

\subsection{Strategy for the numerical solution}

The isospin-breaking parameterizations defined in Sect.~\ref{sec:IB_param} together with their corresponding imaginary parts in Sect.~\ref{sec:imag_inelastic} provide enough information to check whether Roy equations for $\Delta M_\pi\neq 0$ are satisfied. Specifically, by using the imaginary parts as input, the right-hand side (RHS) of Roy equations in Sect.~\ref{sec:roy-beyond} can be computed and then compared with their left-hand sides (LHS), which are directly determined by the real part of the parameterizations. 

Thus, a numerical solution to Roy equations can be obtained by evaluating and minimizing the least-square difference between the LHS and RHS of the Roy equation over a mesh of $N$ points in the seven $S$- and $P$-wave amplitudes in the charge basis, i.e.,  
\begin{equation}\label{eq:LS_def}
\Delta^2_\text{Roy}=\sum_{k,X}\sum_{j=1}^N{\left(\text{Re}\,t_X^{\,k}(s_j)-F\left[t_X^{\,k}\right](s_j)\right)^2}\;,
\end{equation}
where $k$ labels the channels in the charged basis, $X$ denotes the partial wave ($S$ or $P$), and $s_j$ are squared-energy points taken between the $k$-channel threshold and $s_{\mathrm{max}}=(0.975\ \mathrm{GeV})^2$ (slightly below the $K\bar K$ threshold). Here $F[t_X^{\,k}](s)$ denotes the Roy-equation RHS beyond the isospin limit. The numerical Roy solution is then obtained by varying the free parameters to minimize $\Delta_{\mathrm{Roy}}^2$. 

An exact solution of~\eqref{eq:LS_def} corresponds to $\Delta_\text{Roy}=0$, meaning Roy equations are perfectly satisfied. In principle, this renders the specific definition of $\Delta_\text{Roy}$ irrelevant. However, in practice, the input quantities---such as scattering lengths and driving terms that influence the matching conditions---are only known within finite precision. This introduces systematic uncertainties in the solutions. 

Given this scenario, the precise definition of $\Delta_\text{Roy}$ used in the minimization process may influence the quality of an approximated solution. 
Ideally, a $\chi^2$-function would be the most suitable choice for minimization algorithms. However, the lack of well-defined statistical errors precludes its direct application. To address this and to ensure that all partial waves are treated uniformly, regardless of their relative magnitude, we adopt the following function:
\begin{equation}\label{eq:LS_norm_def}
\Delta^2_\text{Roy}=\sum_{k,X}\sum_{j=1}^N{\left(\frac{\text{Re}\,t_X^{\,k}(s_j)-F\left[t_X^{\,k}\right](s_j)}{\text{Re}\,t_X^{\,k}(s_j)}\right)^2}\;,
\end{equation}
where $N$ is varied between 50 to 150 to ensure the stability of the results, and in the end, fixed to 100. 

In this way, using the parametrization parameters in Sect.~\ref{sec:IB_param} as fitting parameters, we look for solutions to Roy equations beyond the isospin limit by minimizing the merit function in~\eqref{eq:LS_norm_def}.

\section{Numerical results}\label{sec:results}

We begin by minimizing the merit function $\Delta_{\mathrm{Roy}}$ up to a maximum energy $s_{\mathrm{max}}=\linebreak(0.975\ \mathrm{GeV})^2$, chosen to lie slightly below the $K\bar K$ threshold and
before non-$\pi\pi$ inelasticities in the $S$-waves $T^n$, $T^c$ and $T^x$ become significant. In addition, we impose matching conditions to the isospin-limit solution at $s_1$, the highest energy at which Roy equations are solved; for $s>s_1$ the $\pi\pi$ input is taken from experimental data. 

The scattering lengths, which appear as subtraction constants in the isospin-breaking Roy equation, are derived from the $\chi$PT$_\gamma$ predictions expanded at ${\cal O}\left(\delta_\pi^2\right)$, whose expressions are explicitly given in Sect.~\ref{sec:chpt}. For their numerical evaluation, we use the PDG values~\cite{ParticleDataGroup:2024cfk} for the charged and neutral pion masses,
\begin{equation}
M_{\pi^\pm}=(139.57039\pm0.00017)\;\mathrm{MeV}\;,\quad
M_{\pi^0}=(134.9768\pm0.0005)\;\mathrm{MeV}\;.
\end{equation}

In addition, for the pion-decay constant, we take the FLAG value~\cite{FlavourLatticeAveragingGroupFLAG:2024oxs}
$$F_\pi=(92.2\pm0.1)\;\mathrm{MeV}\;,$$
and the renormalization scale $\mu$ is set to $\mu=770$ MeV.
For the mesonic low-energy constants (LECs), we use
\begin{equation}
\lbar_1=-0.4\pm0.6\;,\quad\lbar_2=4.3\pm0.1\;,\quad\lbar_3=3.3\pm0.3\;,\quad\lbar_4=4.4\pm0.2\;,
\end{equation}
where the values of $\lbar_1$, $\lbar_2$, and $\lbar_4$ are taken from~\cite{Bijnens:2014lea}, and $\lbar_3$ is obtained as the average between the $N_f=2+1+1$ and $N_f=2+1$ FLAG values~\cite{FlavourLatticeAveragingGroupFLAG:2024oxs}.

Finally, for the electromagnetic LECs ($k_i$) we use the numerical estimates given in~\cite{Haefeli:2007ey},
\begin{align}
\Bar{k}_1=&(1.7\pm1.3)\times 10^{-3}\;,\quad\Bar{k}_2=(4.1\pm1.2)\times 10^{-3}\;,\quad\Bar{k}_3=(2.3\pm2.7)\times 10^{-3}\;,\nonumber\\
\Bar{k}_4=&(3.8\pm1.2)\times 10^{-3}\;,\quad\Bar{k}_6=(4.1\pm1.3)\times 10^{-3}\;,\quad\Bar{k}_8=(2.2\pm2.9)\times 10^{-3}\;.
\end{align}

With these numerical values, we can directly evaluate the $\chi$PT$_\gamma$ expressions for the scattering lengths given in Sect.~\ref{sec:chpt}, obtaining  the following deviations from their isospin-limit values:
\begin{align}
\Delta a_n^{00}=&-5.375\times 10^{-3}\;, \quad\Delta a^{++}=2.918\times 10^{-3}\;,\quad\Delta a_c^{+-}=4.076\times 10^{-3}\;,\nonumber\\
\Delta a_1=&-0.711\times 10^{-3}\;,\quad\Delta a_2=-1.467\times 10^{-3}\;.
\end{align}

In addition, the values for $\Delta a_x^{+-}$ and $\Delta a_c^{+0}$, which are required for evaluating the parameterizations of the $t^{+0}_S$ and $t_S^x$ partial waves, can be computed from $\Delta a_1$ and $\Delta a_2$ using the relations~\eqref{eq:a1rel} and~\eqref{eq:a2rel}, respectively.
But these expressions involve the same parameterizations we aim to compute, which lead to an implicit system of equations and significantly complicating the $\Delta_\text{Roy}$ minimization problem. Instead, since the partial-wave contributions in~\eqref{eq:a1rel} and~\eqref{eq:a2rel} are already suppressed by $\Delta_\pi$ and $\Delta_\pi^2$, respectively, we can initially estimate the parameterization effect using the isospin-limit results and update their values only in a later iteration. In this way, the starting value for the $a_x^{+-}$ and $a_c^{+0}$ scattering length difference reads:
\begin{equation}
\Delta a_x^{+-}\big\vert_0=0.081\times 10^{-3}\;,\quad\Delta a_c^{+0}\big\vert_0=-1.474\times 10^{-3}\;,
\end{equation}
where the subscript 0 indicates that these are starting values for both quantities. These results show only a small correction relative to $\Delta a_2$ for $a_c^{+0}$, but a sizable effect for $a_x^{+-}$ compared to $\Delta a_1$.

The scattering lengths in the isospin limit are taken from the Roy-equation analysis in~\cite{Colangelo:2001df,Caprini:2011ky} and read $a_0^0=0.220$ and $a_0^2=-0.0444$, which in the charge-basis translate to the values
\begin{align}
a_{n,\,\text{IL}}^{00}=0.0437\;,&\quad a^{++}_{\text{IL}}=-0.0444\;,\quad a_{c,\,\text{IL}}^{+-}=0.0659\;,\quad\nonumber\\
a_{x,\,\text{IL}}^{+-}&=0.0881\;,\quad a_{c,\,\text{IL}}^{+0}=-0.0222\;.
\end{align}

Once the scattering length values are fixed, we can begin searching for solutions. To ensure a well-behaved minimum and prevent artificial correlations that might amplify isospin-breaking corrections, we start adiabatically, introducing each parameterization parameter one at a time, stopping when the value of the merit function per number of parameter $\Delta_\text{Roy}^2/N_\text{par}$ no longer improves. 
This procedure results in six free parameters for the $P$ waves and five for the $S$ partial waves. The exceptions are $t_S^n$ and $t_S^c$, which, due to the additional cusp effect, have another 5 and 4 parameters, respectively. Proceeding in this way, we obtain merit function value of  $\Delta_\text{Roy}^2=1.4\cdot 10^{-4}$, achieving a level of consistency between the LHS and RHS of the Roy equations comparable to that in the isospin limit. 

\begin{figure}
    \centering    \includegraphics[width=0.9\linewidth]{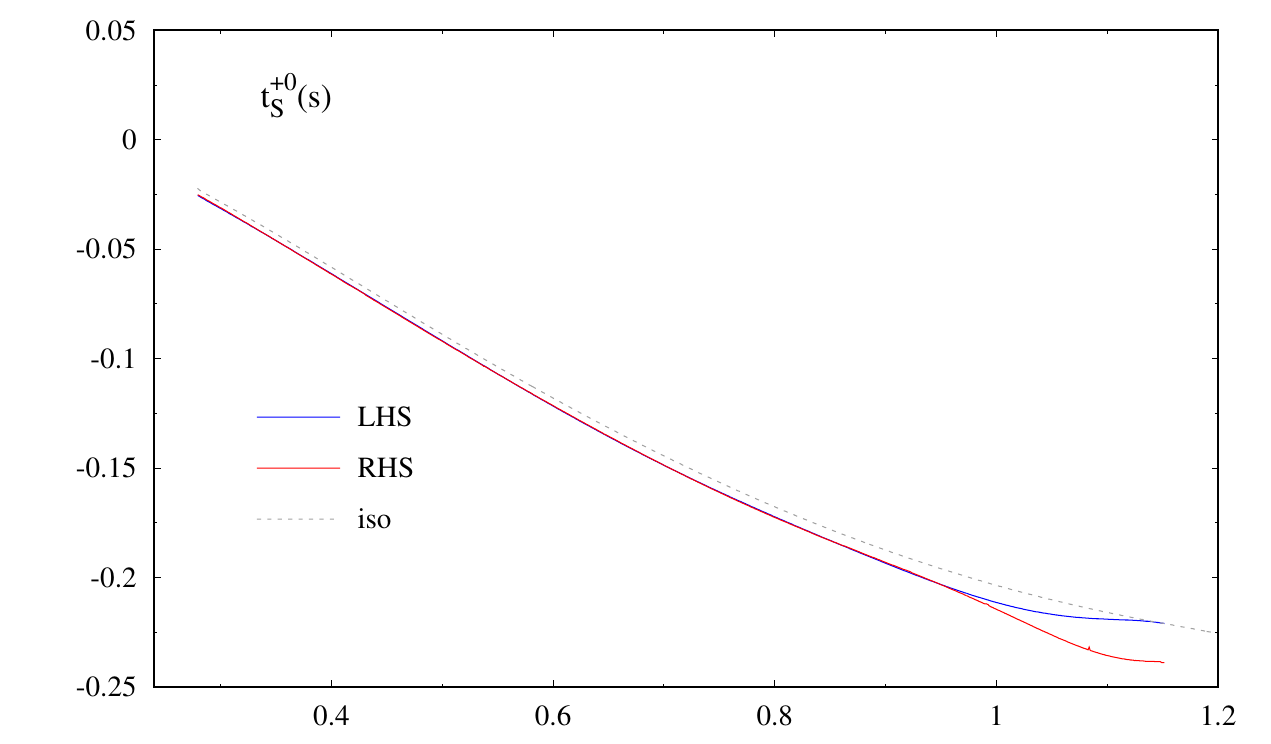}
    \caption{Result for the real part of the $t_S^{+0}(s)$ partial wave, with matching conditions imposed at $s_1=(1.15\,\GeV)^2$ to the isospin-limit result (dashed gray line). Roy equations are imposed only in the $\pi\pi$ elastic region so that the LHS (blue) and RHS (red) curves start deviating above $s_\text{max}=(0.975\,\GeV)^2$.}
    \label{fig:+0S-s1}
\end{figure}

While the results for the $T^n$, $T^c$, and $T^x$ $S$-waves, as well as the $P$-waves, indicate small isospin-breaking corrections---enhanced primarily at the threshold and in the resonance region---the repulsive, non-resonant $T^{++}$ and $T^{+0}$ $S$-waves exhibit a more uniform effect. Notably, in the $T^{+0}$ case, pion-mass difference effects even increase with energy, leading to an unphysical bending enforced by the matching conditions at $s_1$, as shown in Fig.~\ref{fig:+0S-s1}. This result reflects the unique role of the $S2$ wave in the Roy--equations framework: this is the weakest among the leading three $S$ and $P$ waves, and tends to adapt to what the other two stronger waves do. When isospin breaking is introduced, this feature is inherited by the partial waves more strongly linked to the $S2$ wave, the $T^{+0}$ and $T^{++}$ $S$-waves. If one assumes that the isospin-breaking corrections go to zero at $s_1$ in all waves (a reasonable assumption that can only be approximately true), the residual isospin-breaking effects from all waves appear in the weakest wave (in the Roy equation machinery). This results in a significant deviation from the isospin limit solution at $s_1$. The situation is compounded by the fact that the experimental input for the $I=2$ channel is affected by comparatively large uncertainties (see Fig.~5 in Ref.~\cite{Ananthanarayan:2000ht}).
To address this issue, we impose the matching conditions only at higher energies, namely at $s_2=(2\;\GeV)^2$, and parameterize the $t_S^{+0}$ input above $s_1$ using the polynomial,
\begin{equation}\label{eq:+0_matching_input}
 t_S^{+0}(s)\big\vert_\text{input}=t_S^{+0,\,\text{IL}}(s)+b_S^{+0}\frac{s-s_2}{s_1-s_2}\left(1-\frac{s-s_2}
 {s_1-s_2} \right)\;,\quad\text{for}\quad s_1\le s\le s_2\;,
 \end{equation}
which ensures a continuous and differentiable matching with the isospin-limit solution at $s_{2}$, while allowing for a shift at $s_1$ given by 
 \begin{equation}
 \Delta t_S^{+0}(s_1)=t_S^{+0}(s_1)-t_S^{+0,\,\text{IL}}(s_1)=b_S^{+0}\;,
 \end{equation}
where $b_S^{+0}$ is a complex free parameter determined through the $\Delta_\text{Roy}$ minimization.
Note that above $s_1$, we only need as input $\text{Im}\,{t_S^{+0}}$, but below, the isospin-breaking parameterization in~\eqref{eq:+0Sparam} is expressed in terms of its phase shift and elasticity, the latter being fixed from data. Consequently, in order to ensure also a continuous matching at $s_1$, only the real or imaginary part of the parameter $b_s^{+0}$ can be chosen freely. 
In practice, we leave $\text{Re}\,b_S^{+0}$ free so that its imaginary part is given by
\begin{equation}
\text{Im}\, b_S^{+0}=-\text{Im}\,t_S^{+0,\,\text{IL}}+\frac{1-\sqrt{\eta_S^{+0}(s_1)^2-16\,\sigma_{+0}(s_1)^2\left(\text{Re}\,t_S^{+0,\,\text{IL}}(s_1)+\text{Re}\,b_S^{+0}\right)^2}}{4\sigma_{+0}(s_1)}\;.
\end{equation}

Within this new scheme, we minimize the merit function $\Delta_\text{Roy}$ once again, using both the isospin-breaking parameterization parameters and $\text{Re}\,b_S^{+0}$ as free parameters. As a result, we obtain a value for the merit function $\Delta_\text{Roy}^2=1.2\times 10^{-4}$, slightly below our previous results when the matching to the isospin-limit was imposed at $s_1$. The partial waves remain nearly unchanged compared to our previous results, except for $t_S^{+0}$ near $s_1$, where the unphysical bending observed in Fig.~\ref{fig:+0S-s1} now disappears, obtaining at $s_1$ the difference to the isospin limit, 
\begin{equation}
\Delta \text{Re}\,t_S^{+0}(s_1)=\text{Re}\,t_S^{+0}(s_1)-\text{Re}\,t_S^{+0,\,\text{IL}}(s_1)=\text{Re}\,b_S^{+0}=-5.56\times 10^{-3}\;,
\end{equation}
which amounts to a $2.5\%$ effect.
Furthermore, to estimate the impact of the asymptotic matching to the isospin-limit at $s_2$ for the $t_S^{+0}$ partial wave---and how this affects the value of $\text{Re}\,b_S^{+0}$ and hence, the deviation from the isospin-limit at $s_1$---we solve Roy-equations for $\Delta_\pi\neq0$ once again, varying the asymptotic value of the $\text{Im}\,t_0^2$ partial wave within uncertainties. This variation affects not only the matching conditions but also the value of the driving terms, allowing us to quantify the corresponding changes in the solution.

\begin{figure}
    \centering    \includegraphics[width=0.485\linewidth]{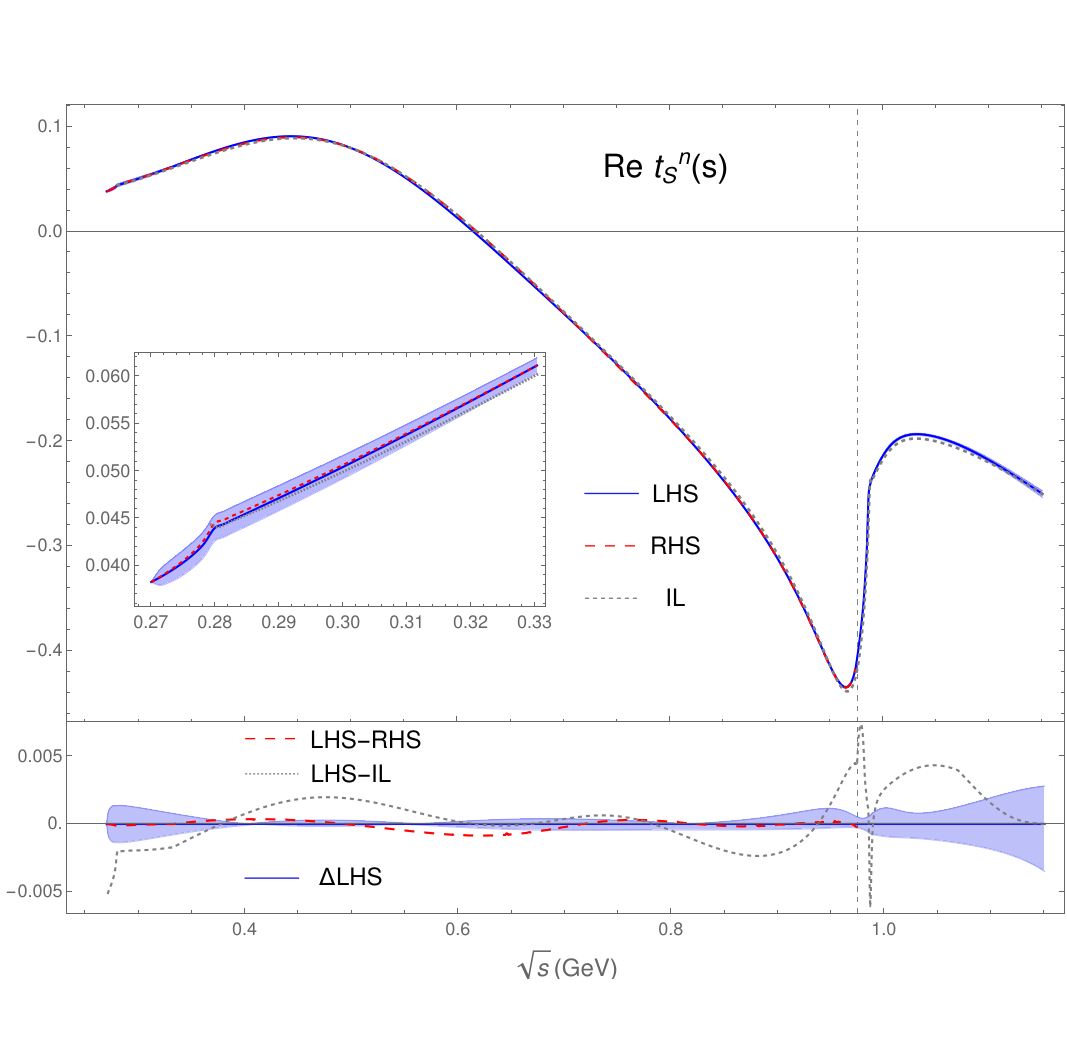}\quad\includegraphics[width=0.485\linewidth]{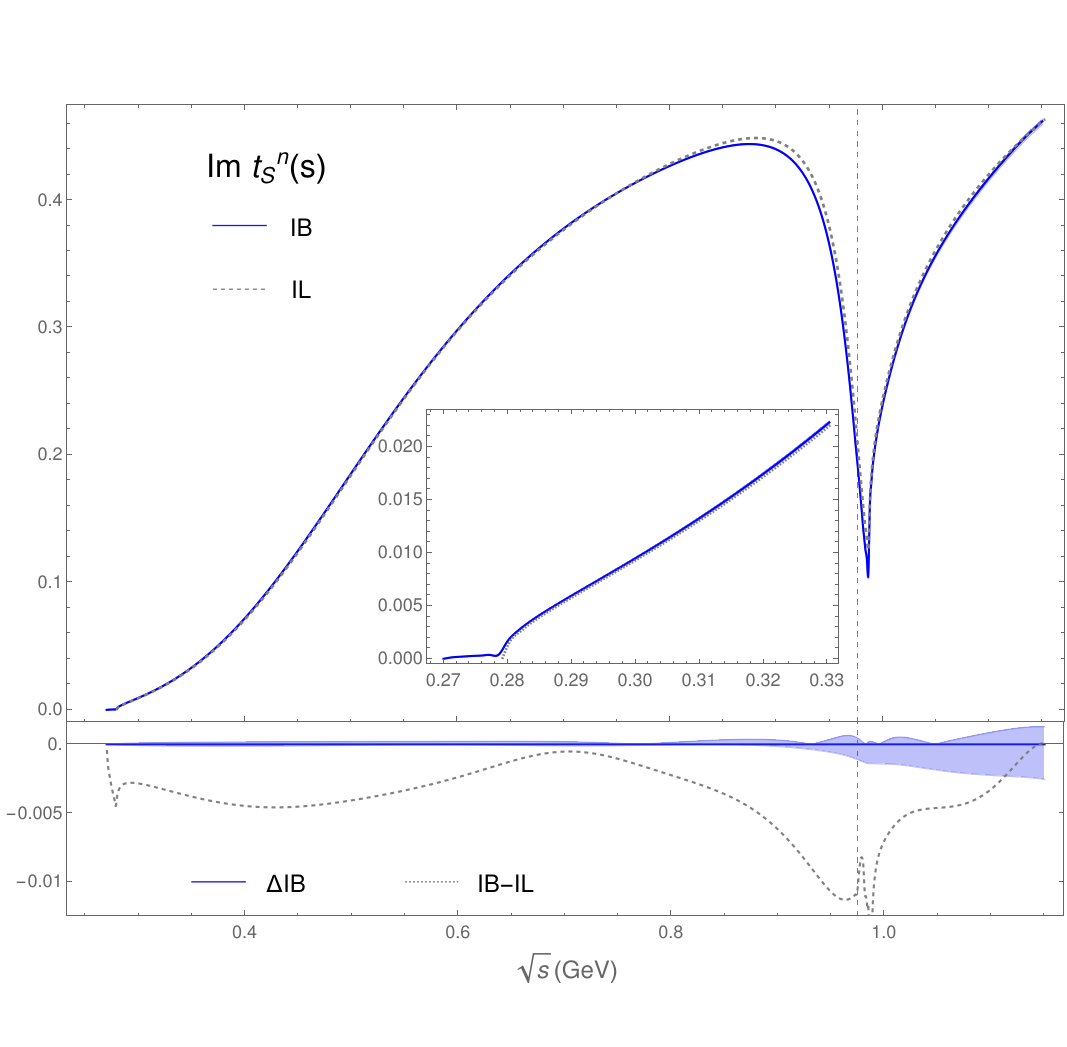}
    \caption{Results for the real (left panel) and imaginary (right panel) parts of the $t_S^{n}(s)$ partial wave. For the real part, we display the isospin-breaking parameterization, i.e., the LHS of Roy equations (solid blue), and the dispersive representation, i.e., the RHS (dashed red), along with the isospin-limit result (gray-dotted line). For the imaginary part, since unitarity is exactly satisfied, we plot only the isospin-breaking parameterization and the isospin-limit result.
    The blue band represents the uncertainty in the isospin-breaking parameterization due to variations in the asymptotic value of $\text{Im}\,t_0^2$. 
    The inset figures in both panels highlight the low-energy region, where the effect of the neutral-pion threshold becomes visible. At the bottom of both panels, we show the difference between the isospin-breaking and isospin limit parameterizations (gray-dotted line), along with the uncertainty band of the isospin-breaking parameterization. The isospin-limit result is evaluated using the variable $\bar s(s)$, defined in~\eqref{eq:sbarmap}, which maps the charged-pion threshold into the neutral one, allowing for a direct comparison of both results at the same energies. For the real part, we also plot at the bottom the difference between the LHS and RHS of Roy equations for $\Delta_\pi\neq 0$.}
    \label{fig:tnS}
\end{figure}

\begin{figure}
    \centering    \includegraphics[width=0.485\linewidth]{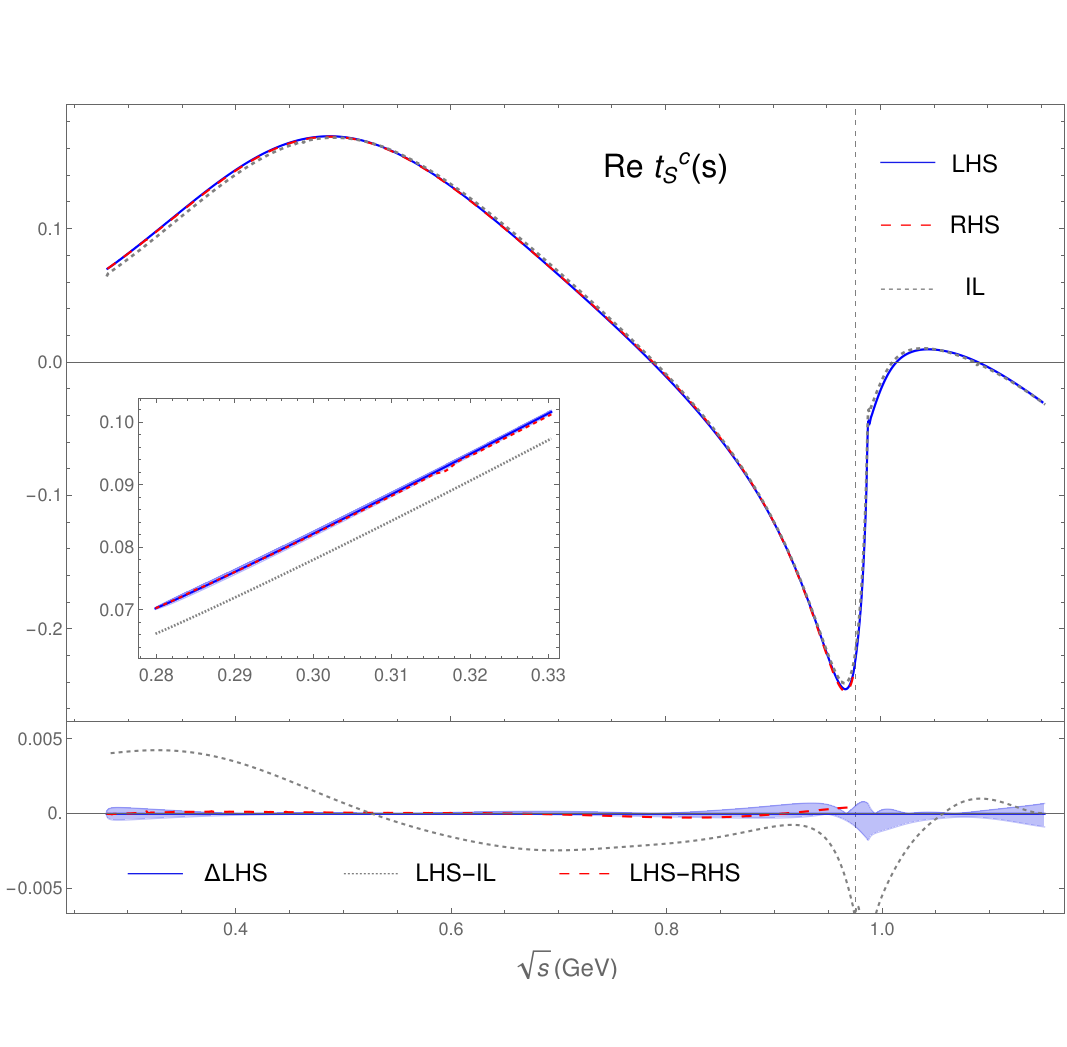}\quad\includegraphics[width=0.485\linewidth]{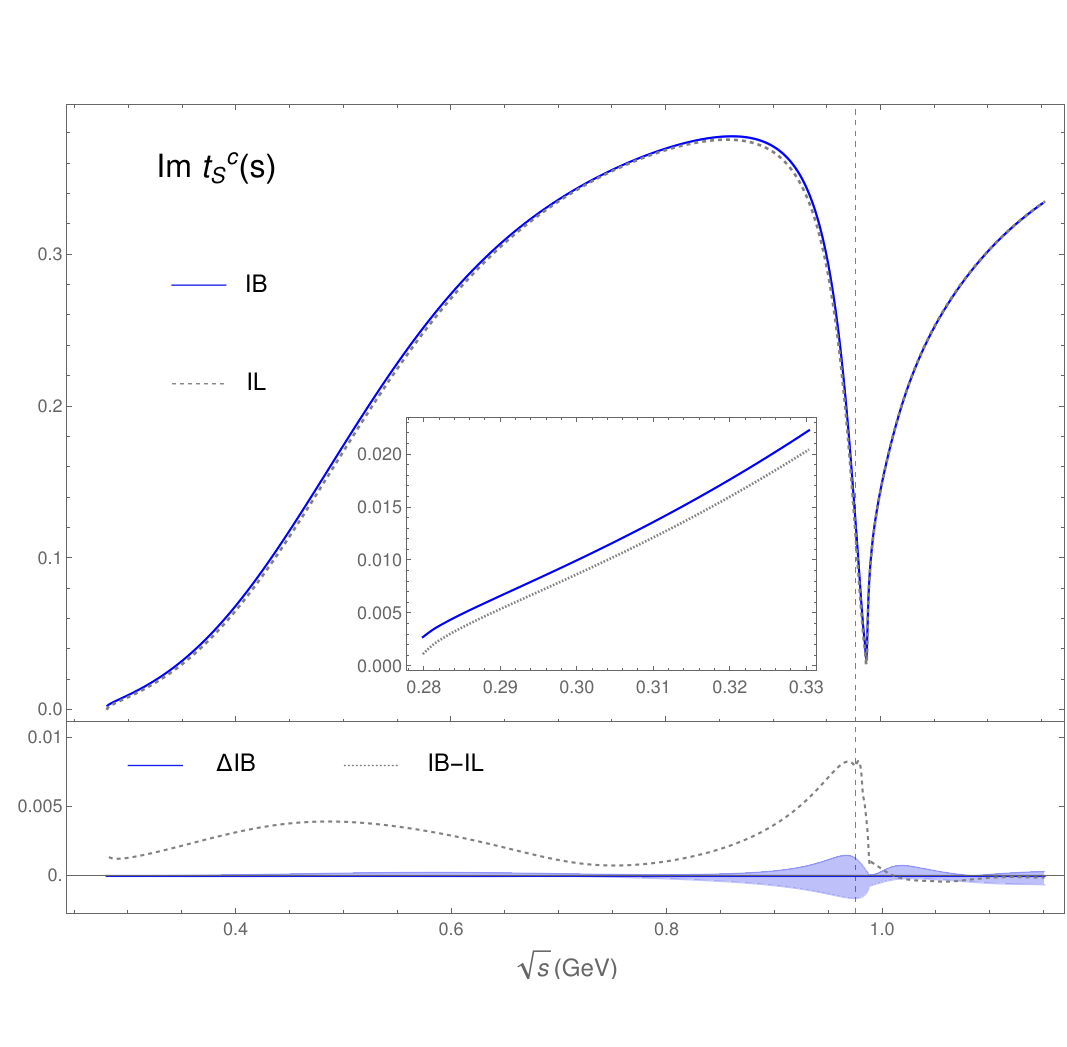}
    \caption{We compare the isospin-breaking and isospin-limit results for the real (left panel) and imaginary part (right) of the $t_S^{c}(s)$ partial wave. The different curves follow the conventions in Fig.~\ref{fig:tnS}. In this case, the physical region for both the isospin-limit and isospin-breaking parameterizations starts at the charged-pion threshold. Thus, the shift observed at low energies in the inset figures originates from the $\chi$PT$_\gamma$ correction to the scattering length.}
    \label{fig:tcS}
\end{figure}

\begin{figure}
    \centering    \includegraphics[width=0.485\linewidth]{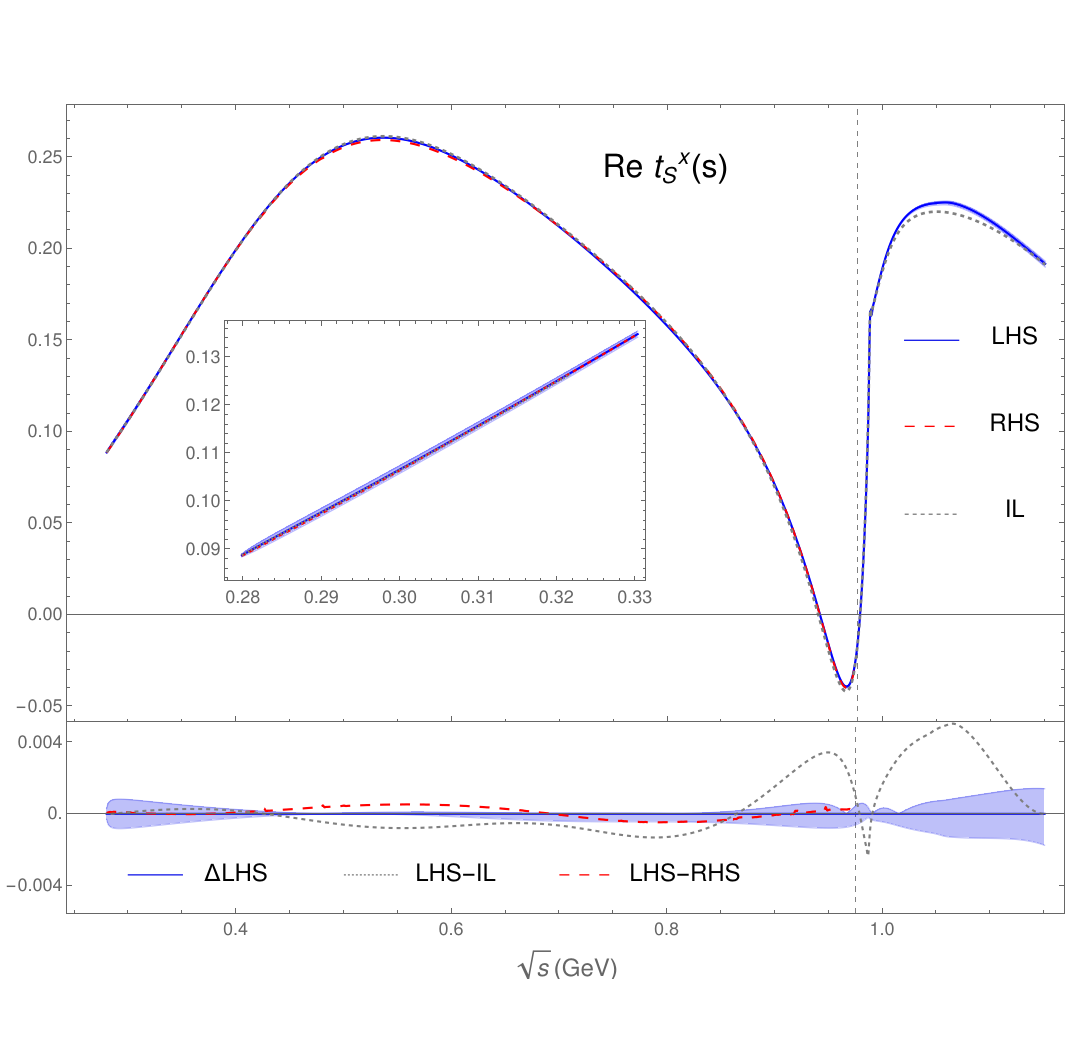}\quad \includegraphics[width=0.485\linewidth]{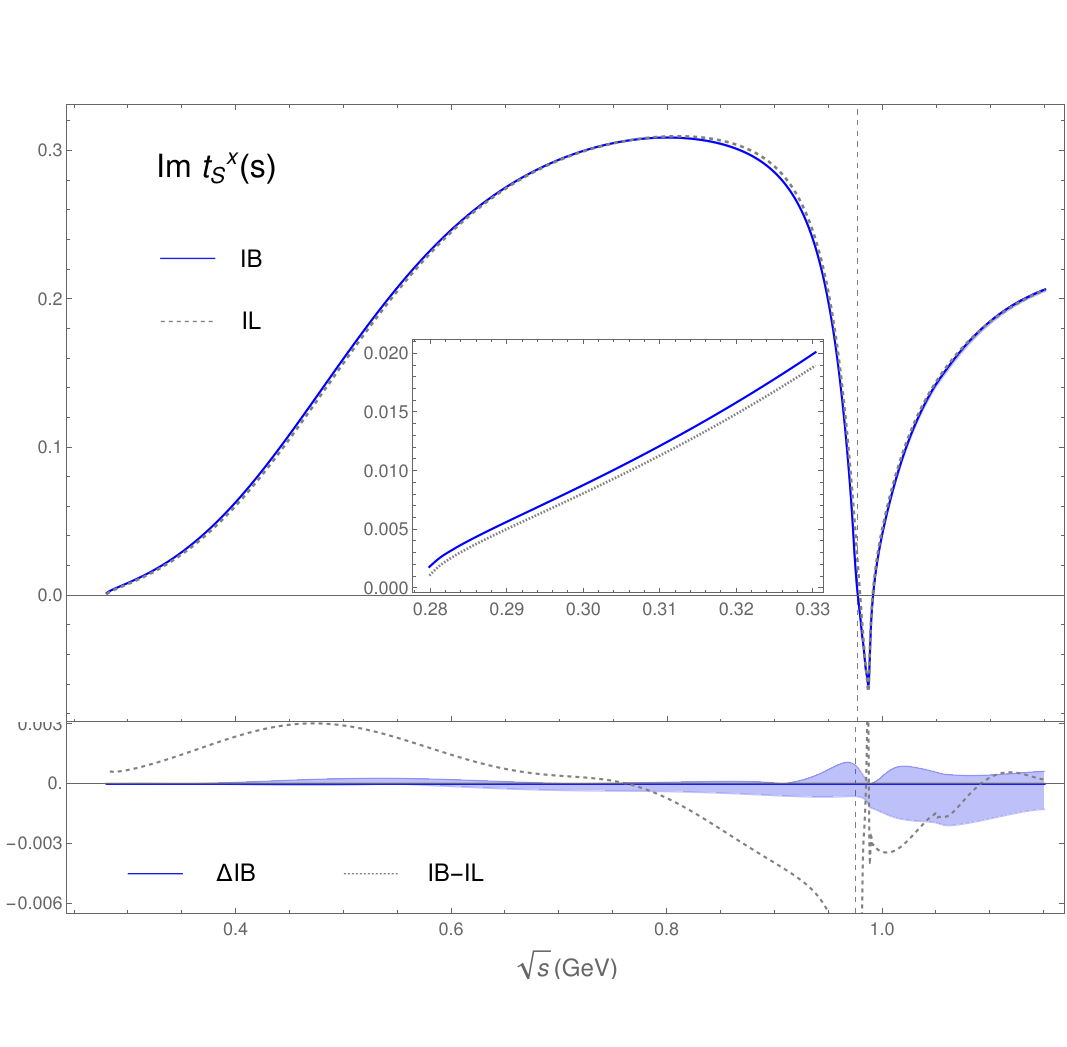}
    \caption{Results for the real (left) and imaginary (right) part of the $t_S^{x}(s)$ partial wave. The curves follow the same conventions as in Fig.~\ref{fig:tnS}. The imaginary part of both the isospin-breaking and isospin-limit partial waves opens at the charged-pion threshold. At this energy, the pion-mass difference correction is given by $\Delta a_x^{+-}$, whose value in \chptg~leads to the small shift observed in the inset figures. }
    \label{fig:txS}
\end{figure}

\begin{figure}
    \centering    \includegraphics[width=0.485\linewidth]{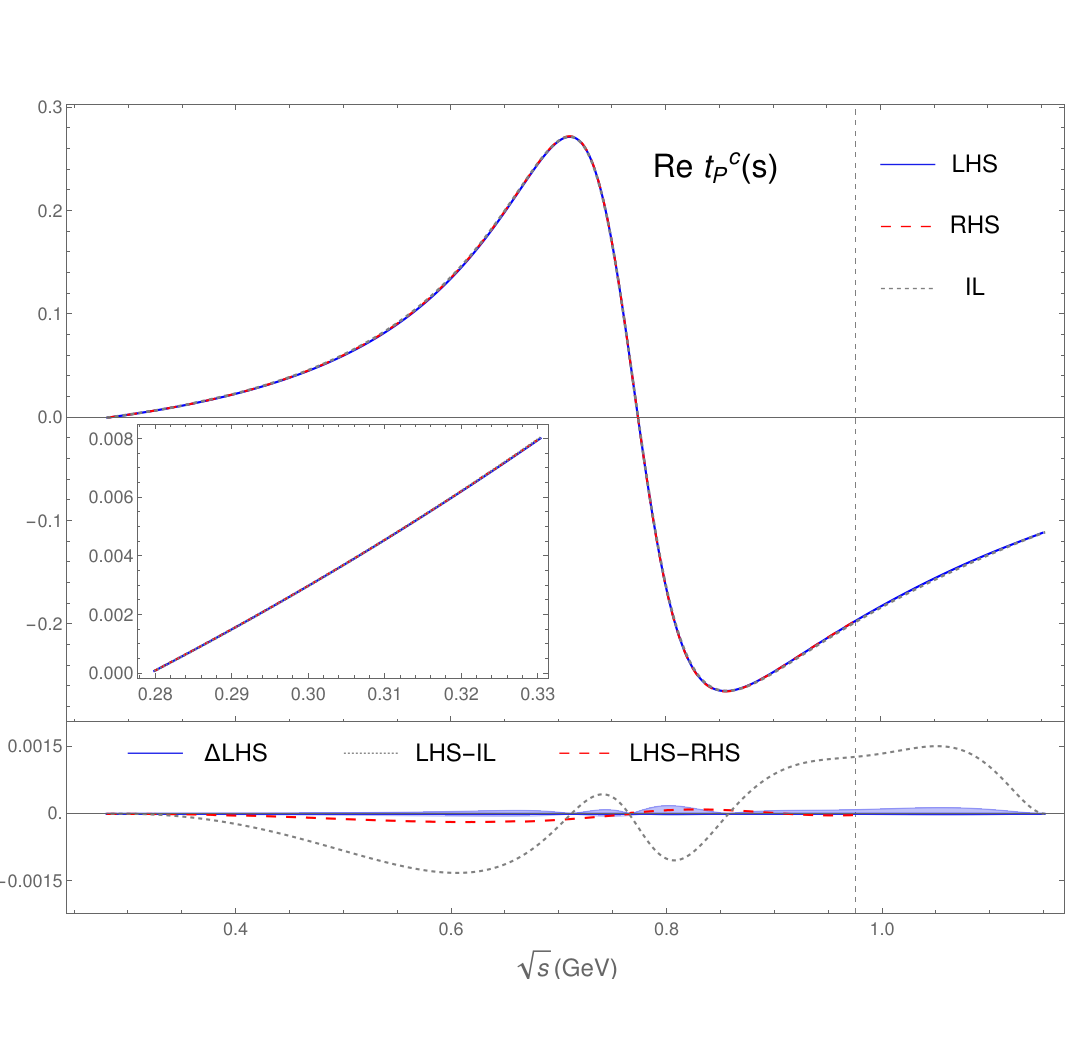}\quad\includegraphics[width=0.485\linewidth]{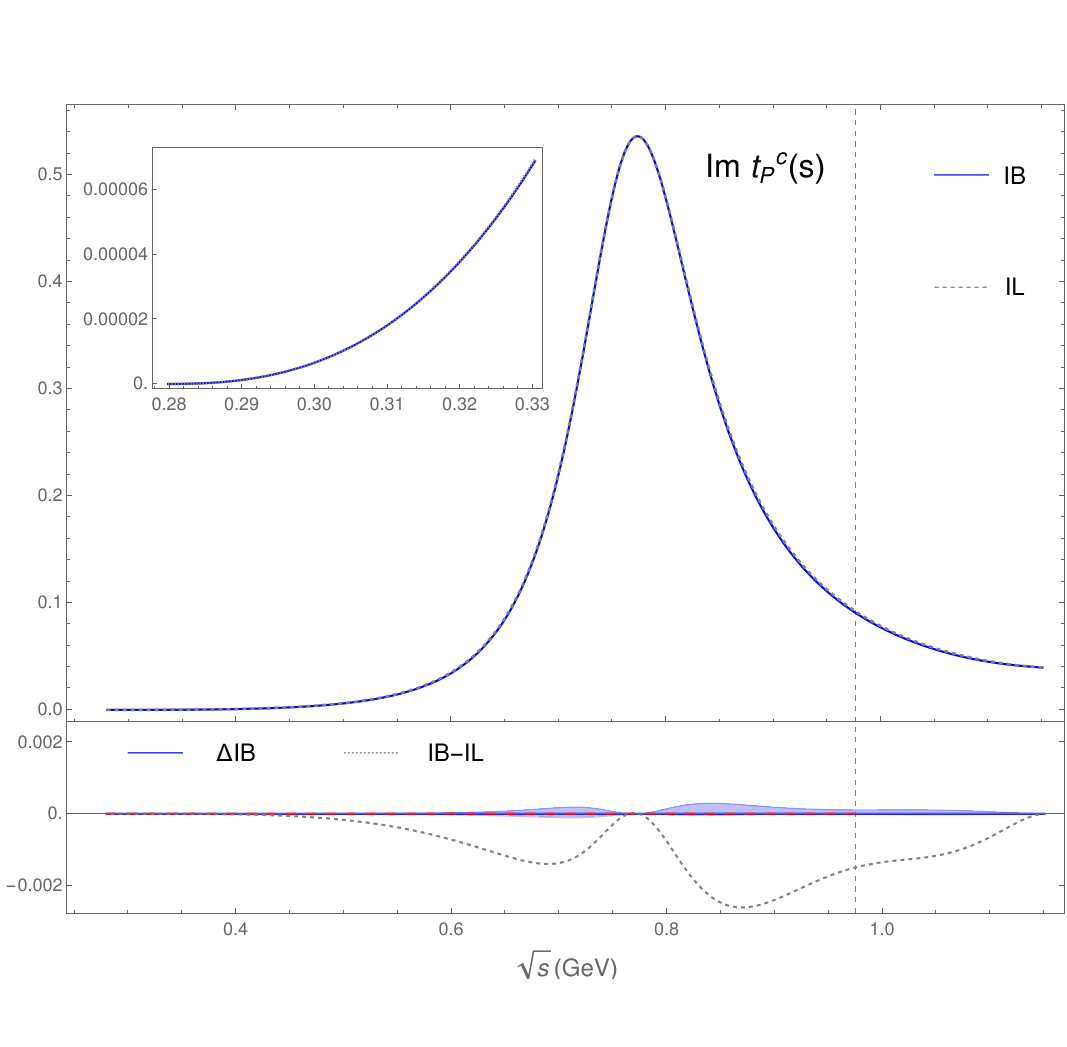}
    \caption{We show the isospin-breaking and isospin-limit results for the real (left panel) and imaginary (right panel) of the $t_P^{c}(s)$ partial wave. The bottom of each figure displays the difference between the isospin-breaking and isospin-limit results, highlighting the minimal impact of pion-mass corrections in this case. For the real part, we also depict the LHS and RHS of Roy equations. Both curves almost coincide in the entire elastic region ($\spi\le s\le s_\text{in}$) with their difference---displayed in the bottom panel---remaining well below the deviation from the isospin-limit result.}
    \label{fig:tcP}
\end{figure}

\begin{figure}
    \centering    \includegraphics[width=0.485\linewidth]{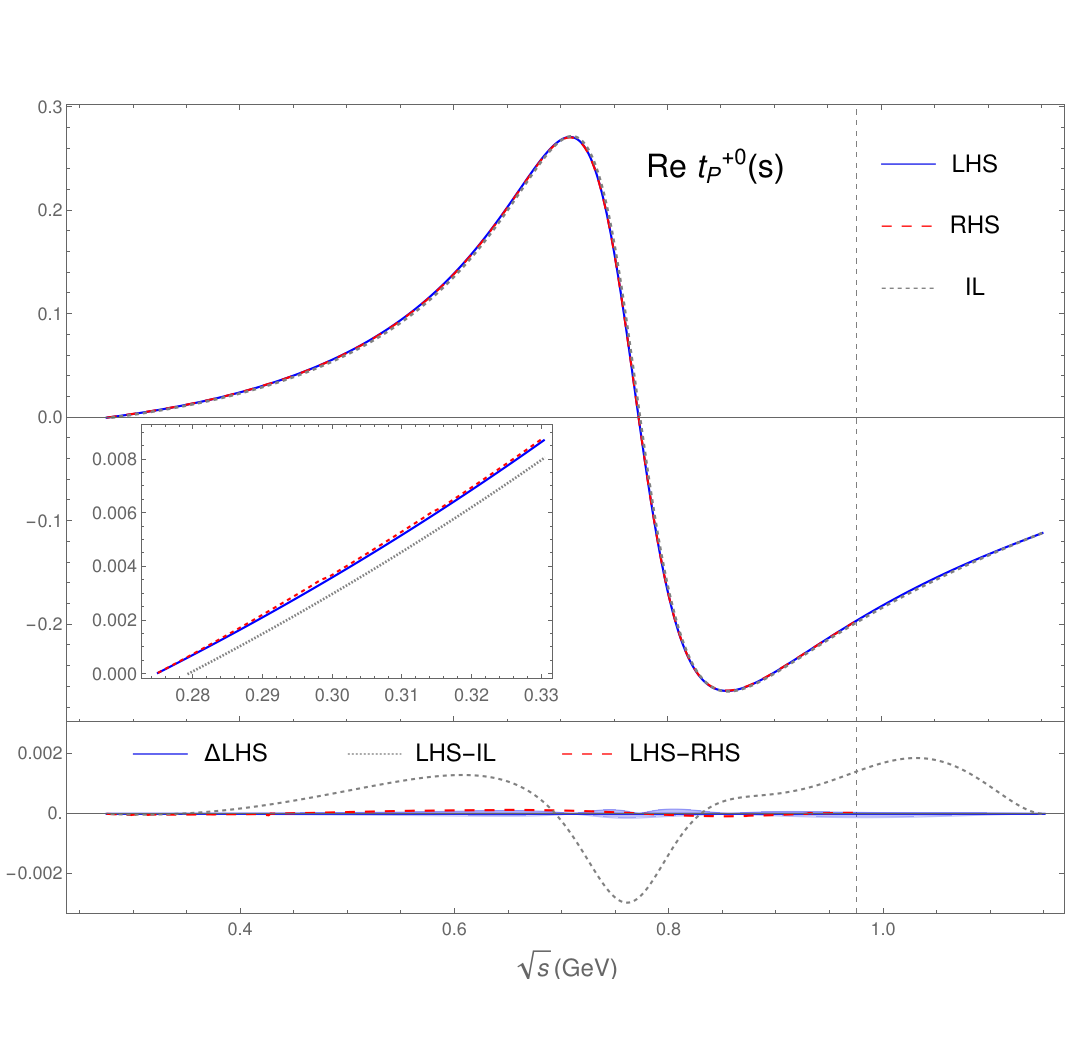}\quad\includegraphics[width=0.485\linewidth]{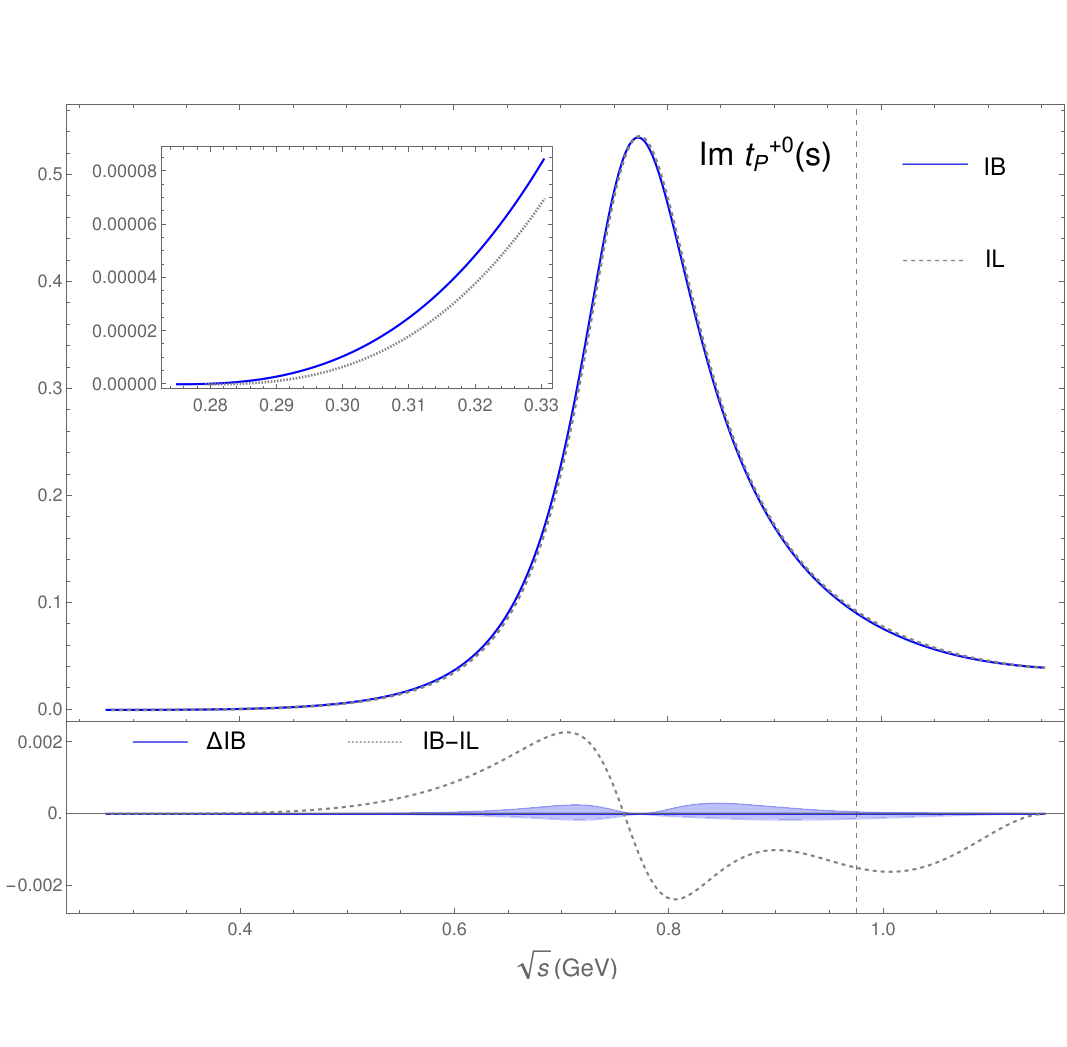}
    \caption{Comparison between the isospin-breaking (blue-solid line) and isospin-limit (gray-dotted) results for the real (left panel) and imaginary (right panel) part of the $t_P^{+0}(s)$ partial wave. In the isospin-breaking case, the physical region starts at $\spipn$, whereas in the isospin limit, it opens at the charged-pion threshold $\spi$. This explains why, despite being a $P$-wave, a shift appears between the isospin-breaking and isospin-limit results at low energies.}
    \label{fig:tp0P}
\end{figure}

\begin{figure}
    \centering    \includegraphics[width=0.485\linewidth]{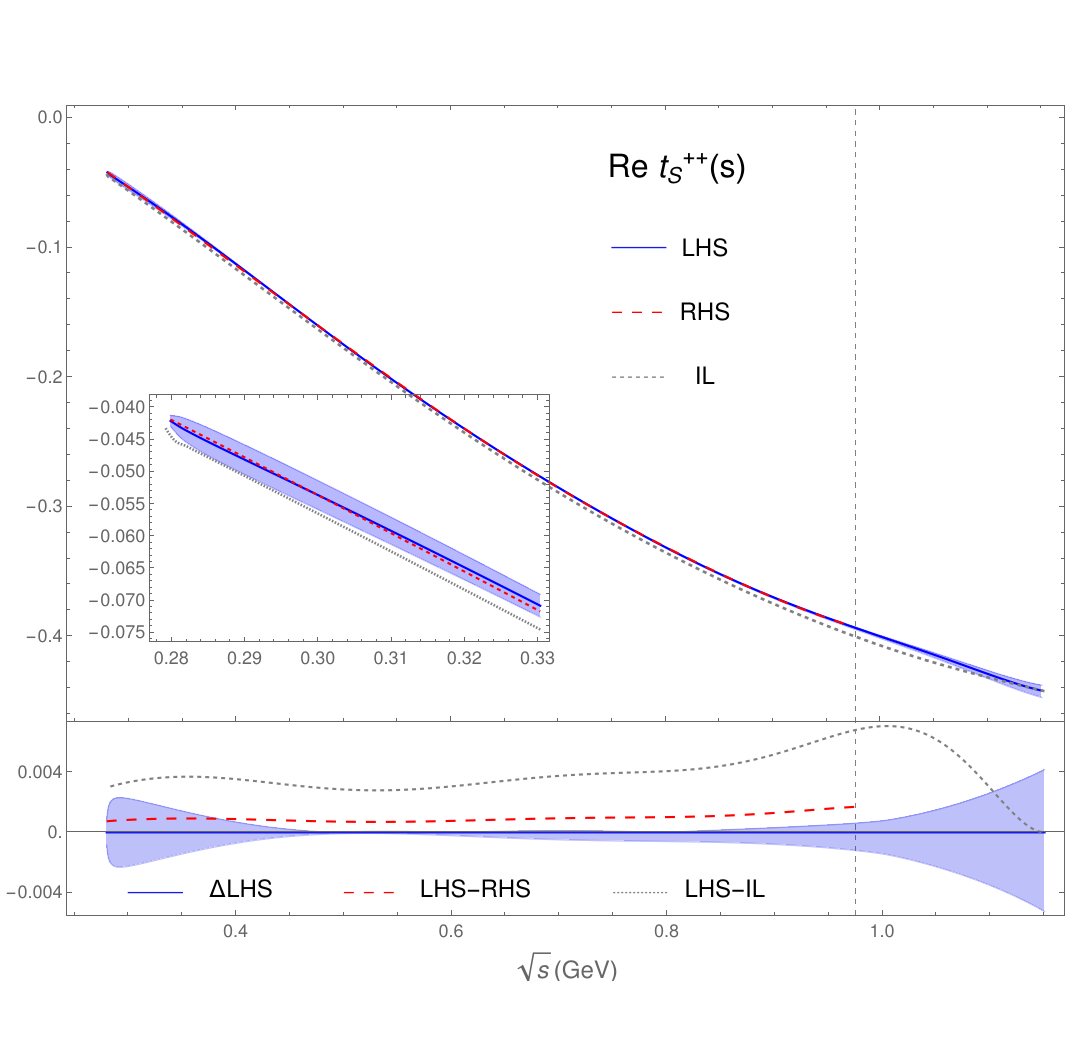}\quad\includegraphics[width=0.485\linewidth]{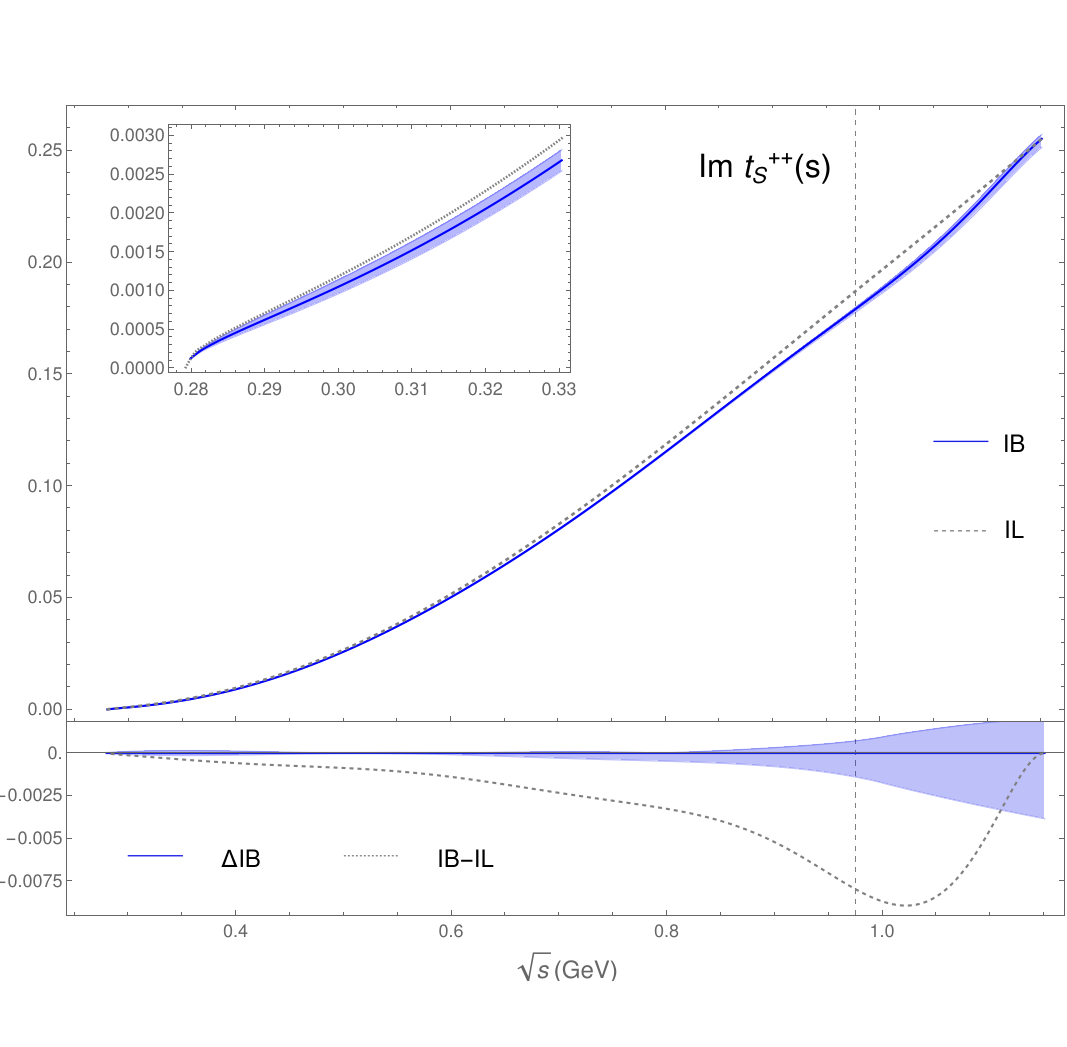}
    \caption{Results for the real (left panel) and imaginary (right) part of the $t_S^{++}(s)$ partial wave. The physical region for both the isospin-breaking (solid-blue line and band) and isospin-limit (gray-dotted curve) partial waves starts at the charged-pion threshold. The pion-mass correction at threshold is given by $\Delta a^{++}$, whose value is determined in \chptg. For the real part of the partial wave, we also include the RHS of the Roy equation (dashed-red line). The RHS$-$LHS difference, shown at the bottom, remains consistently smaller than the pion-mass correction. }
    \label{fig:tppS}
\end{figure}

\begin{figure}
    \centering    \includegraphics[width=0.485\linewidth]{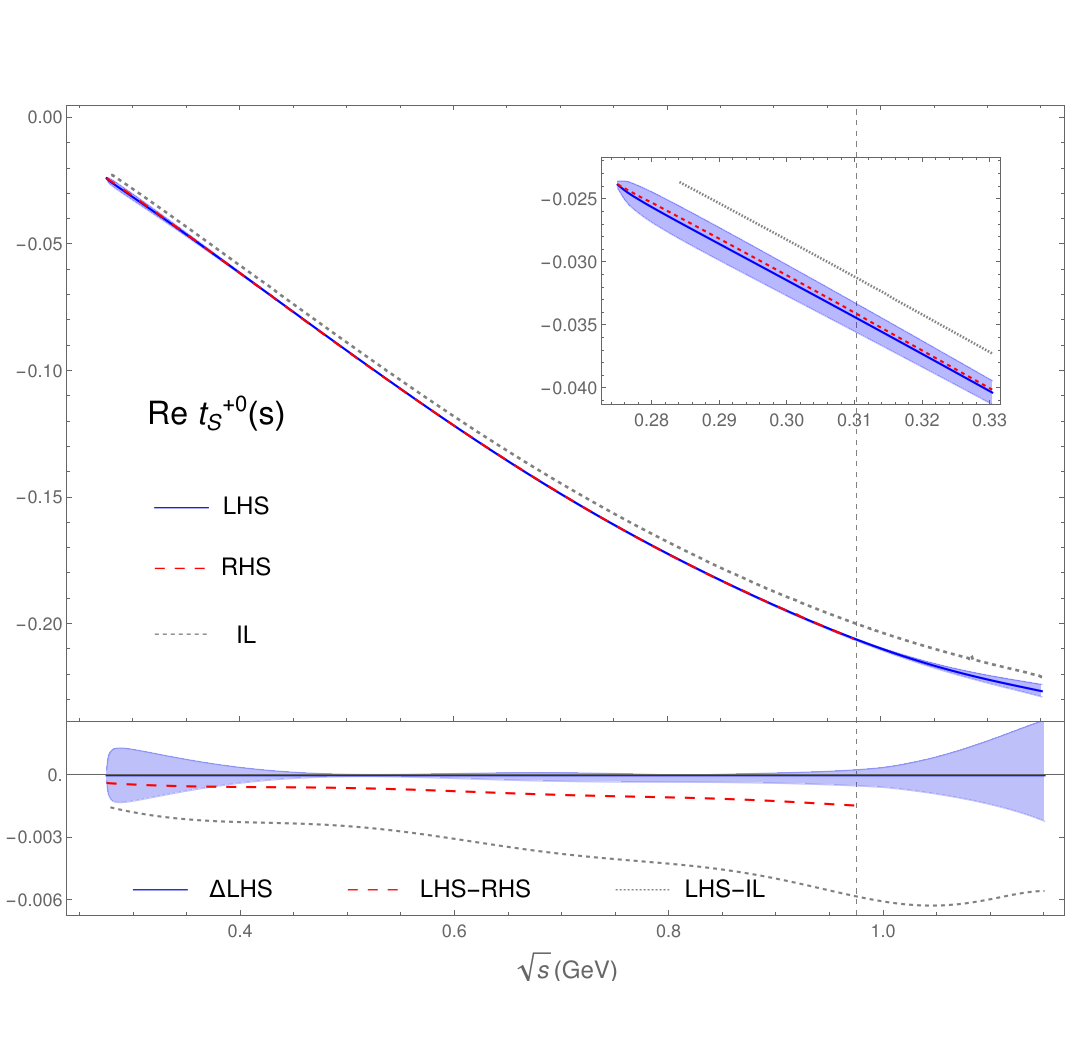}\quad\includegraphics[width=0.45\linewidth]{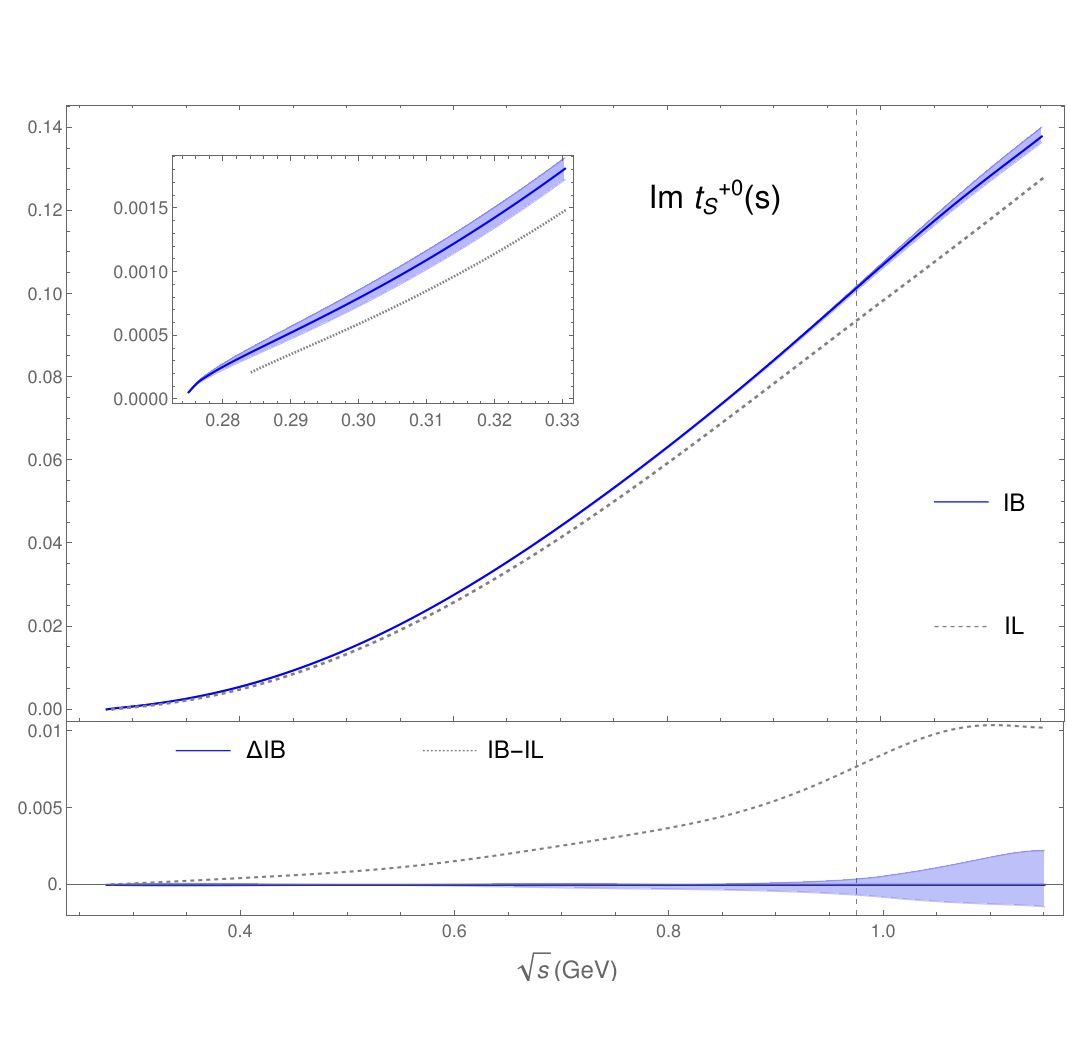}
    \caption{Comparison between the isospin-limit and isospin-breaking results for the part (left panel) and imaginary part (right panel) of the $t_S^{+0}(s)$ partial wave. In the isospin-breaking case, the imaginary part opens at $\spipn$, whereas in the isospin limit, it starts at $\spi$. This shift, together with the \chptg~correction to the $\Delta a_c^{+0}$ scattering length, leads to the significant pion-mass corrections observed in the inset figures at low energy, reaching a maximum of approximately 6\% at threshold. At higher energies, these corrections gradually decrease and stabilize around 3\% at $\sqrt s \sim 0.5$. At $s_1$, and only for this wave, an additional pion-mass difference correction is required in the fit, encoded in the parameter $\text{Re}\,b_S^{+0}=-5.56\times 10^{-3}$. The bottom of both panels shows the difference between the isospin-breaking and isospin-limit parameterizations (gray-dotted line), along with the uncertainty band for the isospin-breaking parameterization. For the real part, we also display the difference between the RHS and LHS of the Roy equation (dashed-red line), which remains significantly smaller than the pion-mass correction.  The isospin-limit result is evaluated using the variable $\hat s(s)$, defined in~\eqref{eq:map+0}, which maps the charged-pion threshold onto the $\pi^+\pi^0$ threshold, enabling a direct comparison between the isospin-breaking and isospin-limit results at the same energies.} 
    \label{fig:tp0S}
\end{figure}

The resulting coefficients in the isospin-breaking parameterization are given in Table~\ref{tab:parameters}  and the final values for the corrections to $a_x^{+-}$ and $a_c^{+0}$ read 
\begin{equation}
\Delta a_x^{+-}=0.032\times 10^{-3}\;,\quad\Delta a_c^{+0}=-1.479\times 10^{-3}\;,
\end{equation}
which highlights a tiny effect for $a_c^{+0}$, but makes the correction for $a_x^{+-}$ even smaller in magnitude than the initial estimate.
\begin{table}
\renewcommand{\arraystretch}{1.2}
    \centering
    \adjustbox{width=\linewidth}{\begin{tabular}{l r r r r r r r r r r r}
                 &
                 $c_0\times10^2$ & $c_1\times10^2$&$c_2\times10^2$&$c_3\times10^3$&$c_4\times10^4$&$c_5\times10^5$ & $\tilde c_0\times10$ & $\tilde c_1\times10^2$&$\tilde c_2\times10^3$&$\tilde c_3\times10^4$&$\tilde c_4\times10^5$\\\toprule
         $t_S^n$ & --- & $35.7$ & $-13.3$ &  $9.95$ & $14.5$ &$-30.9$&$21.9$&$-13.7$&$11.2$&$11.3$&$-6.22$\\\midrule 
         $t_S^{c}$  & --- & $-66.7$ &  $18.5$ & $-28.9$ & $20.6$&$-6.47$ &$-0.953$ & $13.6$& $-10.8$& $12.5$& ---\\\midrule
         $t_P^{c}$ & $4.78$ & $-26.6$ &  $8.16$ & $-5.24$ & $-8.05$ & $13.8$ & --- &  --- & --- &  --- & --- \\\midrule
         $t_S^{x}$ & --- & $-1.02$ & $-0.139$ &  $3.17$ &$-6.86$ & $5.94$ & --- &  --- & --- &  --- & ---\\\midrule
         $t_S^{++}$ & ---   &  $74.6$ & $-26.2$ &  $42.5$ &$-29.6$ & $1.50$ & --- & --- & --- & --- & ---\\\midrule
         $t_S^{+0}$ & ---  &  $-78.7$ & $29.4$ & $-55.2$ &$55.3$ & $-29.1$ &  --- & --- & --- & --- & ---\\\midrule
         $t_P^{+0}$ & $7.91$ & $22.5$ & $-8.01$ &  $9.85$ &$-4.89$ & $0.497$ & --- &  --- & --- & --- & ---\\\bottomrule
    \end{tabular}}
    \caption{Values of the parameter used to describe the isospin-breaking correction in the phases. }
    \label{tab:parameters}
\end{table} 
The real and imaginary parts of the seven $S$- and $P$-wave $\pi\pi$ amplitudes in the charge basis
are shown in Figs.~\ref{fig:tnS}--\ref{fig:tp0S}, where we compare the isospin-breaking parameterizations with the isospin-limit result. The uncertainty band of the isospin-breaking parameterizations reflects the effect of varying the asymptotic value of $\text{Im}\,t_0^2$.
For the real part of the partial waves, we plot both the LHS and the RHS of Roy equations. In all cases, their difference is almost negligible, reflecting how well Roy equations for $\Delta_\pi\neq 0$ are satisfied.

Moreover, at the bottom of each figure, we include the difference between the isospin-breaking and isospin-limit results, illustrating the size of the pion-mass difference corrections. These corrections should be compared with the difference between the LHS and RHS of the Roy equations and the isospin-breaking uncertainty band, which generally remain smaller across the entire energy region. This ensures that the pion-mass difference corrections lie well above our intrinsic uncertainties. 

In more detail, Fig.~\ref{fig:tnS} displays the results for the $t_S^{n}$ partial wave, where one can observe that while the physical region starts at the charged-pion threshold for the isospin-limit parameterization, it opens at the neutral-pion threshold in the isospin-breaking case, thereby amplifying the size of the correction. The corrections are also enhanced in the $f_0(500)$ and $f_0(980)$ resonance regions, while remaining small elsewhere, allowing for a smooth matching with the isospin limit at $s_1$. In more quantitative terms, using the variable $\bar s(s)$ in~\eqref{eq:sbarmap} for the evaluation of the isospin-limit result, the pion-mass difference correction for the real part of the $t_S^n(s)$ partial wave is around 12\% at the neutral pion-threshold, which by construction coincides with the size of $\Delta a_n^{00}$ as computed in $\chi$PT$_\gamma$, below 3\% in the $f_0(500)/\sigma$ region ($\sqrt s\sim 0.5$ GeV), and around 7.5 \% near 1 GeV, the $f_0(980)$ region.

Figs.~\ref{fig:tcS}~and~\ref{fig:txS} show the results for the $t_S^c(s)$ and $t_S^x(s)$ partial waves. In either case, the physical region starts at the charged-pion threshold for both the isospin-limit and isospin-breaking results. Thus, the differences at threshold correspond to the size of the scattering-length differences computed in $\chi$PT$_\gamma$, around 6\% for $t_S^c$ and almost negligible for $t_S^x$. The corrections become smaller at higher energies, with a moderate increase in the $f_0(980)$ resonance region of around 5\%. Near $s_1$, the effect is below 1\%, ensuring a smooth matching with the isospin-limit results.

Figs.~\ref{fig:tcP}~and~\ref{fig:tp0P} depict the results for the $P$-waves. In both cases, the $P$-wave centrifugal barrier effect ensures that they vanish at their corresponding threshold, i.e., $t_P^{c}(\spi)=t_P^{+0}(\spipn)=0$. Nevertheless, while the physical region for the $T^c$ $P$-wave starts at the charged-pion threshold, making the isospin-breaking results indistinguishable from the isospin-limit at low energies, it begins at $\spipn$ for the $t_P^{+0}$, leading to the pion-mass difference corrections observed in Fig.~\ref{fig:tp0P}.
In both cases, pion-mass difference corrections remain small---below 2\% over the entire energy range studied---rising to roughly 3\% only in the $\rho(770)$ resonance region.

Finally, Figs.~\ref {fig:tppS} and~\ref{fig:tp0S} show the results for the repulsive $S$-waves. In this case, the uncertainty bands become significantly larger due to the strong dependence of the $t_S^{++}(s)$ and $t_S^{+0}(s)$ partial waves on the asymptotic value of $\text{Im}\,t_0^2(s)$. While the imaginary part of $t_S^{++}(s)$ opens at the charged-pion threshold---so that its difference from the isospin-limit result stems from the $\chi$PT$_\gamma$ correction to the scattering length---the physical region for $t_S^{+0}(s)$ starts at $\spipn$, introducing an additional correction, as seen in Fig.~\ref{fig:tp0S}. In both cases, the pion-mass difference relative corrections reach a maximum at threshold, slightly exceeding 6\%.  Beyond this maximum, the relative correction decreases until it stabilizes around a plateau at $\sqrt s\sim0.5$ GeV. For the $t_S^{++}$ partial wave, this plateau is around 1\%, while for the $t_S^{+0}$, the relative correction for the real part reaches around 3\%. 
Thus, while the corrections above 1 GeV in the $\pi^+\pi^+\to\pi^+\pi^+$ channel remain relatively small, allowing for a smooth matching to the isospin-limit value at $s_1$, the shift with respect to the isospin-limit value required by the fit in the $\pi^+\pi^0\to\pi^+\pi^0$ $S$-wave is $\text{Re}\,b_S^{+0}=-5.65\times 10^{-3}$, so that the matching is performed only at $s_2$.




\section{\texorpdfstring{Resonance pole parameters for $\Delta_\pi\neq 0$}{Resonance pole parameters}}\label{sec:resonances}

Since our formalism is based on analyticity, the dispersive representation obtained here provides model-independent access to the physical (first) Riemann sheet. An analytic continuation to the second Riemann sheet, which is necessary to determine resonance parameters, is, however, also possible in a model-independent way. Using the Schwarz reflection principle and unitarity, one can show that the $S$-matrix on the second sheet is equal to the inverse of the $S$-matrix on the first. Consequently, the isospin-breaking amplitudes defined in Sec.~\ref{sec:roy-beyond} and App.~\ref{app:kernels} permit the extraction of the poles associated with the elastic $\pi\pi$ resonances---$f_0(500)$, $f_0(980)$, $\rho^+(770)$ and $\rho^0(770)$---since they correspond to zeros of the $S$-matrix on the first sheet. By comparing their pole positions with the isospin-limit values, we are then able to quantify the effects of the pion-mass difference on the resonances.

In the isospin limit, the scalars $f_0(500)$ and $f_0(980)$ are isosinglet ($I=0$) resonances and therefore appear as a single pole in the three scalar amplitudes $T^n$, $T^c$, and $T^x$:
\begin{equation}\label{eq:f0sIL}
\sqrt{s_{f_0(500)}}\big|_{\mathrm{IL}} = (440 - i\,271)\ \mathrm{MeV}, \qquad
\sqrt{s_{f_0(980)}}\big|_{\mathrm{IL}} = (997 - i\,26)\ \mathrm{MeV}\;,
\end{equation}
rounded according to the uncertainties quoted in Refs.~\cite{Caprini:2005zr,Garcia-Martin:2011nna,Pelaez:2022qby,Hoferichter:2023mgy}.

Within our isospin-breaking formalism, this single-pole structure is also preserved thanks to the coupled-channel formalism described in Sect.~\ref{sec:IB_param}. The continuation of the partial waves $t_S^n$, $t_S^c$, and $t_S^x$ to the second sheet yields
\begin{equation}\label{eq:f0sIB}
\sqrt{s_{f_0(500)}}\big|_{\Delta_\pi} = (441 - i\,270)\ \mathrm{MeV}, \qquad
\sqrt{s_{f_0(980)}}\big|_{\Delta_\pi} = (997 - i\,26)\ \mathrm{MeV}\;.
\end{equation}
Comparing Eqs.~\eqref{eq:f0sIL} and \eqref{eq:f0sIB} shows that isospin-breaking shifts are negligible at the present level of precision: for the $f_0(500)$, the pole mass increases by $\sim 1\ \mathrm{MeV}$ and the width decreases by $\sim 2\ \mathrm{MeV}$. In contrast, the $f_0(980)$ pole is essentially unchanged within the quoted rounding.

By contrast, the $\rho(770)$ is an isovector ($I=1$) resonance, so that in the isospin limit the three charge states $\rho^\pm(770)$ and $\rho^0(770)$ are degenerate:
\begin{equation}
\sqrt{s_{\rho(770)}}\big|_{\mathrm{IL}} = (763.29 - i\,71.65)\ \mathrm{MeV}\;.
\end{equation}
Isospin breaking lifts this degeneracy. Analytically continuing the $P$-wave amplitudes $t_P^c$ and $t_P^{+0}$ (see App.~\ref{app:kernels}) gives
\begin{equation}\label{eq:rhoIB}
\sqrt{s_{\rho^0}}\big|_{\Delta_\pi} = (763.28 - i\,71.66)\ \mathrm{MeV}\;, \qquad
\sqrt{s_{\rho^\pm}}\big|_{\Delta_\pi} = (762.29 - i\,71.89)\ \mathrm{MeV}\;.
\end{equation}
These values indicate a mass splitting of order $\sim 1\ \mathrm{MeV}$ and a width difference of order $\sim 0.5\ \mathrm{MeV}$ induced by the pion-mass difference. Taking into account the uncertainties reported in~\cite{Garcia-Martin:2011nna,Hoferichter:2023mgy}, the pion-mass–difference effects on the $\rho$ mass and width are therefore minimal and within their error budget. It is worth noting again that these results do not account for all isospin-breaking corrections to the resonance parameters, but only those directly due to the pion mass difference. All other corrections due to the exchange of virtual and real photons, which do not contribute to the pion mass difference, must still be added and will be calculated in a separate paper.

\section{Conclusions}\label{sec:conclusions}

In this work, we have analyzed the impact of one specific class of isospin-breaking effects in $\pi\pi$ scattering: those due to the charged-neutral pion-mass difference. To quantitatively evaluate these effects in a model-independent way, we have generalized Roy equations to the case where the neutral and charged pions are not degenerate. This significantly complicates the system, even if we consider only the $S$ and $P$ waves: In the isospin limit, there are only three such waves, whereas if isospin is broken, we have to account for and solve coupled equations for seven different $S$ and $P$ waves. This not only makes the mathematical problem more complicated, but also requires additional physical input. Concerning higher waves and the high-energy region of the $S$ and $P$ waves, we worked under the assumption that isospin-breaking effects in these are negligible, particularly given the experimental uncertainties with which these are known. The situation is completely different for the subtraction constants because, at the level of precision we are working, isospin-breaking effects cannot be neglected.
We can, however, rely on \chptg\ to calculate these and apply a matching procedure, much in the same spirit of~\cite{Colangelo:2001df}. The relevant \chptg\ formulae were indeed available in the literature~\cite{Knecht:2002gz}, but not yet in a form which would make them immediately usable for the matching. We have applied to them the necessary manipulations and have provided all the relevant formulae here.
In this setting, we have then solved the modified Roy equations for the seven physical $S$- and $P$-wave partial waves and evaluated the effects due to the pion-mass difference in the region between threshold and $\sqrt s\le 0.975$ GeV.

Our results indicate that the most significant relative corrections occur near the corresponding thresholds, reaching up to 12\% for the neutral channel, with slightly smaller corrections for other partial waves. For the resonant partial waves, these corrections generally diminish as the energy increases, becoming somewhat larger again in the resonant region. In contrast, for the repulsive $S$-waves, the pion-mass difference effects gradually decrease at higher energies, stabilizing around 3\% for the $t_S^{+0}(s)$ partial wave and approximately 1\% for $t_S^{++}(s)$ at an energy $\sqrt s\sim 0.5$ GeV. 
These results suggest that, while pion-mass difference effects are most prominent at low energy for most partial waves and gradually fade at higher energies---allowing for a smooth matching to the isospin limit at $s_1$---they remain at the level of a few percent around 1 GeV. In our approach, they show up most significantly in the $t_S^{+0}$ partial wave, but this might also be due to the fact that uncertainties in this wave around 1 GeV are large, which did not allow us to constrain it more strongly. In other words, as in previous Roy-equation analyses, the results in the region around 1 GeV and above reflect more the input one is using than the mathematics of the Roy equations. Our most relevant results concern the behavior in the resonance region and near threshold. 

On the basis of these Roy-equation solutions away from the isospin limit, we have extracted the pole position of the $f_0(500)$, $f_0(980)$, and $\rho^{\pm,0}(770)$ resonances by analytically continuing the dispersive amplitudes to the second Riemann sheet. We have assessed the impact of the charged–neutral pion mass difference on these pole parameters and find that it manifests itself as very small, charge-dependent shifts in resonance masses and widths, well below the current level of precision of these resonances. It remains to be seen whether the picture remains the same once we add the missing isospin-breaking corrections.

Our results provide a rigorous dispersive representation of the $\pi\pi$ scattering in the presence of a pion mass difference, which is the first step towards a complete analysis of all isospin-breaking effects in this process. The evaluation of further virtual and real photon corrections is underway, both for this process as well as for the phenomenologically relevant vector form factor of the pion. With these, we will be able to assess better the impact of isospin-breaking corrections to the hadronic vacuum polarization contribution to the muon $g-2$, which is our final goal. One particularly relevant application of these studies concerns the estimate of isospin-breaking corrections to be applied to $\tau$-decay data before these can be used to evaluate the hadronic vacuum polarization contribution. A first step in this direction has been performed very recently~\cite{Colangelo:2025iad}.

\section*{Acknowledgments} We thank Joachim Monnard for collaboration at the early stage of this work and Martin Hoferichter for helpful discussions. Financial support by the Swiss National Science Foundation (Project Nos. \ 200020\_200553 and PZ00P2\_174228), the Albert Einstein Center (AEC) for Fundamental Physics, the Ramon y Cajal program (RYC2019-027605-I), and the Grant PID2022-136510NB-C31 funded by
MCIN/AEI/ 10.13039/501100011033 of the Spanish MICIU is gratefully acknowledged. JRE also acknowledges support from the AEC visitor program and GC the kind hospitality at the Universidad Complutense de Madrid.

\appendix
\section{Explicit expressions for the subtraction constants}\label{app:subconst}
\subsection{Neutral channel}
For the $T^n$, the exact, explicit expressions of the subtraction
constants $a_n^{00}$ and $b_n^{00}$ read 
\begin{align}
    a_n^{00}=&\frac{M_{\pi^0}^2}{32 \pi F_\pi^2}
    \Bigg\{1+\xi_0
    \Bigg[4\lbar_1+8\lbar_2-\frac{3}{2}\lbar_3+2\lbar_4-\frac{23}{2}-
    \frac{9-11\dpi}{(1-\dpi)}L_\pi+9 j_0\left(4M_{\pi^0}^2\right)
    \Bigg]\nonumber\\
    &\hspace{1.3cm}+\xi\dpi\left(\frac{\Bar{k}_{31}}{9}-\frac{10}{9}\Bar{k}_2-\Bar{k}_4\right)\Bigg\}\;,\nonumber\\    
    b_n^{00}=&\frac{M_{\pi^0}^2}{384\pi F_\pi^2} \xi
    \left[16\left(\lbar_1+2\lbar_2-3\right) + 2 \dpi(1-24 \lambda_\pi)
    +27j_0\left(4M_{\pi^0}^2\right)\right]\;, 
\end{align}
where $\xi_0:=M_{\pi^0}^2/(16 \pi^2 F_\pi^2)=\xi(1-\dpi)$,
$L_\pi:=-\ln{(1-\dpi)}=\dpi+\cO (\dpi^2)$ and
$\lambda_\pi:=L_\pi/\dpi=1+\cO(\dpi)$. 
\subsection{\texorpdfstring{$\pi^+ \pi^- $ channel}{pi+pi- channel}}
For the $T^c$ amplitude, we get 
\begin{align}
    a^{++}&=-\frac{M_\pi^2}{16 \pi
  F_\pi^2}\Bigg\{1-\dpi-\xi\Bigg[\frac{4}{3}\left(\lbar_1+2\lbar_2\right)-\frac{1}{2}\left(\lbar_3+4\lbar_4\right)\left(1-\dpi\right)^2\nonumber\\
  &-\frac{1}{2}\left(1+3\dpi+\frac{88}{9}\dpi^2\right)-\dpi(1-\dpi)\left(\frac{\Bar{k}_{31}}{9}-4\Bar{k}_{32}+\frac{62}{9}\Bar{k}_2+5\Bar{k}_4\right)\Bigg] \Bigg\}\;,\nonumber\\
b^{++}&=\frac{M_\pi^2}{48\pi F_\pi^2} \xi \left(4\lbar_2-\frac{23}{9}-4\dpi+\dpi^2\right)\;,\nonumber\\
c^{++}&=-\frac{\xi}{864\pi F_\pi^2}\;,\nonumber\\
    a_c^{+-}&=\frac{M_\pi^2}{16 \pi F_\pi^2}\Bigg\{1+\dpi+\xi
    \Bigg[\frac{4}{3}\left(\lbar_1+2\lbar_2\right)-
    \frac{1}{2}\lbar_3\left(1-\dpi\right)^2+2\lbar_4\left(1-\dpi^2\right) \nonumber\\
    &-\frac{27-90\dpi-133\dpi^2+124\dpi^3}{18\left(1-\dpi\right)} +\frac{\left(3+\dpi\right)^2}{4}j_0\left(4M_\pi^2\right)\nonumber\\
    &+\dpi\left(\frac{\Bar{k}_{31}}{9}(1+\dpi)
       +4\Bar{k}_{32}(1-\dpi)
       -\frac{2}{9}\Bar{k}_2\left(5-31\dpi\right)
       +\Bar{k}_4(1+5\dpi)\right)\Bigg]\Bigg\}\;,\nonumber\\    
    b_c^{+-}=&\frac{M_\pi^2}{24 \pi F_\pi^2} \xi
    \left[\lbar_1+\lbar_2 -
      \frac{97-754\dpi+297\dpi^2+144\dpi^3}{144\left(1-\dpi\right)}
      +\frac{3}{32}\left(3+\dpi\right)^2j_0\left(4M_\pi^2\right)\right]\;,\nonumber\\
    c_c^{+-}=&\frac{\xi}{1728 \pi F_\pi^2}\;.
\end{align}
\subsection{\texorpdfstring{$\pi^+ \pi^- \to\pi^0\pi^0$ channel}{neutral channel}}
For the amplitude $T^x$
\begin{align}
a_1&=-\frac{M_\pi^2}{32 \pi F_\pi^2 \left(2-\dpi\right)}\Bigg\{
2\left(3-\dpi\right)
+\frac{\xi}{3}\Bigg[\frac{33+158\dpi-29\dpi^2-36\dpi^3}{3}\nonumber\\
&+ 8\lbar_1
\left(1+\dpi-\dpi^2\right) 
+4\lbar_2 \left(2-\dpi\right)^2-3\lbar_3 \left(1-\dpi\right)^2+12\lbar_4 \left(3-4\dpi+\dpi^2\right)\nonumber\\
&+\frac{3}{2}\left(6-7\dpi+ \dpi^3\right)j_0\left(4 M_\pi^2\right)+2\left(2 -11 \dpi-18\dpi^2+9\dpi^3
\right)\lambda_\pi\nonumber\\
&+\frac{4}{3}\dpi\left(16-21\dpi+2\dpi^2\right)\Bar{j}_{+0}^{(1)}+\frac{3}{2}\dpi^4\Bar{j}_{+0}^{(2)}\nonumber\\
&+\dpi\left(\frac{2}{3}\Bar{k}_{31}(3-\dpi)
  +12\Bar{k}_{32}(1-\dpi)+ \frac{4}{3}\Bar{k}_2(3+5\dpi)
  +12\Bar{k}_4\right)\Bigg]\Bigg\}\;,\nonumber\\ 
b_1&=-\frac{M_\pi^2}{96 \pi F_\pi^2} \xi \Bigg[4\lbar_1+\frac{23}{12}+\frac{51}{4}\dpi+2\dpi^2+2\left(1-3\dpi
\right) \lambda_\pi\nonumber\\
&\qquad\qquad+\frac{3}{8}(1-\dpi)(3+\dpi)j_0\left(4M_\pi^2\right)+\frac{8}{3}\Bar{j}_{+0}^{(1)}\Bigg]\;,\nonumber\\
c_1&=-\frac{\xi}{144 \pi F_\pi^2}\Bar{j}_{+0}^{(1)}\;,\nonumber\\
a_2&=\frac{M_\pi^2}{32 \pi F_\pi^2} \frac{1}{\eta \left(2-\dpi\right)} \Bigg\{2\left(1-\dpi\right)-\xi
\Bigg[\frac{1}{3}\Bigg(\frac{\left(17+10 \dpi-23\dpi^2\right)+8\eta \left(2-\dpi\right)}{3}\nonumber\\
&+8\left(1-\dpi\right)\lbar_1-4\left(2-\dpi\right)
\left(2-\dpi-4\eta\right)\lbar_2-3
\left(1-\dpi\right)^2\left(\lbar_3+4\lbar_4\right)\Bigg)\nonumber\\ 
&+\frac{2}{3}\left[6-8\eta \left(2-\dpi\right)-\dpi\left(7-9\dpi+6\dpi^2\right)\right]\lambda_\pi\nonumber\\
&-\frac{\eta^2
  \dpi^2}{4}\left(2-\dpi\right)\left(1-2\eta^{-1}\right)\Bar{\Bar{j}}_{+0}
\left(4\Bar{M}_\pi^2\right)\nonumber\\
&+(2-\dpi)\eta \left( (1-2\eta^{-1})^2+\frac{\eta^2 \dpi^4}{64}
   \right)\Bar{j}_{+0}\left(4\Bar{M}_\pi^2\right)+\frac{\dpi^4}{8}\left(4-\eta\left(2-\dpi\right)\right)j_{+0}^{(2)}\nonumber\\
&-\left(\left(2-\dpi\right)\left(\frac{16}{9\eta^2}+\frac{\dpi^4
      \eta^2}{16}\right)-\frac{4}{9}\left(8-12
    \dpi+3\dpi^2-4\dpi^3\right)\right)j_{+0}^{(1)}\nonumber\\
    &-\dpi\left(1-\dpi\right)
    \left(\frac{2}{9}\Bar{k}_{31}-4\Bar{k}_{32}+\frac{52}{9}\Bar{k}_2    +4\Bar{k}_4\right)\Bigg]\Bigg\}\;,\nonumber\\   
b_2&=-\frac{M_\pi^2}{96 \pi F_\pi^2} \frac{\xi}{\eta}
\Bigg[2\left(2\lbar_2-2\lambda_\pi+\frac{1}{3}\right)+3\left(1-\eta +\frac{\eta^2}{4} + \frac{ \eta^4
    \dpi^4}{256}\right)\Bar{j}_{+0} \left(4 \Bar{M}_\pi^2\right)\nonumber\\ 
&
\qquad\qquad+\frac{3 \eta^2 \dpi^2}{16}\left(2-\eta\right)\Bar{\Bar{j}}_{+0} \left(4
  \Bar{M}_\pi^2\right)-\left(\frac{4}{3\eta}+3 \eta +\frac{3 \eta^3 \dpi^4}{64}
\right)\Bar{j}_{+0}^{(1)}\nonumber\\
&\qquad\qquad-\frac{3\eta \dpi^2}{32}\left(\eta 
  \dpi^2-16\right)\Bar{j}_{+0}^{(2)}-\frac{\eta \dpi^4}{8}\Bar{j}_{+0}^{(3)}\Bigg]\;,\nonumber\\
c_2&=\frac{\xi}{288 \pi  F_\pi^2\eta}\Bar{j}_{+0}^{(1)}\;,    
\end{align}
where we have defined $\eta:=M_\pi^2/\Bar{M}_\pi^2$ and 
$\Bar{\Bar{j}}_{+0}(s):=\Bar{j}_{+0}(s)-s\Bar{j}_{+0}^{'}(0)$\;, and
$\Bar{j}^{(i)}_{+0}:=\linebreak M_\pi^{2i} \partial_s^i \Bar{j}_{+0}(s)_{|_{s=0}}$\;.
\begin{align}
j_0\left(4M_{\pi^0}^2\right)&=2+\sqrt{\frac{\dpi}{1-\dpi}}\left(-\pi+2\,\mathrm{arctan}\sqrt{\frac{\dpi}{1-\dpi}}\right)\;,\nonumber\\
\Bar{j}_0(4M_\pi^2)&=2-\sqrt{\dpi}\left[\ln\frac{1+\sqrt{\dpi}}{1-\sqrt{\dpi}}+
i \pi \right]\;,\nonumber\\
\Bar{j}_{+0}(4\Bar{M}_\pi^2)&=1+\left(1-\frac{\dpi}{2}-\frac{\eta
    \dpi^2}{8}\right)\lambda_\pi+\frac{\rho}{2} \ln \frac{4-\eta(2-\dpi)-4 \rho}{4-\eta (2-\dpi)+4 \rho}\;,\nonumber\\
\Bar{\Bar{j}}_{+0}(4\Bar{M}_\pi^2)&=1-\frac{2(2-\dpi)}{\eta \dpi^2}+\left(
\frac{4}{\eta \dpi^2}(1-\dpi)+1-\frac{\dpi}{2}-\frac{\eta \dpi^2}{8}
\right) \lambda_\pi \nonumber\\
&\quad+\frac{\rho}{2} \ln \frac{4-\eta(2-\dpi)-4 \rho}{4-\eta (2-\dpi)+4 \rho}\;, \nonumber\\
\Bar{j}_{+0}^{(1)}&=\frac{1}{2 \dpi^3}\left[\dpi (2-\dpi)-2(1-\dpi)L_\pi\right]\;,\nonumber\\
\Bar{j}_{+0}^{(2)}&=\frac{1}{3 \dpi^5}\left[\dpi (12-12 \dpi + \dpi^2)-6(2-\dpi)(1-\dpi)L_\pi \right] \;,\nonumber\\
\Bar{j}_{+0}^{(3)}&=\frac{1}{2 \dpi^7}\left[\dpi(2-\dpi)(30-30\dpi+\dpi^2)-12(1-\dpi)(5-5\dpi+\dpi^2)L_\pi \right]\;,
\end{align}
with
\begin{equation}
\rho:=\sqrt{1-\eta(1-\dpi/2)+\eta^2 \dpi^2/16}\;,
\end{equation}
and with $j(s):=16\pi^2\Bar{J}(s)$\;,
$\Sigma_\pi=M_\pi^2(2-\dpi)$ and
$\Bar{M}_\pi=\frac{M_\pi}{2}(1+\sqrt{1-\dpi})$\;. 

\section{Explicit expressions for the kernels}\label{app:kernels}

\subsection{\texorpdfstring{$\pi^+ \pi^+ $ channel}{++ channel}}
The $S$-wave projection of the $\pi^+ \pi^+ $ channel reads

\begin{align}
    t_S^{++}(s)=&{}\frac{a^{++}s}{\spi}-\frac{a_c^{+-}(s-\spi)}{\spi}+\int_{\spin}^{\spi}\mathrm{d}s'\bigg[K_{c,S}^{++}(s',s)\mathrm{Im}t_{S}^{c,00}(s')\nonumber\\
    &+\int_{\spi}^{s_1}\mathrm{d}s'\bigg[K_S^{++}(s',s)\mathrm{Im}t_{S}^{++}(s')+K_{c,S}^{++}(s',s)\left(\mathrm{Im}t_{S}^{c,00}(s')+\mathrm{Im}t_{S}^{c,+-}(s')\right)\nonumber\\
    &\hspace{1.8cm}+K_{c,P}^{++}(s',s)\mathrm{Im}t_{P}^{c,+-}(s')\bigg]+d_S^{++}(s)\;,
\end{align}
where the kernels are
\begin{align}
    K^{++}_S(s',s)=&{}\frac{1}{\pi}\frac{s(s-\spi)}{s'(s'-\spi)(s'-s)}\;,\nonumber\\
    K_{c,S}^{++}(s',s)=&\frac{1}{\pi}\left[\frac{s-2s'+\spi}{s'(s'-\spi)}+\frac{2}{s-\spi}\ln{\left(\frac{s'+s-\spi}{s'}\right)}\right]\;,\nonumber\\
    K_{c,P}^{++}(s',s)=&\frac{3}{\pi}\left[\frac{3s+2s'-\spi}{s'(s'-\spi)}-\frac{2(2s+s'-\spi)}{(s-\spi)(s'-\spi)}\ln{\left(\frac{s'+s-\spi}{s'}\right)}\right]\;.
\end{align}

\subsection{\texorpdfstring{$\pi^+ \pi^- $ channel}{charged channel}}

In this case, both an $S$- and $P$-wave partial decomposition are present. For the $S$-wave, we obtain
\begin{align}
    t_S^{c}(s)=&{}-\frac{a^{++}(s-\spi)}{2\spi}+\frac{a_c^{+-}(s+\spi)}{2\spi}+\int_{\spin}^{\spi}\mathrm{d}s' K_{s,S}^{+-}(s',s)\mathrm{Im}t_{S}^{c,00}(s')\nonumber\\
    &+\int_{\spi}^{s_1}\mathrm{d}s'\bigg[K_{s,S}^{+-}(s',s)\left(\mathrm{Im}t_{S}^{c,00}(s')+\mathrm{Im}t_{S}^{c,+-}(s')\right)+K_{s,P}^{+-}(s',s)\mathrm{Im}t_{P}^{c,+-}(s')\nonumber\\
    &\hspace{1.8cm}+K_{+-,S}^{++}(s',s)\mathrm{Im}t_{S}^{++}(s')\bigg]+d_S^c(s)\;,
\end{align}
with
\begin{align}
     K_{s,S}^{+-}(s',s)=&{}\frac{1}{\pi}\left[\frac{1}{s'-s}-\frac{s'+s+3(s'-\spi)}{2s'(s'-\spi)}+\frac{1}{s-\spi}\ln{\left(\frac{s'+s-\spi}{s'}\right)}\right]\;,\nonumber\\
    K_{s,P}^{+-}(s',s)=&\frac{3}{\pi}\left[-\frac{3s+2s'-\spi}{2s'(s'-\spi)}+\frac{2s+s'-\spi}{(s-\spi)(s'-\spi)}\ln{\left(\frac{s'+s-\spi}{s'}\right)}\right]\;,\nonumber\\
    K_{+-,S}^{+-}(s',s)=&\frac{1}{\pi}\left[\frac{s-2s'+\spi}{2s'(s'-\spi)}+\frac{1}{s-\spi}\ln{\left(\frac{s'+s-\spi}{s'}\right)}\right]\;,
\end{align}
while for the $P$-wave we have
\begin{align}
    t_P^{c}(s)=&{}\frac{a_c^{+-}-a^{++}}{6\spi}(s-\spi)+\int_{\spin}^{\spi}\mathrm{d}s' K_{s,P}^{+-}(s',s)\mathrm{Im}t_{S}^{c,00}(s')\nonumber\\
    &+\int_{\spi}^{s_1}\mathrm{d}s'\bigg[K_{s,P}^{+-}(s',s)\left(\mathrm{Im}t_{S}^{c,00}(s')+\mathrm{Im}t_{S}^{c,+-}(s')-\mathrm{Im}t_{S}^{++}(s')\right)\nonumber\\
    &\qquad\qquad+K_{p,P}^{+-}(s',s)\mathrm{Im}t_{P}^{c,+-}(s')\bigg]+d_P^c(s)\;,
\end{align}
where the kernels are
\begin{align}
    K_{p,S}^{+-}(s',s)=&{}\frac{1}{\pi}\left[-\frac{s-\spi}{6s'(s'-\spi)}-\frac{2}{s-\spi}+\frac{s+2s'-\spi}{(s-\spi)^2}\ln{\left(\frac{s'+s-\spi}{s'}\right)}\right]\;,\nonumber\\
    K_{p,P}^{+-}(s',s)=&\frac{1}{\pi}\bigg[\frac{s(s-\spi)}{s'(s'-\spi)(s'-s)}-\frac{s-\spi}{2s'(s'-\spi)}-\frac{6(2s+s'-\spi)}{(s-\spi)(s'-\spi)}\nonumber\\
    &\hspace{0.4cm}+\frac{3(s+2s'-\spi)(2s+s'-\spi)}{(s-\spi)^2(s'-\spi)}\ln{\left(\frac{s'+s-\spi}{s'}\right)}\bigg]\;.
\end{align}

\subsection{\texorpdfstring{$\pi^+ \pi^- \to\pi^0\pi^0$ channel}{X channel}}

The $S$-wave projection for the $\pi^+ \pi^- \to\pi^0\pi^0$ scattering reads
\begin{align}
    t_S^x(s)=&{}\frac{a_x^{+-}s}{\spi}-\frac{a_c^{+0}\left[s-2\Sigma_\pi\right]}{s_{\pi^{\pm 0}}}+\int_{\spin}^{\spi}\mathrm{d}s'K_{x,S}(s',s)\mathrm{Im}t_{S}^{x,00}(s')\nonumber\\
    &+\int_{\spi}^{s_1}\mathrm{d}s'K_{x,S}(s',s)\left(\mathrm{Im}t_{S}^{x,00}(s')+\mathrm{Im}t_{S}^{x,+-}(s')\right)\nonumber\\
    &+\int_{\spipn}^{s_1}\mathrm{d}s'\left[K_{x,S}^{0+}(s',s)\mathrm{Im}t_{S}^{0+}(s')+K_{x,P}^{0+}(s',s)\mathrm{Im}t_{P}^{0+}(s')\right]\Bigg\}+d_S^x(s)\;,
\end{align}
where the kernels read
\begin{align}
    K_{x,S}(s',s)=&{}\frac{1}{\pi}\frac{s(s-\spi)}{s'(s'-\spi)(s'-s)}\;,\nonumber\\
    K_{x,S}^{0+}(s',s)=&\frac{1}{\pi}\Bigg\{\frac{s-2s'+4M_\pi M_{\pi^0}}{s'(s'-\spipn)}+\frac{1}{2q(s,\spi)q(s,\spin)}\times\nonumber\\
    &\hspace{0.5cm}\left[\ln{\left(1+\frac{4q(s,\spi)q(s,\spin)}{s+2s'-2\Sigma_\pi}\right)}-\ln{\left(1-\frac{4q(s,\spi)q(s,\spin)}{s+2s'-2\Sigma_\pi}\right)}\right]\Bigg\}\;,\nonumber\\
    K_{x,P}^{0+}(s',s)=&\frac{-3}{\pi\lambda_{\pm 0}(s')}\Bigg\{3s+2s'-2\Sigma_\pi-\frac{s'\left(s'+2s-2\Sigma_\pi\right)+\Delta_\pi^2}{2q(s,\spi)q(s,\spin)}\times\nonumber\\
    &\left[\ln{\left(1+\frac{4q(s,\spi)q(s,\spin)}{s+2s'-2\Sigma_\pi}\right)}-\ln{\left(1-\frac{4q(s,\spi)q(s,\spin)}{s+2s'-2\Sigma_\pi}\right)}\right]\Bigg\}\;,
\end{align}
where $q(x,y)=\frac{1}{2}\sqrt{x-y}$ and 
\begin{equation}
    \lambda_{\pm 0}(s)=\lambda(s,M_\pi,M_{\pi^0})=\left[s-(M_{\pi}+M_{\pi^0})^2\right]\left[s-(M_{\pi}-M_{\pi^0})^2\right]\;.
\end{equation}

\subsection{\texorpdfstring{$\pi^+ \pi^0\to\pi^+\pi^0$ channel}{+0 channel}}
The last $\pi^+ \pi^0 \to\pi^+\pi^0$ channel can be decomposed in both an $S$- and $P$-wave. For the $S$-wave, we get
\begin{align}
    t_S^+(s)=&{}\frac{a_c^{+0}\left(s+(M_\pi-M_{\pi^0})^2\right)\left(s+\spipn\right)}{2s\spipn}-\frac{a_x^{+-}\lambda_{\pm 0}(s)}{2s\spi}\nonumber\\
    &+\int_{\spin}^{\spi}\mathrm{d}s'K_{x,S}^{0+}(s',s)\mathrm{Im}t_{S}^{x,00}(s')+\int_{\spi}^{s_1}\mathrm{d}s'K_{x,S}^{0+}(s',s)\left(\mathrm{Im}t_{S}^{x,00}(s')+\mathrm{Im}t_{S}^{x,+-}(s')\right)\nonumber\\
    &+\int_{\spipn}^{s_1}\mathrm{d}s'\left[K_{s,S}^{0+}(s',s)\mathrm{Im}t_{S}^{0+}(s')+K_{s,P}^{0+}(s',s)\mathrm{Im}t_{P}^{0+}(s')\right]\Bigg\}+d_S^+(s)\;,
\end{align}
with
\begin{align}
    K_{x,S}^{0+}(s',s)=&{}\frac{1}{\pi}\Bigg[\frac{\lambda_{\pm 0}(s)}{2ss'(s'-\spi)}-\frac{1}{s'}+\frac{s}{\lambda_{\pm 0}(s)}\ln{\left(\frac{ss'+\lambda_{\pm 0}(s)}{ss'}\right)}\Bigg]\;,\nonumber\\
    K_{s,S}^{0+}(s',s)=&\frac{1}{\pi}\Bigg\{\frac{1}{s'-s}-\frac{2}{s'}-\frac{1}{2ss'}\frac{\left(s+\spipn\right)\left(s+(M_\pi-M_{\pi^0})^2\right)}{s'-\spipn}\nonumber\\
    &\qquad+\frac{s}{\lambda_{\pm 0}(s)}\ln{\left(\frac{s\left(s+s'-2\Sigma_\pi\right)}{ss'-\Delta_\pi^2}\right)}\Bigg\}\;,\nonumber\\
    K_{p,P}^{0+}(s',s)=&\frac{3}{\pi\lambda_{\pm 0}(s')}\Bigg\{s+s'+\frac{\lambda_{\pm 0}(s)}{2s}\nonumber\\
    &\qquad\qquad-\frac{s\left(\lambda_{\pm 0}(s')+2ss'-2\Delta_\pi^2\right)}{\lambda_{\pm 0}(s)}\ln{\left(\frac{s\left(s+s'-2\Sigma_\pi\right)}{ss'-\Delta_\pi^2}\right)}\Bigg\}\;,
\end{align}
while the $P$-wave decomposition reads
\begin{align}
    t_P^+(s)=&{}-\frac{a_c^{+0}\lambda_{\pm 0}(s)}{6s\spipn}+\frac{a_x^{+-}\lambda_{\pm 0}(s)}{6s\spi}+\int_{\spin}^{\spi}\mathrm{d}s'K_{x,P}^{0+}(s',s)\mathrm{Im}t_{S}^{x,00}(s')\nonumber\\
    &+\int_{\spi}^{s_1}\mathrm{d}s'K_{x,P}^{0+}(s',s)\left(\mathrm{Im}t_{S}^{x,00}(s')+\mathrm{Im}t_{S}^{x,+-}(s')\right)\nonumber\\
    &+\int_{\spipn}^{s_1}\mathrm{d}s'\left[K_{p,S}^{0+}(s',s)\mathrm{Im}t_{S}^{0+}(s')+K_{p,P}^{0+}(s',s)\mathrm{Im}t_{P}^{0+}(s')\right]\Bigg\}+d_P^+(s)\;,
\end{align}
where the kernels are
\begin{align}
    K_{x,P}^{0+}(s',s)&=\frac{1}{\pi}\Bigg[-\frac{\lambda_{\pm 0}(s)}{6ss'(s'-\spi)}-\frac{2s}{\lambda_{\pm 0}(s)}+\frac{s}{\lambda_{\pm 0}(s)}\left(1+\frac{2ss'}{\lambda_{\pm 0}(s)}\right)\ln{\left(\frac{ss'+\lambda_{\pm 0}(s)}{ss'}\right)}\Bigg]\;,\nonumber\\
    K_{p,S}^{0+}(s',s)&=\frac{1}{\pi}\Bigg\{\frac{2s}{\lambda_{\pm 0}(s)}+\frac{\lambda_{\pm 0}(s)}{6ss'\left(s'-\spipn\right)}\nonumber\\
    &\qquad+\frac{s}{\lambda_{\pm 0}(s)}\left[1-\frac{2s\left(s+s'-2\Sigma_\pi\right)}{\lambda_{\pm 0}(s)}\right]\ln{\left(\frac{s\left(s+s'-2\Sigma_\pi\right)}{ss'-\Delta_\pi^2}\right)}\Bigg\}\;,\nonumber\\
    K_{p,P}^{0+}(s',s)&=\frac{1}{\pi\lambda_{\pm 0}(s')}\Bigg\{\left(\frac{1}{s'-s}-\frac{1}{2s}\right)\lambda_{\pm 0}(s)+\frac{6s\left[\Delta_\pi^2-s'\left(s'+2s-2\Sigma_\pi\right)\right]}{\lambda_{\pm 0}(s)}\nonumber\\
    &+\frac{3s\left[\Delta_\pi^2-2s'\left(s-\Sigma_\pi)\right)-s'^2\right]\left[\lambda_{\pm 0}(s)-2s\left(s+s'-2\Sigma_\pi\right)\right]}{\lambda_{\pm 0}(s)^2}\nonumber\\
    &\times\ln{\left(\frac{s\left(s+s'-2\Sigma_\pi\right)}{ss'-\Delta_\pi^2}\right)}\Bigg\}\;.
\end{align}

\section{Expressions of the coefficients fixed by boundary conditions}\label{app:bcs1}
Continuity conditions at $s_1$ determine the two highest coefficients in the polynomial that parametrizes the isospin-breaking effects. 
These conditions can be translated into explicit analytic expressions for these coefficients, like the ones given in the text in Eq.~\eqref{eq:matching_conditions}. Here we provide the remaining ones. First, for the $t^{+0}_S$ partial wave:
\begin{align}\label{eq:match0+S}
c_6^{+0,S}=&-r_{0+}^{-6}\Biggl[\sum_{k=0}^5{(7-k)\,c_k^{+0,S}\;r_{0+}^k} \nonumber\\ 
&+\frac{i}{2\dpi\delta^2_0(s_1)}\left(7+(s_1- s_{0+})\frac{\delta_0^{2\, \prime}(s_1)}{\delta^2_0(s_1)}\right)\log\left(\frac{1+2i\sigma_{+0}(s_1)t_0^2(s_1)}{\eta_0^2(s_1)}\right)\nonumber\\
&+\frac{\left(s_1-\spipn\right)^2}{\dpi\delta_0^2(s_1)\left(1+2i\sigma_{+0}(s_1)t_0^2(s_1)\right)}\Bigg(\frac{s_1\Sigma_\pi-\Delta_\pi^2}{s_1^3(s_1- \spi)\sigma_{+0}(s_1)}\,t_0^2(s_1)+\frac{\sigma_{+0}(s_1)}{s_1-\spi}t_0^{2 \, \prime}(s_1)\nonumber\\
&\hspace{5cm}+\left(\frac{7\delta_0^2(s_1)}{s_1-\spipn}+\frac{i\eta_0^{2 \, \prime}(s_1)}{2\eta_0^2(s_1)}\right)\frac{\left(1+2i\sigma_{+0}(s_1)t_0^2(s_1)\right)}{s_1-\spipn}
\Bigg)\Bigg]\;,\nonumber\\
c_7^{+0,S}=&-r_{0+}^{-7}\Bigg[\sum_{k=0}^{6}{c_k^{+0,S}r_{0+}^k} 
+\frac{1}{\dpi}\left(1+\frac{i}{2\delta_0^2(s_1)}\log\left(\frac{1+2i\sigma_{+0}(s_1)t_0^2(s_1)}{\eta_0^2(s_1)}\right)\right)\Bigg]\;,
\end{align}
where $r_{0+} \equiv s_1/s_{0+}-1$, the prime denotes the derivative with respect to $s$, and the map defined in~\eqref{eq:map+0} implies that both coefficients become complex.

Second, for the $t^{+0}_P$ partial wave:
\begin{align}\label{eq:match0+P}
c_6^{+0,P}=&-r_{0+}^{-6}\Biggl[\sum_{k=0}^5{(7-k)\,c_k^{+0,P}r_{0+}^k}\nonumber\\
&+\frac{i}{2\dpi\delta^1_1(s_1)}\left(7+\left(s_1-\spipn\right)\frac{\delta_1^{1 \, \prime}(s_1)}{\delta^1_1(s_1)}\right)\left\{\log\left(\frac{1+2i\sigma_{+0}(s_1)t_1^1(s_1)}{\eta_1^1(s_1)}\right)-2\pi i\right\}\nonumber\\
&+\frac{\left(s_1-\spipn\right)^2}{\dpi\delta_1^1(s_1)\left(1+2i\sigma_{+0}(s_1)t_1^1(s_1)\right)}\Bigg(\frac{s_1\Sigma_\pi-\Delta_\pi^2}{s_1^3(s_1- \spi)\sigma_{+0}(s_1)}\,t_1^1(s_1)+\frac{\sigma_{+0}(s_1)}{s_1-\spi}t_1^{1 \, \prime}(s_1)\nonumber\\
&+\left(\frac{7\delta_1^1(s_1)}{s_1-\spipn}+\frac{i\eta_1^{1 \, \prime}(s_1)}{2\eta_1^1(s_1)}\right)\frac{\left(1+2i\sigma_{+0}(s_1)t_1^1(s_1)\right)}{s_1-\spipn}
\Bigg)\Bigg]\;,\\
c_7^{+0,P}=&-r_{0+}^{-7}\Bigg\{\sum_{k=0}^{6}{c_k^{+0,P}r_{0+}^k} +\frac{1}{\dpi}\left[1+\frac{i}{2\delta_1^1(s_1)}\left(\log\left(\frac{1+2i\sigma_{+0}(s_1)t_1^1(s_1)}{\eta_1^1(s_1)}\right)-2\pi i\right)\right]\Bigg\}\;. \nonumber
\end{align}

Next, we provide the expressions for the S-waves linked by a coupled-channel problem. First, the $t^n_S$ wave
\begin{align}
    c_6^n=&-r_{+-}^{-6}\;\Bigg[\sum_{k=0}^5\;(7-k)\,c_k^n\,r_{+-}^k
    +\frac{\sigma(s_1)}{2}\sum_{k=0}^4\;\tilde c_{\,k}^{\,n}\;r_{+-}^k\left(14-2k-\frac{\spi}{s_1}\right)\nonumber\\
    &\qquad\qquad+\frac{i}{\delta_S^{n,\IL}(s_1)}\Bigg\{\frac{4}{(s_1-\spi)}\left(2-\frac{1+\tilde\eta_S^{x}(s)^2}{S_n^{\IL}(s_1)}\right)-\left(\frac{1-\tilde\eta_S^{x}(s)^2}{2S_n^{\IL}(s_1)}\right)\frac{\tilde\eta_S^{x\, \prime}(s)}{\tilde\eta_S^{x}(s)}\nonumber\\
    &\qquad\qquad-\frac{1+\tilde\eta_S^{x}(s)^2}{2\;S_n^{\IL}(s_1)}\frac{\eta_S^{n \, \prime}(s_1)}{\eta_S^n(s_1)}\;\\
    &\qquad\qquad+\frac{\delta_S^{n,\IL \, \prime}(s_1)}{\delta_S^{n,\IL}(s_1)}\left(\left(1-\frac{1+\tilde\eta_S^{x}(s)^2}{2\,S_n^{\IL}(s_1)}\right)-4\,i\,\delta_S^{n,\IL}(s_1)\left(1+\frac{1+\tilde\eta_S^{x}(s)^2}{4\,S_n^{\IL}(s_1)}\right)\right)\Bigg\}\Bigg]\;,\nonumber\\
    c_7^n=&-r_{+-}^{-7}\;\Bigg[\sum_{k=0}^6\;c_k^n\;r_{+-}^k+\sigma(s_1)\sum_{k=0}^4\,\tilde c_k^n\;r_{+-}^k  +\frac{i}{r_{+-}\;\spi\;\delta_S^{n,\IL}(s_1)}\left(1-\frac{1+\tilde\eta_S^{x}(s)^2}{2\,S_n^{\IL}(s_1)}\right) \Bigg]\;, \notag 
\end{align}
where $r_{+-} \equiv s_1/\spi-1$ and 
\begin{align}\label{eq:iso_defs_in_S}
S_n^{\IL}(s)=&1+2i\,\sigma(s)\left(t_0^0(s)+2t_0^2(s)\right)/3\;,\nonumber\\ 
\tilde\eta_S^{x}(s)=&\sqrt{ 1-8\ \sigma_0(s)\sigma(s)\,\left\vert\left(t_0^0(s)-t_0^2(s)\right)/3\right\vert^2}\;,
\end{align}
and, once again, the prime denotes the derivative with respect to $s$.

Then, for the $t^c_S$ wave:
\begin{align}
c_6^c=&-r_{+-}^{-6}\;\Bigg[\sum_{k=0}^5\;(7-k)\,c_k^c\;r_{+-}^k+\frac{\sigma(s_1)}{2}\sum_{k=0}^3\;\tilde c_{\,k}^{\,c}\;r_{+-}^k\left(14-2k-\frac{\spi}{s_1}\right)\nonumber\\
&\qquad\qquad+\frac{i}{\delta_S^{c,\text{IL}}(s_1)}\Bigg\{\left(\frac{4}{s_1-s_{+-}}+\frac{\delta_S^{c,\text{IL}\, \prime}(s_1)}{2\delta_S^{c,\text{IL}}(s_1)}+i\delta_S^{n,\text{IL}}(s_1)'+\frac{\eta_S^{n\, \prime}(s_1)}{2\;\eta_S^n(s_1)}\right)\frac{1-\tilde\eta_S^{x}(s)^2}{S_n^{\text{IL}}(s_1)}\nonumber\\
&\qquad\qquad+\frac{1+\tilde\eta_S^{x}(s)^2}{2\,S_n^{\text{IL}}(s_1)}\frac{\tilde\eta_S^{x\, \prime}(s)}{\tilde\eta_S^{x}(s)}\Bigg\}\Bigg]\;,\nonumber \\
c_7^c=&-r_{+-}^{-7}\;\Bigg[\sum_{k=0}^6\;c_k^c\;r_{+-}^k+\sigma(s_1)\sum_{k=0}^3\,\tilde c_{\,k}^{\,c}\,r_{+-}^k +\frac{i}{r_{+-}\;\spi\; \delta_S^{c,\text{IL}}(s_1)}\frac{1-\tilde\eta_S^{x}(s)^2}{2\,S_n^{\text{IL}}(s_1)}\Bigg]\;.
\end{align}
Finally, for the $t^x_S$ wave
\begin{align}
c_6^x=&-r_{+-}^{-6}\Bigg[\sum_{k=0}^5\;(7-k)\,c_k^x\;r_{+-}^k +\frac{1}{r_{+-}\,\spi} \left\{\left(\frac{9}{r_{+-}\,\spi}+2i\delta_S^{n,\text{IL} \, \prime}(s_1)+\frac{\eta_S^{n\, \prime}(s_1)}{\eta_S^n(s_1)}\right)\frac{1-\tilde\eta_S^{x}(s)^2}{S_n^{\text{IL}}(s_1)} \right. \nonumber\\
&\left. \qquad \qquad +\frac{1+\tilde\eta_S^{x}(s)^2}{S_n^{\text{IL}}(s_1)}\frac{\tilde\eta_S^{x\, \prime}(s)}{\tilde\eta_S^{x}(s)}\right\} \Bigg]\;,\nonumber\\
c_7^{x}=&-r_{+-}^{-7}\;\Bigg[\sum_{k=0}^6\;c_k^x\;r_{+-}^k+\frac{1}{(s_1-s_{+-})^2}\frac{1-\tilde\eta_S^{x}(s)^2}{S_n^{\text{IL}}(s_1)}\Bigg]\;.
\end{align}

\bibliographystyle{apsrev4-1_mod_2}
\bibliography{Literature}

\begin{thebibliography}{39}%
\makeatletter
\providecommand \@ifxundefined [1]{%
 \@ifx{#1\undefined}
}%
\providecommand \@ifnum [1]{%
 \ifnum #1\expandafter \@firstoftwo
 \else \expandafter \@secondoftwo
 \fi
}%
\providecommand \@ifx [1]{%
 \ifx #1\expandafter \@firstoftwo
 \else \expandafter \@secondoftwo
 \fi
}%
\providecommand \natexlab [1]{#1}%
\providecommand \enquote  [1]{``#1''}%
\providecommand \bibnamefont  [1]{#1}%
\providecommand \bibfnamefont [1]{#1}%
\providecommand \citenamefont [1]{#1}%
\providecommand \href@noop [0]{\@secondoftwo}%
\providecommand \href [0]{\begingroup \@sanitize@url \@href}%
\providecommand \@href[1]{\@@startlink{#1}\@@href}%
\providecommand \@@href[1]{\endgroup#1\@@endlink}%
\providecommand \@sanitize@url [0]{\catcode `\\12\catcode `\$12\catcode
  `\&12\catcode `\#12\catcode `\^12\catcode `\_12\catcode `\%12\relax}%
\providecommand \@@startlink[1]{}%
\providecommand \@@endlink[0]{}%
\providecommand \url  [0]{\begingroup\@sanitize@url \@url }%
\providecommand \@url [1]{\endgroup\@href {#1}{\urlprefix }}%
\providecommand \urlprefix  [0]{URL }%
\providecommand \Eprint [0]{\href }%
\providecommand \doibase [0]{http://dx.doi.org/}%
\providecommand \selectlanguage [0]{\@gobble}%
\providecommand \bibinfo  [0]{\@secondoftwo}%
\providecommand \bibfield  [0]{\@secondoftwo}%
\providecommand \translation [1]{[#1]}%
\providecommand \BibitemOpen [0]{}%
\providecommand \bibitemStop [0]{}%
\providecommand \bibitemNoStop [0]{.\EOS\space}%
\providecommand \EOS [0]{\spacefactor3000\relax}%
\providecommand \BibitemShut  [1]{\csname bibitem#1\endcsname}%
\let\auto@bib@innerbib\@empty
\bibitem [{\citenamefont {Ananthanarayan}\ \emph {et~al.}(2001)\citenamefont
  {Ananthanarayan}, \citenamefont {Colangelo}, \citenamefont {Gasser},\ and\
  \citenamefont {Leutwyler}}]{Ananthanarayan:2000ht}%
  \BibitemOpen
  \bibfield  {author} {\bibinfo {author} {\bibfnamefont {B.}~\bibnamefont
  {Ananthanarayan}}, \bibinfo {author} {\bibfnamefont {G.}~\bibnamefont
  {Colangelo}}, \bibinfo {author} {\bibfnamefont {J.}~\bibnamefont {Gasser}},
  and \bibinfo {author} {\bibfnamefont {H.}~\bibnamefont {Leutwyler}},\ }\href
  {\doibase 10.1016/S0370-1573(01)00009-6} {\bibfield  {journal} {\bibinfo
  {journal} {Phys. Rept.}\ }\textbf {\bibinfo {volume} {353}},\ \bibinfo
  {pages} {207} (\bibinfo {year} {2001})},\ \Eprint
  {http://arxiv.org/abs/hep-ph/0005297} {arXiv:hep-ph/0005297
  [hep-ph]}\BibitemShut {NoStop}%
\bibitem [{\citenamefont {Colangelo}\ \emph {et~al.}(2001)\citenamefont
  {Colangelo}, \citenamefont {Gasser},\ and\ \citenamefont
  {Leutwyler}}]{Colangelo:2001df}%
  \BibitemOpen
  \bibfield  {author} {\bibinfo {author} {\bibfnamefont {G.}~\bibnamefont
  {Colangelo}}, \bibinfo {author} {\bibfnamefont {J.}~\bibnamefont {Gasser}},
  and \bibinfo {author} {\bibfnamefont {H.}~\bibnamefont {Leutwyler}},\ }\href
  {\doibase 10.1016/S0550-3213(01)00147-X} {\bibfield  {journal} {\bibinfo
  {journal} {Nucl. Phys. B}\ }\textbf {\bibinfo {volume} {603}},\ \bibinfo
  {pages} {125} (\bibinfo {year} {2001})},\ \Eprint
  {http://arxiv.org/abs/hep-ph/0103088} {arXiv:hep-ph/0103088}\BibitemShut
  {NoStop}%
\bibitem [{\citenamefont {Caprini}\ \emph {et~al.}(2012)\citenamefont
  {Caprini}, \citenamefont {Colangelo},\ and\ \citenamefont
  {Leutwyler}}]{Caprini:2011ky}%
  \BibitemOpen
  \bibfield  {author} {\bibinfo {author} {\bibfnamefont {I.}~\bibnamefont
  {Caprini}}, \bibinfo {author} {\bibfnamefont {G.}~\bibnamefont {Colangelo}},
  and \bibinfo {author} {\bibfnamefont {H.}~\bibnamefont {Leutwyler}},\ }\href
  {\doibase 10.1140/epjc/s10052-012-1860-1} {\bibfield  {journal} {\bibinfo
  {journal} {Eur. Phys. J. C}\ }\textbf {\bibinfo {volume} {72}},\ \bibinfo
  {pages} {1860} (\bibinfo {year} {2012})},\ \Eprint
  {http://arxiv.org/abs/1111.7160} {arXiv:1111.7160 [hep-ph]}\BibitemShut
  {NoStop}%
\bibitem [{\citenamefont {Garcia-Martin}\ \emph
  {et~al.}(2011{\natexlab{a}})\citenamefont {Garcia-Martin}, \citenamefont
  {Kaminski}, \citenamefont {Pelaez}, \citenamefont {Ruiz~de Elvira},\ and\
  \citenamefont {Yndurain}}]{GarciaMartin:2011cn}%
  \BibitemOpen
  \bibfield  {author} {\bibinfo {author} {\bibfnamefont {R.}~\bibnamefont
  {Garcia-Martin}}, \bibinfo {author} {\bibfnamefont {R.}~\bibnamefont
  {Kaminski}}, \bibinfo {author} {\bibfnamefont {J.~R.}\ \bibnamefont
  {Pelaez}}, \bibinfo {author} {\bibfnamefont {J.}~\bibnamefont {Ruiz~de
  Elvira}}, and \bibinfo {author} {\bibfnamefont {F.~J.}\ \bibnamefont
  {Yndurain}},\ }\href {\doibase 10.1103/PhysRevD.83.074004} {\bibfield
  {journal} {\bibinfo  {journal} {Phys. Rev.}\ }\textbf {\bibinfo {volume}
  {D83}},\ \bibinfo {pages} {074004} (\bibinfo {year} {2011}{\natexlab{a}})},\
  \Eprint {http://arxiv.org/abs/1102.2183} {arXiv:1102.2183
  [hep-ph]}\BibitemShut {NoStop}%
\bibitem [{\citenamefont {Pel{\'a}ez}\ \emph {et~al.}(2025)\citenamefont
  {Pel{\'a}ez}, \citenamefont {Rab{\'a}n},\ and\ \citenamefont
  {de~Elvira}}]{Pelaez:2024uav}%
  \BibitemOpen
  \bibfield  {author} {\bibinfo {author} {\bibfnamefont {J.~R.}\ \bibnamefont
  {Pel{\'a}ez}}, \bibinfo {author} {\bibfnamefont {P.}~\bibnamefont
  {Rab{\'a}n}}, and \bibinfo {author} {\bibfnamefont {J.~R.}\ \bibnamefont
  {de~Elvira}},\ }\href {\doibase 10.1103/PhysRevD.111.074003} {\bibfield
  {journal} {\bibinfo  {journal} {Phys. Rev. D}\ }\textbf {\bibinfo {volume}
  {111}},\ \bibinfo {pages} {074003} (\bibinfo {year} {2025})},\ \Eprint
  {http://arxiv.org/abs/2412.15327} {arXiv:2412.15327 [hep-ph]}\BibitemShut
  {NoStop}%
\bibitem [{\citenamefont {Gasser}(2009)}]{Gasser:2009zz}%
  \BibitemOpen
  \bibfield  {author} {\bibinfo {author} {\bibfnamefont {J.}~\bibnamefont
  {Gasser}},\ }\href {\doibase 10.22323/1.069.0029} {\bibfield  {journal}
  {\bibinfo  {journal} {PoS}\ }\textbf {\bibinfo {volume} {EFT09}},\ \bibinfo
  {pages} {029} (\bibinfo {year} {2009})}\BibitemShut {NoStop}%
\bibitem [{\citenamefont {Hyams}\ \emph {et~al.}(1973)\citenamefont {Hyams}
  \emph {et~al.}}]{Hyams:1973zf}%
  \BibitemOpen
  \bibfield  {author} {\bibinfo {author} {\bibfnamefont {B.}~\bibnamefont
  {Hyams}}  \emph {et~al.},\ }\href {\doibase 10.1016/0550-3213(73)90618-4}
  {\bibfield  {journal} {\bibinfo  {journal} {Nucl. Phys. B}\ }\textbf
  {\bibinfo {volume} {64}},\ \bibinfo {pages} {134} (\bibinfo {year}
  {1973})}\BibitemShut {NoStop}%
\bibitem [{\citenamefont {Pislak}\ \emph {et~al.}(2001)\citenamefont {Pislak}
  \emph {et~al.}}]{BNL-E865:2001wfj}%
  \BibitemOpen
  \bibfield  {author} {\bibinfo {author} {\bibfnamefont {S.}~\bibnamefont
  {Pislak}}  \emph {et~al.} (\bibinfo {collaboration} {BNL-E865}),\ }\href
  {\doibase 10.1103/PhysRevLett.105.019901} {\bibfield  {journal} {\bibinfo
  {journal} {Phys. Rev. Lett.}\ }\textbf {\bibinfo {volume} {87}},\ \bibinfo
  {pages} {221801} (\bibinfo {year} {2001})},\ \bibinfo {note} {[Erratum:
  Phys.Rev.Lett. 105, 019901 (2010)]},\ \Eprint
  {http://arxiv.org/abs/hep-ex/0106071} {arXiv:hep-ex/0106071}\BibitemShut
  {NoStop}%
\bibitem [{\citenamefont {Batley}\ \emph {et~al.}(2006)\citenamefont {Batley}
  \emph {et~al.}}]{NA482:2005wht}%
  \BibitemOpen
  \bibfield  {author} {\bibinfo {author} {\bibfnamefont {J.~R.}\ \bibnamefont
  {Batley}}  \emph {et~al.} (\bibinfo {collaboration} {NA48/2}),\ }\href
  {\doibase 10.1016/j.physletb.2005.11.087} {\bibfield  {journal} {\bibinfo
  {journal} {Phys. Lett. B}\ }\textbf {\bibinfo {volume} {633}},\ \bibinfo
  {pages} {173} (\bibinfo {year} {2006})},\ \Eprint
  {http://arxiv.org/abs/hep-ex/0511056} {arXiv:hep-ex/0511056}\BibitemShut
  {NoStop}%
\bibitem [{\citenamefont {Batley}\ \emph {et~al.}(2008)\citenamefont {Batley}
  \emph {et~al.}}]{NA482:2007xvj}%
  \BibitemOpen
  \bibfield  {author} {\bibinfo {author} {\bibfnamefont {J.~R.}\ \bibnamefont
  {Batley}}  \emph {et~al.} (\bibinfo {collaboration} {NA48/2}),\ }\href
  {\doibase 10.1140/epjc/s10052-008-0547-0} {\bibfield  {journal} {\bibinfo
  {journal} {Eur. Phys. J. C}\ }\textbf {\bibinfo {volume} {54}},\ \bibinfo
  {pages} {411} (\bibinfo {year} {2008})}\BibitemShut {NoStop}%
\bibitem [{\citenamefont {Adeva}\ \emph {et~al.}(2011)\citenamefont {Adeva}
  \emph {et~al.}}]{Adeva:2011tc}%
  \BibitemOpen
  \bibfield  {author} {\bibinfo {author} {\bibfnamefont {B.}~\bibnamefont
  {Adeva}}  \emph {et~al.},\ }\href {\doibase 10.1016/j.physletb.2011.08.074}
  {\bibfield  {journal} {\bibinfo  {journal} {Phys. Lett. B}\ }\textbf
  {\bibinfo {volume} {704}},\ \bibinfo {pages} {24} (\bibinfo {year} {2011})},\
  \Eprint {http://arxiv.org/abs/1109.0569} {arXiv:1109.0569
  [hep-ex]}\BibitemShut {NoStop}%
\bibitem [{\citenamefont {Barberio}\ and\ \citenamefont
  {Was}(1994)}]{Barberio:1993qi}%
  \BibitemOpen
  \bibfield  {author} {\bibinfo {author} {\bibfnamefont {E.}~\bibnamefont
  {Barberio}} and \bibinfo {author} {\bibfnamefont {Z.}~\bibnamefont {Was}},\
  }\href {\doibase 10.1016/0010-4655(94)90074-4} {\bibfield  {journal}
  {\bibinfo  {journal} {Comput. Phys. Commun.}\ }\textbf {\bibinfo {volume}
  {79}},\ \bibinfo {pages} {291} (\bibinfo {year} {1994})}\BibitemShut
  {NoStop}%
\bibitem [{\citenamefont {Nanava}\ and\ \citenamefont
  {Was}(2007)}]{Nanava:2006vv}%
  \BibitemOpen
  \bibfield  {author} {\bibinfo {author} {\bibfnamefont {G.}~\bibnamefont
  {Nanava}} and \bibinfo {author} {\bibfnamefont {Z.}~\bibnamefont {Was}},\
  }\href {\doibase 10.1140/epjc/s10052-007-0316-5} {\bibfield  {journal}
  {\bibinfo  {journal} {Eur. Phys. J. C}\ }\textbf {\bibinfo {volume} {51}},\
  \bibinfo {pages} {569} (\bibinfo {year} {2007})},\ \Eprint
  {http://arxiv.org/abs/hep-ph/0607019} {arXiv:hep-ph/0607019}\BibitemShut
  {NoStop}%
\bibitem [{\citenamefont {Colangelo}\ \emph {et~al.}(2009)\citenamefont
  {Colangelo}, \citenamefont {Gasser},\ and\ \citenamefont
  {Rusetsky}}]{Colangelo:2008sm}%
  \BibitemOpen
  \bibfield  {author} {\bibinfo {author} {\bibfnamefont {G.}~\bibnamefont
  {Colangelo}}, \bibinfo {author} {\bibfnamefont {J.}~\bibnamefont {Gasser}},
  and \bibinfo {author} {\bibfnamefont {A.}~\bibnamefont {Rusetsky}},\ }\href
  {\doibase 10.1140/epjc/s10052-008-0818-9} {\bibfield  {journal} {\bibinfo
  {journal} {Eur. Phys. J. C}\ }\textbf {\bibinfo {volume} {59}},\ \bibinfo
  {pages} {777} (\bibinfo {year} {2009})},\ \Eprint
  {http://arxiv.org/abs/0811.0775} {arXiv:0811.0775 [hep-ph]}\BibitemShut
  {NoStop}%
\bibitem [{\citenamefont {Bernard}\ \emph {et~al.}(2013)\citenamefont
  {Bernard}, \citenamefont {Descotes-Genon},\ and\ \citenamefont
  {Knecht}}]{Bernard:2013faa}%
  \BibitemOpen
  \bibfield  {author} {\bibinfo {author} {\bibfnamefont {V.}~\bibnamefont
  {Bernard}}, \bibinfo {author} {\bibfnamefont {S.}~\bibnamefont
  {Descotes-Genon}}, and \bibinfo {author} {\bibfnamefont {M.}~\bibnamefont
  {Knecht}},\ }\href {\doibase 10.1140/epjc/s10052-013-2478-7} {\bibfield
  {journal} {\bibinfo  {journal} {Eur. Phys. J. C}\ }\textbf {\bibinfo {volume}
  {73}},\ \bibinfo {pages} {2478} (\bibinfo {year} {2013})},\ \Eprint
  {http://arxiv.org/abs/1305.3843} {arXiv:1305.3843 [hep-ph]}\BibitemShut
  {NoStop}%
\bibitem [{\citenamefont {Bernard}\ \emph {et~al.}(2015)\citenamefont
  {Bernard}, \citenamefont {Descotes-Genon},\ and\ \citenamefont
  {Knecht}}]{Bernard:2015vqa}%
  \BibitemOpen
  \bibfield  {author} {\bibinfo {author} {\bibfnamefont {V.}~\bibnamefont
  {Bernard}}, \bibinfo {author} {\bibfnamefont {S.}~\bibnamefont
  {Descotes-Genon}}, and \bibinfo {author} {\bibfnamefont {M.}~\bibnamefont
  {Knecht}},\ }\href {\doibase 10.1140/epjc/s10052-015-3359-z} {\bibfield
  {journal} {\bibinfo  {journal} {Eur. Phys. J. C}\ }\textbf {\bibinfo {volume}
  {75}},\ \bibinfo {pages} {145} (\bibinfo {year} {2015})},\ \Eprint
  {http://arxiv.org/abs/1501.07102} {arXiv:1501.07102 [hep-ph]}\BibitemShut
  {NoStop}%
\bibitem [{\citenamefont {Gasser}\ and\ \citenamefont
  {Leutwyler}(1984)}]{Gasser:1983yg}%
  \BibitemOpen
  \bibfield  {author} {\bibinfo {author} {\bibfnamefont {J.}~\bibnamefont
  {Gasser}} and \bibinfo {author} {\bibfnamefont {H.}~\bibnamefont
  {Leutwyler}},\ }\href {\doibase 10.1016/0003-4916(84)90242-2} {\bibfield
  {journal} {\bibinfo  {journal} {Annals Phys.}\ }\textbf {\bibinfo {volume}
  {158}},\ \bibinfo {pages} {142} (\bibinfo {year} {1984})}\BibitemShut
  {NoStop}%
\bibitem [{\citenamefont {Gasser}\ and\ \citenamefont
  {Leutwyler}(1985)}]{Gasser:1984gg}%
  \BibitemOpen
  \bibfield  {author} {\bibinfo {author} {\bibfnamefont {J.}~\bibnamefont
  {Gasser}} and \bibinfo {author} {\bibfnamefont {H.}~\bibnamefont
  {Leutwyler}},\ }\href {\doibase 10.1016/0550-3213(85)90492-4} {\bibfield
  {journal} {\bibinfo  {journal} {Nucl. Phys. B}\ }\textbf {\bibinfo {volume}
  {250}},\ \bibinfo {pages} {465} (\bibinfo {year} {1985})}\BibitemShut
  {NoStop}%
\bibitem [{\citenamefont {Gasser}\ \emph {et~al.}(2003)\citenamefont {Gasser},
  \citenamefont {Rusetsky},\ and\ \citenamefont {Scimemi}}]{Gasser:2003hk}%
  \BibitemOpen
  \bibfield  {author} {\bibinfo {author} {\bibfnamefont {J.}~\bibnamefont
  {Gasser}}, \bibinfo {author} {\bibfnamefont {A.}~\bibnamefont {Rusetsky}},
  and \bibinfo {author} {\bibfnamefont {I.}~\bibnamefont {Scimemi}},\ }\href
  {\doibase 10.1140/epjc/s2003-01383-1} {\bibfield  {journal} {\bibinfo
  {journal} {Eur. Phys. J. C}\ }\textbf {\bibinfo {volume} {32}},\ \bibinfo
  {pages} {97} (\bibinfo {year} {2003})},\ \Eprint
  {http://arxiv.org/abs/hep-ph/0305260} {arXiv:hep-ph/0305260}\BibitemShut
  {NoStop}%
\bibitem [{\citenamefont {Colangelo}\ \emph {et~al.}(2018)\citenamefont
  {Colangelo}, \citenamefont {Lanz}, \citenamefont {Leutwyler},\ and\
  \citenamefont {Passemar}}]{Colangelo:2018jxw}%
  \BibitemOpen
  \bibfield  {author} {\bibinfo {author} {\bibfnamefont {G.}~\bibnamefont
  {Colangelo}}, \bibinfo {author} {\bibfnamefont {S.}~\bibnamefont {Lanz}},
  \bibinfo {author} {\bibfnamefont {H.}~\bibnamefont {Leutwyler}}, and \bibinfo
  {author} {\bibfnamefont {E.}~\bibnamefont {Passemar}},\ }\href {\doibase
  10.1140/epjc/s10052-018-6377-9} {\bibfield  {journal} {\bibinfo  {journal}
  {Eur. Phys. J. C}\ }\textbf {\bibinfo {volume} {78}},\ \bibinfo {pages} {947}
  (\bibinfo {year} {2018})},\ \Eprint {http://arxiv.org/abs/1807.11937}
  {arXiv:1807.11937 [hep-ph]}\BibitemShut {NoStop}%
\bibitem [{\citenamefont {Colangelo}\ \emph {et~al.}(2017)\citenamefont
  {Colangelo}, \citenamefont {Lanz}, \citenamefont {Leutwyler},\ and\
  \citenamefont {Passemar}}]{Colangelo:2016jmc}%
  \BibitemOpen
  \bibfield  {author} {\bibinfo {author} {\bibfnamefont {G.}~\bibnamefont
  {Colangelo}}, \bibinfo {author} {\bibfnamefont {S.}~\bibnamefont {Lanz}},
  \bibinfo {author} {\bibfnamefont {H.}~\bibnamefont {Leutwyler}}, and \bibinfo
  {author} {\bibfnamefont {E.}~\bibnamefont {Passemar}},\ }\href {\doibase
  10.1103/PhysRevLett.118.022001} {\bibfield  {journal} {\bibinfo  {journal}
  {Phys. Rev. Lett.}\ }\textbf {\bibinfo {volume} {118}},\ \bibinfo {pages}
  {022001} (\bibinfo {year} {2017})},\ \Eprint
  {http://arxiv.org/abs/1610.03494} {arXiv:1610.03494 [hep-ph]}\BibitemShut
  {NoStop}%
\bibitem [{\citenamefont {Khuri}\ and\ \citenamefont
  {Treiman}(1960)}]{Khuri:1960zz}%
  \BibitemOpen
  \bibfield  {author} {\bibinfo {author} {\bibfnamefont {N.~N.}\ \bibnamefont
  {Khuri}} and \bibinfo {author} {\bibfnamefont {S.~B.}\ \bibnamefont
  {Treiman}},\ }\href {\doibase 10.1103/PhysRev.119.1115} {\bibfield  {journal}
  {\bibinfo  {journal} {Phys. Rev.}\ }\textbf {\bibinfo {volume} {119}},\
  \bibinfo {pages} {1115} (\bibinfo {year} {1960})}\BibitemShut {NoStop}%
\bibitem [{\citenamefont {Kambor}\ \emph {et~al.}(1996)\citenamefont {Kambor},
  \citenamefont {Wiesendanger},\ and\ \citenamefont {Wyler}}]{Kambor:1995yc}%
  \BibitemOpen
  \bibfield  {author} {\bibinfo {author} {\bibfnamefont {J.}~\bibnamefont
  {Kambor}}, \bibinfo {author} {\bibfnamefont {C.}~\bibnamefont
  {Wiesendanger}}, and \bibinfo {author} {\bibfnamefont {D.}~\bibnamefont
  {Wyler}},\ }\href {\doibase 10.1016/0550-3213(95)00676-1} {\bibfield
  {journal} {\bibinfo  {journal} {Nucl. Phys. B}\ }\textbf {\bibinfo {volume}
  {465}},\ \bibinfo {pages} {215} (\bibinfo {year} {1996})},\ \Eprint
  {http://arxiv.org/abs/hep-ph/9509374} {arXiv:hep-ph/9509374}\BibitemShut
  {NoStop}%
\bibitem [{\citenamefont {Anisovich}\ and\ \citenamefont
  {Leutwyler}(1996)}]{Anisovich:1996tx}%
  \BibitemOpen
  \bibfield  {author} {\bibinfo {author} {\bibfnamefont {A.~V.}\ \bibnamefont
  {Anisovich}} and \bibinfo {author} {\bibfnamefont {H.}~\bibnamefont
  {Leutwyler}},\ }\href {\doibase 10.1016/0370-2693(96)00192-X} {\bibfield
  {journal} {\bibinfo  {journal} {Phys. Lett. B}\ }\textbf {\bibinfo {volume}
  {375}},\ \bibinfo {pages} {335} (\bibinfo {year} {1996})},\ \Eprint
  {http://arxiv.org/abs/hep-ph/9601237} {arXiv:hep-ph/9601237}\BibitemShut
  {NoStop}%
\bibitem [{\citenamefont {Gasser}\ and\ \citenamefont
  {Rusetsky}(2018)}]{Gasser:2018qtg}%
  \BibitemOpen
  \bibfield  {author} {\bibinfo {author} {\bibfnamefont {J.}~\bibnamefont
  {Gasser}} and \bibinfo {author} {\bibfnamefont {A.}~\bibnamefont
  {Rusetsky}},\ }\href {\doibase 10.1140/epjc/s10052-018-6378-8} {\bibfield
  {journal} {\bibinfo  {journal} {Eur. Phys. J. C}\ }\textbf {\bibinfo {volume}
  {78}},\ \bibinfo {pages} {906} (\bibinfo {year} {2018})},\ \Eprint
  {http://arxiv.org/abs/1809.06399} {arXiv:1809.06399 [hep-ph]}\BibitemShut
  {NoStop}%
\bibitem [{\citenamefont {Knecht}\ and\ \citenamefont
  {Urech}(1998)}]{Knecht:1997jw}%
  \BibitemOpen
  \bibfield  {author} {\bibinfo {author} {\bibfnamefont {M.}~\bibnamefont
  {Knecht}} and \bibinfo {author} {\bibfnamefont {R.}~\bibnamefont {Urech}},\
  }\href {\doibase 10.1016/S0550-3213(98)00044-3} {\bibfield  {journal}
  {\bibinfo  {journal} {Nucl. Phys. B}\ }\textbf {\bibinfo {volume} {519}},\
  \bibinfo {pages} {329} (\bibinfo {year} {1998})},\ \Eprint
  {http://arxiv.org/abs/hep-ph/9709348} {arXiv:hep-ph/9709348}\BibitemShut
  {NoStop}%
\bibitem [{\citenamefont {Knecht}\ and\ \citenamefont
  {Nehme}(2002)}]{Knecht:2002gz}%
  \BibitemOpen
  \bibfield  {author} {\bibinfo {author} {\bibfnamefont {M.}~\bibnamefont
  {Knecht}} and \bibinfo {author} {\bibfnamefont {A.}~\bibnamefont {Nehme}},\
  }\href {\doibase 10.1016/S0370-2693(02)01499-5} {\bibfield  {journal}
  {\bibinfo  {journal} {Phys. Lett.}\ }\textbf {\bibinfo {volume} {B532}},\
  \bibinfo {pages} {55} (\bibinfo {year} {2002})},\ \Eprint
  {http://arxiv.org/abs/hep-ph/0201033} {arXiv:hep-ph/0201033
  [hep-ph]}\BibitemShut {NoStop}%
\bibitem [{\citenamefont {Gasser}\ and\ \citenamefont
  {Leutwyler}(1982)}]{Gasser:1982ap}%
  \BibitemOpen
  \bibfield  {author} {\bibinfo {author} {\bibfnamefont {J.}~\bibnamefont
  {Gasser}} and \bibinfo {author} {\bibfnamefont {H.}~\bibnamefont
  {Leutwyler}},\ }\href {\doibase 10.1016/0370-1573(82)90035-7} {\bibfield
  {journal} {\bibinfo  {journal} {Phys. Rept.}\ }\textbf {\bibinfo {volume}
  {87}},\ \bibinfo {pages} {77} (\bibinfo {year} {1982})}\BibitemShut {NoStop}%
\bibitem [{\citenamefont {Buettiker}\ \emph {et~al.}(2004)\citenamefont
  {Buettiker}, \citenamefont {Descotes-Genon},\ and\ \citenamefont
  {Moussallam}}]{Buettiker:2003pp}%
  \BibitemOpen
  \bibfield  {author} {\bibinfo {author} {\bibfnamefont {P.}~\bibnamefont
  {Buettiker}}, \bibinfo {author} {\bibfnamefont {S.}~\bibnamefont
  {Descotes-Genon}}, and \bibinfo {author} {\bibfnamefont {B.}~\bibnamefont
  {Moussallam}},\ }\href {\doibase 10.1140/epjc/s2004-01591-1} {\bibfield
  {journal} {\bibinfo  {journal} {Eur. Phys. J. C}\ }\textbf {\bibinfo {volume}
  {33}},\ \bibinfo {pages} {409} (\bibinfo {year} {2004})},\ \Eprint
  {http://arxiv.org/abs/hep-ph/0310283} {arXiv:hep-ph/0310283}\BibitemShut
  {NoStop}%
\bibitem [{\citenamefont {Hoferichter}\ \emph {et~al.}(2016)\citenamefont
  {Hoferichter}, \citenamefont {Ruiz~de Elvira}, \citenamefont {Kubis},\ and\
  \citenamefont {Meißner}}]{Hoferichter:2015hva}%
  \BibitemOpen
  \bibfield  {author} {\bibinfo {author} {\bibfnamefont {M.}~\bibnamefont
  {Hoferichter}}, \bibinfo {author} {\bibfnamefont {J.}~\bibnamefont {Ruiz~de
  Elvira}}, \bibinfo {author} {\bibfnamefont {B.}~\bibnamefont {Kubis}}, and
  \bibinfo {author} {\bibfnamefont {U.-G.}\ \bibnamefont {Meißner}},\ }\href
  {\doibase 10.1016/j.physrep.2016.02.002} {\bibfield  {journal} {\bibinfo
  {journal} {Phys. Rept.}\ }\textbf {\bibinfo {volume} {625}},\ \bibinfo
  {pages} {1} (\bibinfo {year} {2016})},\ \Eprint
  {http://arxiv.org/abs/1510.06039} {arXiv:1510.06039 [hep-ph]}\BibitemShut
  {NoStop}%
\bibitem [{\citenamefont {Navas}\ \emph {et~al.}(2024)\citenamefont {Navas}
  \emph {et~al.}}]{ParticleDataGroup:2024cfk}%
  \BibitemOpen
  \bibfield  {author} {\bibinfo {author} {\bibfnamefont {S.}~\bibnamefont
  {Navas}}  \emph {et~al.} (\bibinfo {collaboration} {Particle Data Group}),\
  }\href {\doibase 10.1103/PhysRevD.110.030001} {\bibfield  {journal} {\bibinfo
   {journal} {Phys. Rev. D}\ }\textbf {\bibinfo {volume} {110}},\ \bibinfo
  {pages} {030001} (\bibinfo {year} {2024})}\BibitemShut {NoStop}%
\bibitem [{\citenamefont {Aoki}\ \emph {et~al.}(2024)\citenamefont {Aoki} \emph
  {et~al.}}]{FlavourLatticeAveragingGroupFLAG:2024oxs}%
  \BibitemOpen
  \bibfield  {author} {\bibinfo {author} {\bibfnamefont {Y.}~\bibnamefont
  {Aoki}}  \emph {et~al.} (\bibinfo {collaboration} {Flavour Lattice Averaging
  Group (FLAG)}),\ }\href@noop {} {\  (\bibinfo {year} {2024})},\ \Eprint
  {http://arxiv.org/abs/2411.04268} {arXiv:2411.04268 [hep-lat]}\BibitemShut
  {NoStop}%
\bibitem [{\citenamefont {Bijnens}\ and\ \citenamefont
  {Ecker}(2014)}]{Bijnens:2014lea}%
  \BibitemOpen
  \bibfield  {author} {\bibinfo {author} {\bibfnamefont {J.}~\bibnamefont
  {Bijnens}} and \bibinfo {author} {\bibfnamefont {G.}~\bibnamefont {Ecker}},\
  }\href {\doibase 10.1146/annurev-nucl-102313-025528} {\bibfield  {journal}
  {\bibinfo  {journal} {Ann. Rev. Nucl. Part. Sci.}\ }\textbf {\bibinfo
  {volume} {64}},\ \bibinfo {pages} {149} (\bibinfo {year} {2014})},\ \Eprint
  {http://arxiv.org/abs/1405.6488} {arXiv:1405.6488 [hep-ph]}\BibitemShut
  {NoStop}%
\bibitem [{\citenamefont {Haefeli}\ \emph {et~al.}(2008)\citenamefont
  {Haefeli}, \citenamefont {Ivanov},\ and\ \citenamefont
  {Schmid}}]{Haefeli:2007ey}%
  \BibitemOpen
  \bibfield  {author} {\bibinfo {author} {\bibfnamefont {C.}~\bibnamefont
  {Haefeli}}, \bibinfo {author} {\bibfnamefont {M.~A.}\ \bibnamefont {Ivanov}},
  and \bibinfo {author} {\bibfnamefont {M.}~\bibnamefont {Schmid}},\ }\href
  {\doibase 10.1140/epjc/s10052-007-0493-2} {\bibfield  {journal} {\bibinfo
  {journal} {Eur. Phys. J. C}\ }\textbf {\bibinfo {volume} {53}},\ \bibinfo
  {pages} {549} (\bibinfo {year} {2008})},\ \Eprint
  {http://arxiv.org/abs/0710.5432} {arXiv:0710.5432 [hep-ph]}\BibitemShut
  {NoStop}%
\bibitem [{\citenamefont {Caprini}\ \emph {et~al.}(2006)\citenamefont
  {Caprini}, \citenamefont {Colangelo},\ and\ \citenamefont
  {Leutwyler}}]{Caprini:2005zr}%
  \BibitemOpen
  \bibfield  {author} {\bibinfo {author} {\bibfnamefont {I.}~\bibnamefont
  {Caprini}}, \bibinfo {author} {\bibfnamefont {G.}~\bibnamefont {Colangelo}},
  and \bibinfo {author} {\bibfnamefont {H.}~\bibnamefont {Leutwyler}},\ }\href
  {\doibase 10.1103/PhysRevLett.96.132001} {\bibfield  {journal} {\bibinfo
  {journal} {Phys. Rev. Lett.}\ }\textbf {\bibinfo {volume} {96}},\ \bibinfo
  {pages} {132001} (\bibinfo {year} {2006})},\ \Eprint
  {http://arxiv.org/abs/hep-ph/0512364} {arXiv:hep-ph/0512364}\BibitemShut
  {NoStop}%
\bibitem [{\citenamefont {Garcia-Martin}\ \emph
  {et~al.}(2011{\natexlab{b}})\citenamefont {Garcia-Martin}, \citenamefont
  {Kaminski}, \citenamefont {Pelaez},\ and\ \citenamefont {Ruiz~de
  Elvira}}]{Garcia-Martin:2011nna}%
  \BibitemOpen
  \bibfield  {author} {\bibinfo {author} {\bibfnamefont {R.}~\bibnamefont
  {Garcia-Martin}}, \bibinfo {author} {\bibfnamefont {R.}~\bibnamefont
  {Kaminski}}, \bibinfo {author} {\bibfnamefont {J.~R.}\ \bibnamefont
  {Pelaez}}, and \bibinfo {author} {\bibfnamefont {J.}~\bibnamefont {Ruiz~de
  Elvira}},\ }\href {\doibase 10.1103/PhysRevLett.107.072001} {\bibfield
  {journal} {\bibinfo  {journal} {Phys. Rev. Lett.}\ }\textbf {\bibinfo
  {volume} {107}},\ \bibinfo {pages} {072001} (\bibinfo {year}
  {2011}{\natexlab{b}})},\ \Eprint {http://arxiv.org/abs/1107.1635}
  {arXiv:1107.1635 [hep-ph]}\BibitemShut {NoStop}%
\bibitem [{\citenamefont {Pelaez}\ \emph {et~al.}(2023)\citenamefont {Pelaez},
  \citenamefont {Rodas},\ and\ \citenamefont {de~Elvira}}]{Pelaez:2022qby}%
  \BibitemOpen
  \bibfield  {author} {\bibinfo {author} {\bibfnamefont {J.~R.}\ \bibnamefont
  {Pelaez}}, \bibinfo {author} {\bibfnamefont {A.}~\bibnamefont {Rodas}}, and
  \bibinfo {author} {\bibfnamefont {J.~R.}\ \bibnamefont {de~Elvira}},\ }\href
  {\doibase 10.1103/PhysRevLett.130.051902} {\bibfield  {journal} {\bibinfo
  {journal} {Phys. Rev. Lett.}\ }\textbf {\bibinfo {volume} {130}},\ \bibinfo
  {pages} {051902} (\bibinfo {year} {2023})},\ \bibinfo {note} {[Erratum:
  Phys.Rev.Lett. 132, 239901 (2024)]},\ \Eprint
  {http://arxiv.org/abs/2206.14822} {arXiv:2206.14822 [hep-ph]}\BibitemShut
  {NoStop}%
\bibitem [{\citenamefont {Hoferichter}\ \emph {et~al.}(2024)\citenamefont
  {Hoferichter}, \citenamefont {de~Elvira}, \citenamefont {Kubis},\ and\
  \citenamefont {Mei{\ss}ner}}]{Hoferichter:2023mgy}%
  \BibitemOpen
  \bibfield  {author} {\bibinfo {author} {\bibfnamefont {M.}~\bibnamefont
  {Hoferichter}}, \bibinfo {author} {\bibfnamefont {J.~R.}\ \bibnamefont
  {de~Elvira}}, \bibinfo {author} {\bibfnamefont {B.}~\bibnamefont {Kubis}},
  and \bibinfo {author} {\bibfnamefont {U.-G.}\ \bibnamefont {Mei{\ss}ner}},\
  }\href {\doibase 10.1016/j.physletb.2024.138698} {\bibfield  {journal}
  {\bibinfo  {journal} {Phys. Lett. B}\ }\textbf {\bibinfo {volume} {853}},\
  \bibinfo {pages} {138698} (\bibinfo {year} {2024})},\ \Eprint
  {http://arxiv.org/abs/2312.15015} {arXiv:2312.15015 [hep-ph]}\BibitemShut
  {NoStop}%
\bibitem [{\citenamefont {Colangelo}\ \emph {et~al.}(2025)\citenamefont
  {Colangelo}, \citenamefont {Cottini}, \citenamefont {Hoferichter},\ and\
  \citenamefont {Holz}}]{Colangelo:2025iad}%
  \BibitemOpen
  \bibfield  {author} {\bibinfo {author} {\bibfnamefont {G.}~\bibnamefont
  {Colangelo}}, \bibinfo {author} {\bibfnamefont {M.}~\bibnamefont {Cottini}},
  \bibinfo {author} {\bibfnamefont {M.}~\bibnamefont {Hoferichter}}, and
  \bibinfo {author} {\bibfnamefont {S.}~\bibnamefont {Holz}},\ }\href@noop {}
  {\  (\bibinfo {year} {2025})},\ \Eprint {http://arxiv.org/abs/2510.26871}
  {arXiv:2510.26871 [hep-ph]}\BibitemShut {NoStop}%
\end{thebibliography}%

\end{document}